\documentclass[11pt]{article}
 \usepackage{titlesec}
\usepackage{hyperref}
\usepackage{graphicx}
\usepackage{stackrel}
\usepackage[T1]{fontenc}
\usepackage{lmodern}

\usepackage{blindtext}
\usepackage{subfiles} 

\titleclass{\subsubsubsection}{straight}[\subsection]

\usepackage[utf8]{inputenc}
\usepackage{amsmath,amsfonts,amssymb,stackengine,graphicx}

\usepackage[utf8]{inputenc}
\newif\iffigs\figstrue

\usepackage{cite}
\usepackage[title]{appendix}
\usepackage{epsfig,latexsym}
\usepackage{hyperref}
\usepackage{amsmath}
\usepackage{verbatim}
\usepackage{color}
\usepackage{mathrsfs}
\usepackage{slashed}
\usepackage{amssymb}

\def\sla{\raise.15ex\hbox{$/$}\kern-.57em}

\DeclareMathAlphabet{\mathpzc}{OT1}{pzc}{m}{it}

 \csname
@addtoreset\endcsname{equation}{section}

\def\gz0{\gamma^{0}}

\def\sign{\rm sign}

 \def\det{{\rm det\,}}

\def\beq{\begin{equation}}
\def\eeq{\end{equation}}
\def\be{\begin{equation}}
\def\ee{\end{equation}}
\def\bea{\begin{eqnarray}}
\def\eea{\end{eqnarray}}
\def\bec{\begin{center}}
\def\ec{\end{center}}
\def\beal{\begin{align}}
\def\enal{\end{align}}

\def\12{\frac{1}{2}}

\def\pr{\partial}

\newcounter{subsubsubsection}[subsubsection]
\renewcommand\thesubsubsubsection{\thesubsubsection.\arabic{subsubsubsection}}

\titleformat{\subsubsubsection}
  {\normalfont\normalsize\bfseries}{\thesubsubsubsection}{1em}{}
\titlespacing*{\subsubsubsection}
{0pt}{3.25ex plus 1ex minus .2ex}{1.5ex plus .2ex}

\makeatletter
\renewcommand\paragraph{\@startsection{paragraph}{5}{\z@}%
  {3.25ex \@plus1ex \@minus.2ex}%
  {-1em}%
  {\normalfont\normalsize\bfseries}}
\renewcommand\subparagraph{\@startsection{subparagraph}{6}{\parindent}%
  {3.25ex \@plus1ex \@minus .2ex}%
  {-1em}%
  {\normalfont\normalsize\bfseries}}
\def\toclevel@subsubsubsection{4}
\def\toclevel@paragraph{5}
\def\toclevel@paragraph{6}
\def\l@subsubsubsection{\@dottedtocline{4}{7em}{4em}}
\def\l@paragraph{\@dottedtocline{5}{10em}{5em}}
\def\l@subparagraph{\@dottedtocline{6}{14em}{6em}}
\makeatother

\usepackage{makeidx}
\makeindex

\begin{document}

\begin{flushright}
{\today}
\end{flushright}

\vspace{10pt}

\begin{center}


{\Large\sc On Cosmologies and Vacua Driven by Tension and Curvatures}\vskip 12pt


\vspace{15pt}
{\sc J.~Mourad${}^{\; a}$  \ and \ A.~Sagnotti${}^{\; b}$\\[15pt]

${}^a$\sl\small APC, UMR 7164-CNRS, Universit\'e   Paris Cit\'e  \\
10 rue Alice Domon et L\'eonie Duquet \\75205 Paris Cedex 13 \ FRANCE
\\ e-mail: {\small \it
mourad@apc.univ-paris7.fr}\vspace{10
pt}

{${}^b$\sl\small
Scuola Normale Superiore and INFN\\
Piazza dei Cavalieri, 7\\ 56126 Pisa \ ITALY \\
e-mail: {\small \it sagnotti@sns.it}}\vspace{10pt}
}

\vspace{15pt} {\sc\large Abstract}\end{center}
\vskip 8pt
We investigate the effects of the exponential potentials typical of non-supersymmetric strings in cosmologies whose spatial and internal slices are maximally symmetric spaces with curvatures labeled by a pair of integers $k$ and $k'$ ($=\pm 1$). We classify the solutions according to their singularity structure and asymptotic behavior and present a semi-quantitative picture of the generic dynamics in the physically most relevant cases with flat spatial slices. The analysis relies on exact solutions emerging when one of the effects dominates, special solutions arising when two or more effects are comparable, scaling asymptotics, and some numerical tests. 
\setcounter{page}{1}

\pagebreak

\tableofcontents

\pagebreak
\baselineskip=20pt
\section{\sc  Introduction}\label{sec:intro}

The tools currently available in String Theory~\cite{stringtheory} encounter severe conceptual and technical difficulties in the absence of supersymmetry. Yet, in ten dimensions, in addition to the five well-known supersymmetric models, there are other string theories that satisfy all known consistency conditions, where supersymmetry is broken or absent~\cite{nonsusytach1,nonsusytach2,so1616,u32,usp32}.  For a recent review of different realizations of supersymmetry breaking in String Theory and a more complete list of references, see~\cite{dms_review}. The spectra of non-supersymmetric strings generally contain tachyons, and to date we still lack general ways to deal with them. However, recently uncovered duality links open up new perspectives~\cite{recentdualities} and could eventually shed light on their non-perturbative features. 

For the time being, the low-energy effective field theory continues to provide some interesting indications, and in this respect the three non-tachyonic non-supersymmetric models~\cite{u32,usp32,so1616} appear to be a natural point of attack. The first of them~\cite{so1616} is the $SO(16)\times SO(16)$ heterotic model, whose vacuum is first affected by supersymmetry breaking at the torus level, while the others~\cite{u32,usp32} are orientifolds~\cite{orientifolds}, whose vacua are first affected at the (projective-)disk level. The last model is particularly intriguing, since supersymmetry is not absent in it, but it is non-linearly realized~\cite{nonlinear_susy}. It is the simplest manifestation of a general phenomenon, usually termed ``brane supersymmetry breaking''~\cite{bsb}.

The three different setups entail the same type of leading effect, the emergence of exponential potentials
\beq
\Delta{\cal S} \ = \ - \ T \, \int d^{10} x \,\sqrt{-G} \ e^{\gamma_s\,\phi} 
\eeq
in the string frame, with $\gamma_s=0$ or $\gamma_s=-1$, where $T$ is a positive one-loop vacuum energy in the heterotic case and an overall positive brane-orientifold tension, with different values for the two orientifolds. Equivalently
\beq
\Delta{\cal S} \ = \ - \ T \, \int d^{10} x \, \sqrt{-g} \ e^{\gamma\,\phi}
\eeq
in the Einstein frame, with $\gamma=\frac{5}{2}$ or $\gamma=\frac{3}{2}$, for closed strings or orientifolds. The critical value $\gamma_c=\frac{4 \sqrt{D-1}}{D-2}$, which in ten dimensions coincides with the orientifold value $\frac{3}{2}$, will play an important role in the subsequent discussion, where we often leave $D$ unspecified. In the heterotic model of~\cite{so1616}, $T$ originates from a torus contribution, while in the two orientifold models of~\cite{u32,usp32} it reflects the overall tension of branes and orientifolds present in the vacuum, but for the rest the two types of expressions are very similar. 

The isotropic cosmologies with flat spatial slices driven by $\Delta {\cal S}$, which were first explored in~\cite{dm_vacuum} and then in~\cite{russo}, manifest a similar behavior in the three models, and the string coupling never overcomes a finite upper bound during the cosmological evolution. This peculiarity was termed  in~\cite{climbing} a ``climbing'' behavior, since the dilaton emerges from the initial singularity and climbs the potential up to a turning point, in contrast to solutions where it emerges from the top and descends all along. The descending option becomes available only for $\gamma< \gamma_c$. Hence, the two ten-dimensional orientifold models, for which $\gamma$ is precisely $\gamma_c$, lie at the transition between the two regimes.

In non-critical strings~\cite{noncritical,lindil1,lindil2}, there is another special value for $\gamma$,
\beq
\gamma_0\ = \ \frac{4}{D-2} \ ,
\eeq
and, in this more general context, the Einstein-frame corrections are accompanied by factors of $e^{\left(\gamma_0+n\right)\phi}$ where $n$ is an integer. $\gamma_0$ plays an important role in the presence of maximally symmetric curved spatial and internal slices, which are the main focus of this work. Denoting by $k$ and $k'$ the spatial and internal curvatures, we discuss general exact solutions that emerge when only one of the three ingredients $(T,k,k')$ is present, together with special exact solutions that emerge when two or more effects are comparable. In these latter cases $\gamma_0$, rather than $\gamma_c$, separates different regimes. The solutions we discuss contain zero, one, or two singularities. Overall, positive curvature and negative tension tend to drive the system toward collapse, whereas negative curvature and positive tension favor expansion.
When only the internal curvature is present, the asymptotics of these cosmological models is typically free Kasner at the two ends for $k' \geq 0$ (with a finite past and a finite future), while Milne-like options emerge for $k'<0$. Similar results are obtained if only the spatial curvature $k$ is present. For all special solutions, stability is a central concern, since they rely on fine-tuned initial conditions. We thus examine their stability properties, and these ingredients are then used to study generic cases for which exact solutions are not available. 
Nevertheless, analytical considerations and numerical tests combine to provide a semi-quantitative picture of the available options, linking early-time behaviors captured by simple exact solutions to their late-time asymptotics.

The results collected in this paper can also describe compactifications with internal intervals. 

The correspondence simply amounts to reversing the signs of the basic ingredients, so that a cosmological solution with $(T,k,k')$ maps to a compactification with $(-T,-k,-k')$. This correspondence links the present results to the analysis in~\cite{mrs1}.

The remainder of this paper is organized as follows. In Section~\ref{sec:harmonic} we characterize the relevant class of metric tensors and the corresponding equations in the presence of tension and curvatures, but in the absence of fluxes. We mainly focus on the harmonic gauge, which leads to several classes of exact solutions and is very convenient to compare them in regimes where one of the three driving inputs, $(T,k,k')$, dominates over the others. However, we also discuss other gauge choices, primarily that based on cosmic time. We conclude the section by outlining the type of stability analysis used throughout the paper, which is tailored to the dynamical-system viewpoint rather than to gravity itself. In Section~\ref{sec:t0curv0} we briefly review the important class of Kasner solutions that emerge in the absence of tension or curvatures, following the treatment in~\cite{ms21_susy}. These settings dominate the cosmological dynamics at early or late times in most cases. Section~\ref{sec:t0curvnot0} is devoted to cosmologies whose dynamics is dominated by one of the curvatures, be it $k'$ or $k$. There we also discuss special exact solutions that emerge when their effects are comparable, extending the results in~\cite{mrs1}, and address their stability. In Section~\ref{sec:tnot0curv0} we discuss in detail cosmologies driven by the tension $T$ alone, extending to general values of $D$ the treatment in~\cite{ms21_nonsusy} and providing a detailed classification of possible regimes. Section~\ref{sec:Tensionkp} is devoted to special solutions that emerge when the tension $T$ and one of the curvatures have comparable effects, and to their stability. There is some overlap with~\cite{mrs1}, where a special solution of this type was described in the context of interval compactifications.  In Section~\ref{sec:Tkkp} we discuss special solutions where all three ingredients $(T,k,k')$ have comparable effects and elaborate on their stability.
Section 8 explores approximate treatments of cases where two of the three ingredients are present but one dominates the other. In Section~\ref{sec:log_asympt}, working in the cosmic gauge, we explore asymptotic limits and new types of attractor solutions. In the Conclusions, we summarize the semi-quantitative picture of the general case that emerges from the preceding analytical results and is supported by numerical tests.
The paper ends with appendices containing tables that summarize the wide variety of general and special solutions, together with the relation between cosmologies and compactifications of this type. More details on the classification of Section~\ref{sec:tnot0curv0} can also be found there.

\section{\sc  Action Principle and Field Equations}\label{sec:harmonic}

In the string frame, the relevant bosonic contributions to the low-energy effective actions of interest are encoded in
 \beq
 {\cal S}_S \ = \ \frac{1}{2\kappa_{D}^2} \ \int d^{D}x\,\sqrt{-\,G} \, \Big\{ e^{-2\phi}\Big[\, \mathcal{R}\, + \,4(\partial\phi)^2 \Big] \, - \, T \, e^{\,\gamma_s\,\phi} \Big\} \ . \label{eqs1}
 \eeq
The values of $\gamma_s$ of direct relevance for ten-dimensional String Theory and their Einstein-frame counterparts $\gamma$ to $D$ dimensions can be found in Table~\ref{table:tab_1}, where
\beq
\gamma_0 \ = \ \frac{4}{D-2} \ . \label{gamma_0}
\eeq
\begin{table}[!ht]
 \begin{center}
\begin{tabular}{ ||c||c|c||}
 \hline\hline
  Model & $\gamma_s$ & $\gamma$ \\ [0.5ex]
  \hline\hline
   USp(32) & $-1$ & $\gamma_0\,+\,1$ \\ [0.5ex]
  \hline
  U(32) & $-1$ & $\gamma_0\,+\,1$ \\ [0.5ex]
  \hline
 SO(16) $\times$ SO(16) & 0 & $\gamma_0\,+\,2$  \\ [0.5ex] \hline
 Non-critical string  & -2 & $\gamma_0$  \\ [0.5ex]
 \hline\hline
\end{tabular}
 \end{center}
 \caption{\small String-frame and Einstein-frame parameters (extended to $D$ dimensions) for the tachyon-free ten-dimensional string models and for non-critical strings.}\vskip 12pt
 \label{table:tab_1}
 \end{table}
$T$ is positive for the ten-dimensional non-supersymmetric strings of~\cite{so1616,u32,usp32}, while for non-critical strings
\beq
{T} \ = \ \frac{10-D}{2\,\alpha'} \ , \label{T9}
\eeq
so that it is positive for $D<10$ and negative for $D>10$.

In the Einstein frame, for generic values of $D$, the action becomes
 \beq
{\cal S}_E  \ = \  \frac{1}{2\kappa_{D}^2}\int d^{D}x \, \sqrt{-g}\left[\mathcal{R}\ - \ \frac{4}{D-2}\ (\partial\phi)^2\ - \ T \, e^{\,\gamma\,\phi} \
\right] \ , \label{eqs4}
\eeq
and the corresponding equations read
\bea
\mathcal{R}_{MN} \,-\, \frac{1}{2}\ g_{MN}\, \mathcal{R} &=& \!\!\frac{4}{D-2}\  \pr_M\phi\, \pr_N\phi  \ - \ \frac{1}{2}\,g_{MN}\Big[\frac{4\,(\pr\phi)^2}{D-2}\,+\, V(\phi)\Big] \ ,  \nonumber
\\
\frac{8}{D-2} \ \Box\phi &=& V^\prime(\phi)   \ . \label{eqsbeta}
\eea
We work, in general, with exponential potentials of the form
\beq \label{eq:tadpole_potential}
V(\phi) \ = \ T \, e^{\gamma\,\phi} \ ,
\eeq
with the values of $\gamma$ and $T$ that we discussed above.
Equivalently, one can replace the Einstein equation with
\bea
{\cal R}_{MN}  &=& \frac{4}{D-2}\  \pr_M\phi\, \pr_N\phi \ + \  g_{MN}\,\frac{V(\phi)}{D-2} \ . \label{eqsnotlagbeta}
\eea

Our starting point is the class of metric tensors and dilaton profiles
\beq
ds^{\,2} \ = \ - \ e^{2B(\tau)}\,d\tau^2 \ +\ \ell^2 \left(e^{2A(\tau)}\, ds_{p+1,k}^2 \ + \ e^{2C(\tau)}\, ds_{D-p-2,k'}^2 \right)  \ ,\label{metric_sym}
\eeq
where $\ell$ is an overall scale. Here, $ds_{p+1,k}^2$ describes a spatial slice of dimension $p+1$ and curvature $k$, while $ds_{D-p-2,k'}^2$ describes an internal space of dimension $D-p-2$ and curvature $k'$. Both spaces are maximally symmetric and are generally curved: we denote by $k=(\pm 1,0)$ and $k'=(\pm 1,0)$ the corresponding curvatures, with the understanding that for $k=0$ the spatial slice is Euclidean, while for $k'=0$ the internal space is a torus. In all these cases, the internal space can be compact. This is directly true for the sphere ($k=1$), but can also be true for the hyperbolic space ($k=-1$), after quotienting by a discrete subgroup of its isometry group, which is $SO(1,D-p-2)$ in this case. There is an extensive mathematical literature on these constructions. These quotients can give rise to orientable or non-orientable manifolds or orbifolds, depending on the choice of the discrete subgroup used to implement the projection. See, for example, the books listed in~\cite{math_projection}.

The equations for $A$, $C$ and the scalar profile $\phi$ in the presence of a general potential $V$ are
 \bea
 A'' + A'\, F' \!\!\!&=&  \frac{V(\phi)}{(D-2)} \ e^{2\,B}\ - \ \frac{k\,p}{\ell^2}\ e^{2(B-A)}  \ ,  \\
C'' +  C '\, F'\!\!\!&=& \frac{V(\phi)}{(D-2)} \ e^{2\,B} \ - \ \frac{k'(D-p-3)}{\ell^2}\ e^{2(B-C)} \ ,  \nonumber  \\
   \phi'' +  \phi'\, F' \!\!\!&=& - \ \frac{V'(\phi)\,(D-2)}{8}\ e^{2\,B} \label{Eqs_back} \ , \nonumber 
  \eea
  where
  \beq
F\ =\ (p+1)A\ -\ B\ +\ (D-p-2)C \ ,  \label{eqs6}
\eeq
and ``primes'' denote derivatives with respect to the parametric time $\tau$ in eq.~\eqref{metric_sym}.~\footnote{Note that identifying with $\ell$ the two scales of spatial and internal slices is not a loss of generality. If they were different, one could reduce the system to this form by shifting $A$ or $C$, and indeed the introduction of $\ell$ is simply giving the curvature terms the proper dimension.}
The initial values cannot be given independently for all fields, since the allowed choices must satisfy the ``Hamiltonian constraint''
 \begin{align}
&p(p+1)\left(A'\right)^2 \,+\, 2(p+1)(D-p-2) A'C'\,+\, (D-p-2)(D-p-3)\left(C'\right)^2 \,-\, \frac{4\,(\phi')^2}{D-2}  \nonumber \\
 &- \, {V(\phi)} \, e^{\, 2\,B}\, + \, \frac{k\,p(p+1)}{\ell^2}\ e^{2(B-A)}\, + \, \frac{k'(D-p-3)(D-p-2)}{\ell^2}\ e^{2(B-C)} \,= \, 0 \ . \label{EqB_red}
 \end{align}

In the presence of a general potential $V(\phi)$, static solutions thus exist for a given $\phi_0$ if
\bea
V'(\phi_0) &=& 0 \ , \nonumber \\
\frac{V(\phi_0)}{(D-2)} &=& \frac{k\,p}{\ell^2}\ e^{-2A} \ = \ \frac{k'(D-p-3)}{\ell^2}\ e^{-2C} \ ,
\eea
so that $\phi_0$ must be an extremum of $V$, the two curvatures $k$ and $k'$ must have the same sign and the spatial and internal scales must be comparable. Alternatively, both curvatures must vanish, together with $V(\phi_0)$. In the following, we shall focus on exponential potentials, which provide the leading contributions in string theory, and these types of solution will only emerge for $\gamma=0$, where the first condition is identically satisfied.

 Note that the choice of leaving $B$ unspecified has endowed the system with a reparametrization invariance. Two special choices will prove to be convenient in the following:
 \begin{itemize}
 \item  \textbf{the ``harmonic gauge''} $F=0$, for which
\beq
B \ = \ (p+1) A \ + \ (D-p-2) C  \ . \label{harm_gauge}
\eeq
This gauge choice simplifies the preceding equations somewhat, which become
 \bea
 && A'' \ = \  \frac{T}{(D-2)} \ e^{2\,B\,+\,\gamma\,\phi}\ - \ \frac{k\,p}{\ell^2}\ e^{2(B-A)}  \label{Eqs_back_F} \ ,  \\
&& C'' \ = \ \frac{T}{(D-2)} \ e^{2\,B\,+\,\gamma\,\phi}\ - \ \frac{k'(D-p-3)}{\ell^2}\ e^{2(B-C)} \ ,  \nonumber  \\
  &&  \phi'' \ = \ - \ \frac{T\,\gamma\,(D-2)}{8}\ e^{2\,B\,+\,\gamma\,\phi}  \ , \label{EqB_back_F} 
  \eea
together with the Hamiltonian constraint~\eqref{EqB_red}, where $B$ is as in eq.~\eqref{harm_gauge}. The harmonic gauge leads to several classes of exact solutions, at the price of some subtleties in their interpretation. 
\item \textbf{the ``cosmic gauge''} $B=0$, in which the time variable $\tau$ affords a simple physical interpretation as ``cosmic time'', while the equations become
 \bea
&& A'' + A'\left[ (p+1) A' \ + \ (D-p-2) C'\right] \ = \   \frac{T}{(D-2)} \ e^{\gamma\,\phi}\ - \ \frac{k\,p}{\ell^2}\ e^{-2A}  \ ,  \nonumber \\
&& C'' +  C '\left[ (p+1) A' \ + \ (D-p-2) C'\right]\ = \  \frac{T}{(D-2)} \ e^{\gamma\,\phi}\ - \ \frac{k'(D-p-3)}{\ell^2}\ e^{-2C} \ ,  \nonumber  \\
 &&  \phi'' +  \phi'\left[ (p+1) A' \ + \ (D-p-2) C'\right] \ = \ - \ \frac{T\,\gamma\,(D-2)}{8}\ e^{\gamma\,\phi} \label{Eqs_back_C} \ , \\
   && p(p+1)\left(A'\right)^2 \,+\, 2(p+1)(D-p-2) A'C'\,+\, (D-p-2)(D-p-3)\left(C'\right)^2 \,-\, \frac{4\,(\phi')^2}{D-2} \nonumber\\
 && - \, {T} \, e^{\gamma\,\phi}\, + \, \frac{k\,p(p+1)}{\ell^2}\ e^{-2A}\, + \, \frac{k'(D-p-3)(D-p-2)}{\ell^2}\ e^{-2C} \,= \, 0 \ . \label{EqB_back_C}
  \eea

\end{itemize}

The most relevant cases for this setup are $D=10$, $p=2$, with $\gamma=\frac{3}{2}$ or $\gamma=\frac{5}{2}$. For non-critical strings, $D \neq 10$ and the most relevant case is again $p=2$, with $\gamma=\gamma_0$ of eq.~\eqref{gamma_0}, up to leading correction terms determined by $\gamma_0+1$ or $\gamma_0+2$.

A different gauge choice led to the first exact solutions of this type~\cite{dm_vacuum}, the isotropic Dudas-Mourad vacua in the presence of exponential potentials with $\gamma=\frac{3}{2}$ and $\gamma=\frac{5}{2}$. This option eliminates one exponential from the field equations, and also allows one to find exact solutions when only one of the three ingredients $(T,k,k')$ is present. 
\begin{itemize}
\item If only $T \neq 0$ and $\gamma\neq 0$, this choice translates into the condition
\beq
2 B \ + \ \gamma\,\phi \ = \ 0 \ ;
\eeq
\item if only $k \neq 0$, this choice translates into the condition
\beq
B \ - \ A \ = \ 0 \ ,
\eeq
which is the conformal gauge in spacetime; 
\item finally, if only $k'\neq 0$ this choice translates into the condition
\beq
B \ - \ C \ = \ 0 \ ,
\eeq
which is the conformal gauge in the internal space. In all cases, the Hamiltonian constraint reduces to an algebraic relation among the first derivatives $A'$, $C'$ and $\phi'$, which describes a hyperboloid and can be solved in terms of two independent functions.
\end{itemize}
In the following, we work in the harmonic gauge, which provides a general framework for all cases.

The choices $p=D-3$ and $p=D-2$ identify interesting special cases which were left out in the preceding discussion, but to which we can now return briefly. For  $p=D-3$ there is no internal curvature, but the system is still described in terms of $(A,C,\phi)$, and it becomes
 \bea
 A'' + A'\, F' \!\!\!&=&  \frac{T}{(D-2)} \ e^{2\,B\,+\,\gamma\,\phi}\ - \ \frac{k\left(D-3\right)}{\ell^2}\ e^{2(B-A)}  \ ,  \\
C'' +  C '\, F'\!\!\!&=& \frac{T}{(D-2)} \ e^{2\,B\,+\,\gamma\,\phi} \ ,  \nonumber  \\
   \phi'' +  \phi'\, F' \!\!\!&=& - \ \frac{T\,\gamma\,(D-2)}{8}\ e^{2\,B\,+\,\gamma\,\phi} \label{Eqs_back_d3} \ , \nonumber 
  \eea
  where
  \beq
  F \ = \ (D-2)A \ + \ C \ - \ B \ , 
  \eeq
  while the Hamiltonian constraint reduces to
   \begin{align}
&(D-2)(D-3)\left(A'\right)^2 \,+\, 2(D-2) A'C' \,-\, \frac{4\,(\phi')^2}{D-2}  \nonumber \\
 &- \, {T} \, e^{\, 2\,B\,+\,\gamma\,\phi}\, + \, \frac{k\,(D-2)(D-3)}{\ell^2}\ e^{2(B-A)} \,= \, 0 \ . 
 \end{align}
For $p=D-2$ there is no internal space altogether, and the system is described in terms of $A$ and $\phi$ only and reduces to
 \bea
 A'' + A'\, F' \!\!\!&=&  \frac{T}{(D-2)} \ e^{2\,B\,+\,\gamma\,\phi}\ - \ \frac{k\left(D-2\right)}{\ell^2}\ e^{2(B-A)}  \ ,  \\
   \phi'' +  \phi'\, F' \!\!\!&=& - \ \frac{T\,\gamma\,(D-2)}{8}\ e^{2\,B\,+\,\gamma\,\phi} \label{D8_sol} \ , 
  \eea
  where
    \beq
  F \ = \ (D-1)A  \ - \ B \ , 
  \eeq
  while the Hamiltonian constraint becomes
 \beq
(D-1)(D-2)\left(A'\right)^2  \,-\, \frac{4\,(\phi')^2}{D-2} \,- \, {T} \, e^{\, 2\,B\,+\,\gamma\,\phi}\, + \, \frac{k\,(D-1)(D-2)}{\ell^2}\ e^{2(B-A)} \,= \, 0 \ . \label{EqB_red_d2}
 \eeq

\subsection{\sc General Stability Issues} \label{sec:gen_stab}

The perturbative stability of solutions can be addressed by linearizing the Einstein equations~\eqref{eqsbeta} around the backgrounds of interest while allowing for perturbations that generally depend on all coordinates. In the cases under scrutiny, the individual modes obey linear time-dependent equations. The issue is then to follow their evolution, as in~\cite{ivano}, in order to identify modes that can overcome the background~\footnote{For a review of further developments along these lines, see~\cite{dms_review}.}. If this occurs in the future, it reveals a \emph{future instability}, while if it occurs in the past, it reveals a \emph{past instability}. If no such relative growth occurs, the solution can be regarded as stable. 

Here we limit ourselves to a first look at the problem obtained by restricting attention to homogeneous perturbations. This is much simpler, since it amounts to perturbing eqs.~\eqref{EqB_back_F} or \eqref{EqB_back_C}, but misses the effects of nonzero modes and cannot address localized perturbations, which are the physically relevant sources of instabilities. For example, flat space is a solution of eqs.~\eqref{EqB_back_F} or \eqref{EqB_back_C} if $T=0$ and $k=k'=0$ with vanishing values for $A$, $C$ and $\phi$. There are clearly homogeneous perturbations that grow linearly in time, signaling future or past instabilities. However, they correspond to spin-2 modes of infinite wavelength that are usually left out as incompatible with asymptotic flatness. However, the simplified analysis is still motivated from the perspective of the dynamical systems~\eqref{EqB_back_F} or \eqref{EqB_back_C}, and it has relevant indications for their numerical analysis, where unstable solutions of this type are difficult to detect.

The general exact solutions that will emerge when two of the three inputs $(T,k,k')$ vanish will depend on some free parameters, and this dynamical-systems-motivated stability analysis will then reduce to ascertaining the effects of their relative variations. We ignore overall time translations, since the dynamical systems of interest have no explicit time dependence.

\section{\sc Cosmologies without Tension or Curvature} \label{sec:t0curv0}

The first setting that we discuss concerns cosmologies in which tension and curvature are absent. Although simple, this case plays a central role in the asymptotics of the complete system.

In the harmonic gauge $F=0$ the solutions
\beq
A \ = \ A_1\,\tau\ + \ A_0 \ , \qquad C \ = \ C_1\,\tau\ + \ C_0 \ , \qquad \phi \ = \ \phi_1\,\tau\ + \ \phi_0 \ ,
\eeq
are linear in the parametric time $\tau$, with $A_i$, $C_i$ and $\phi_i$ arbitrary constants, 
and consequently
\beq
B \ = \ \left[ (p+1)A_1 \ + \ (D-p-2) C_1\right] \tau  \ + \ (p+1)A_0 \ + \ (D-p-2) C_0 \ ,
\eeq
However, rescaling the $\vec{x}$ and $\vec{y}$ coordinates and the parametric time $\tau$, one can modify the constant terms $A_0$ and $C_0$, so that the solution can be cast in the form
\bea
ds^2 &=& - \ d\tau^2\ e^{2 \left[ (p+1)A_1 \ + \ (D-p-2) C_1\right] \tau} \ + \ e^{2 A_1\tau} d\vec{x}^2 \ + \ e^{2 C_1\tau} d\vec{y}^2 \ , \nonumber \\
e^\phi &=& e^{\phi_1\tau + \phi_0} \ ,
\eea
while the Hamiltonian constraint reduces to
 \beq
\mu^2 \ = (p+1) A_1^2  \ + \ (D-p-2) C_1^2 \ + \ \frac{4}{D-2}\, \phi_1^2 \ , \label{eqmu2}
 \eeq
where
\beq
\mu \ = \ (p+1)A_1 \ + \ (D-p-2) C_1 \ . \label{ACmu}
\eeq
If $\mu=0$, $A_1$, $C_1$ and $\phi_1$ must all vanish, and one is simply describing a flat Minkowski space (solution of type I in Table~\ref{tab:solsTkp11}). New interesting options emerge if $\mu \neq 0$. The cosmic time $t$ can then be linked to $\tau$ via
\beq
\mu \,t \ = \ e^{\mu\,\tau} \ ,
\eeq
so that $\mu$ and $t$ have identical signs,
and the solutions finally read
\bea
ds^2 &=& - \ dt^2 \ + \ \left( \mu t\right)^{2 \alpha_A} d\vec{x}^2 \ + \ \left( \mu t\right)^{2 \alpha_C} d\vec{y}^2 \ , \nonumber \\
e^\phi &=& e^{\phi_0} \left( \mu t\right)^{\alpha_\phi} \ . \label{free_metric_phi}
\eea
This is the solution of type II in Table~\ref{tab:solsTkp11}.

The integration constants satisfy
\bea
&& \!\!\!(p+1) \alpha_A \ + \ (D-p-2) \alpha_C \ = \ 1 \ , \label{free} \\
&& \!\!\!(p+1) \alpha_A^2 \ + \ (D-p-2) \alpha_C^2 \ + \ \frac{4}{D-2}\, \alpha_\phi^2 \ = \ 1  \ . \nonumber
\eea
so that the independent choices of the $\alpha$'s correspond to the intersection between a plane and an ellipsoid.
An alternative expression for the quadratic constraint,
\beq{}{}{}{}
\frac{(p+1)(D-1)^2}{(D-2)(D-p-2)} \left(\alpha_A \ - \ \frac{1}{D-1}\right)^2  \ + \  \left(\frac{\gamma_c}{2}\, \alpha_\phi\right)^2 \ = 1 \ . \label{ellipse}
\eeq
suggests a convenient parametrization of the independent solutions, 
\bea{}{}{}{}
\alpha_A &=& \frac{1}{D-1}\left[1 \ + \ \sqrt{\frac{(D-2)(D-p-2)}{(p+1)}} \ \cos\theta\right] \ , \nonumber \\
\alpha_C &=& \frac{1}{D-1}\left[1 \ - \ \sqrt{\frac{(D-2)(p+1)}{(D-p-2)}} \ \cos\theta\right] \ , \nonumber \\
\alpha_\phi &=&  \frac{2}{\gamma_c} \ \sin \theta \ , \label{param_theta}
\eea
where
\beq
\gamma_c \ = \ 4 \, \frac{\sqrt{D-1}}{D-2}\ .  \label{gammac}
\eeq
In particular, starting from ten dimensions, and for the special case of four-dimensional cosmologies $(p=2)$, the preceding expressions become
\beq
\alpha_A \,=\, \frac{1}{9}\ \Big(1 \ + \ 4 \,\cos\theta\Big) \ , \quad
\alpha_C \,=\, \frac{1}{9}\ \Big(1 \ - \ 2\, \cos\theta\Big) \ , \quad
\alpha_\phi \,=\, \frac{4}{3} \ \sin \theta \ . \label{param_theta2}
\eeq

In summary, the independent choices of $A$, $B$ and $C$ depend on the angle $\theta$, while $\phi$ also depends on $\phi_0$. 

One can address the stability of these solutions, from the vantage point of the dynamical system, allowing for a small deformation of the free parameter $\theta$ and estimating the three relative deviations
\beq
\frac{\delta\,\alpha_A(\theta)}{\alpha_A(\theta)} \ , \qquad \frac{\delta\,\alpha_C(\theta)}{\alpha_C(\theta)} \ , \qquad \frac{\delta\,\alpha_\phi(\theta)}{\alpha_\phi(\theta)} \ .
\eeq
For small values of $\delta\,\theta$, these are generically small, with the exception of points where one of the $\alpha$'s vanishes. These solutions are thus stable under these perturbations away from the special points.

\section{\sc Cosmologies with Only Curvature}  \label{sec:t0curvnot0}

There are several options for the $(k,k')$ pair within this class of cosmologies. We begin by discussing the exact solutions for the cases $(k,k')=(0,\pm 1)$. These results will also determine the exact 
solutions with $(k,k')=(\pm 1,0)$, which can be simply deduced by swapping $p$ and $D-p-3$ in the resulting expressions. If both $k$ and $k'$ are not zero, we have found a special class of exact solutions, but we do not have general ones.
\subsection
[\texorpdfstring{{\mdseries\textsc{Solutions with $k=0$ and $k' \neq 0$}}}{Solutions with k=0 and k' ≠ 0}]
{{\mdseries\textsc{Solutions with $k=0$ and $k' \neq 0$}}}\label{sec:kzerokpnotzero}

Referring for definiteness to the case $k=0$, one can work directly in terms of 
\beq
X \ = \  B \ - \ C \ = \ (p+1)A \ + \ (D-p-3)C \ ,
\eeq
$A$ and $\phi$, and in the harmonic gauge $F=0$ the system reduces to
    \bea
A''&=&0   \ , \nonumber  \\
X''&=&- \ k'\ \frac{\left(D-p-3\right)^2}{\ell^2}\ e^{2X} \ ,  \nonumber \\
\phi''&=&0 \ , \label{onlykp}
\eea
so that
\beq
A \ = \   A_1\,\tau  \ + \ A_0 \ , \qquad \phi \ = \ \phi_1\,\tau \ + \ \phi_0  \label{dilA}
\eeq
involve four integration constants. $A_0$ can be removed again by rescaling the spatial coordinates, $\tau$ and $\ell$, while the equation for $X$ can be turned into
\beq
\left(X'\right)^2 \ = - \  k'\ \frac{\left(D-p-3\right)^2}{\ell^2}\ e^{2X} \ + \ \frac{1}{\rho^2}  \ , \label{eq_X}
\eeq
and the Hamiltonian constraint links $\rho$ to $A_1$ and $\phi_1$ according to
\beq \label{eq:XW_ham_constrt0}
\frac{1}{\rho^2} \ = \ \frac{(p+1)(D-2)}{(D-p-2)}\, \left(A_1\right)^2 \ + \ \frac{4(D-p-3)}{(D-2)(D-p-2)} \, \left(\phi_1\right)^2 \ .
\eeq
The solutions thus depend on $\ell$, the scale of the internal space, and on the three integration constants $\rho$, ${\theta}$ and $\phi_0$, with
\bea
A_1 &=& \frac{\cos{\Theta}}{\rho}\, \sqrt{\frac{(D-p-2)}{(p+1)(D-2)}} \ , \nonumber \\
\phi_1 &=& \frac{\sin{\Theta}}{2\,\rho }\, \sqrt{\frac{(D-2)(D-p-2)}{(D-p-3)}} \ . \label{Aphi}
\eea
As a result
\bea \label{eq:XW_ABCphi}
B &=& \frac{(D-p-2) \,X \ - \ (p+1)\left(A_1\tau+A_0\right)}{(D-p-3)}\ , \nonumber \\
C &=& \frac{X \ - \ (p+1)\left(A_1\tau+A_0\right)}{D-p-3} \ ,
\eea
and one must distinguish two cases for $k'$:
\begin{enumerate}
    \item If $k'=1$ the internal space is a sphere, $\frac{1}{\rho^2}$ must be strictly positive in view of eq.~\eqref{eq:XW_ham_constrt0} and the solution for $X$ takes the form
    \beq
{X}\ = \ -  \log \left[\frac{(D-p-3)\,\rho}{\ell} \, \cosh \left(\frac{\tau}{\rho}\right) \right] \ , \label{Xkppos}
\eeq
with $-\infty<\tau<+\infty$. In this case the metric is
\bea
ds^2 &=& - \ d\tau^2 \, e^{2 B_0}\, e^{-\, \frac{2(p+1) \left(A_1\,\tau\,+\,A_0\right)}{D-p-3}} \left[ \frac{\ell\,e^{-B_0}}{\rho(D-p-3)\cosh\left(\frac{\tau}{\rho}\right)}\right]^{2\,\frac{(D-p-2)}{(D-p-3)}}\!\!\! +\, e^{2\left(A_1\tau+A_0+\widetilde{A}_0\right)}\, d\vec{x}^2 \nonumber \\
&+& \ell^2 \, ds_{D-p-2,k'=1}^2 \ e^{-\, \frac{2(p+1)\left(A_1\,\tau\,+\,A_0\right)}{D-p-3}} \left[ \frac{\ell\,e^{-B_0}}{\rho(D-p-3)\cosh\left(\frac{\tau}{\rho}\right)}\right]^\frac{2}{D-p-3} \ , \nonumber \\
e^\phi &=& e^{\phi_1\,\tau\,+\,\phi_0} \ .
\label{coshtau}
\eea
This is the type-III solution in Table~\ref{tab:solsTkp11}. 

$B_0$ reflects a gauge freedom in the definition of harmonic coordinates, which only requires that $F'=0$, while $\widetilde{A}_0$ reflects the scaling symmetry present for the flat $\vec{x}$ coordinates. There is a choice of $B_0$ that removes the $\ell$-dependence altogether, so that the solution only depends on $\rho$. Moreover, $A_0$ and $\widetilde{A}_0$ can be removed from the spatial part, but play a role in the internal and $\tau$ directions. Here and in the following sections, for simplicity, we refrain from taking $B_0$ into account and set $A_0+\widetilde{A}_0=0$, while retaining $A_0$ and the $\tau$ and internal directions. As we shall see, $A_0$ will play a role in granting a perturbative window for the adiabatic analysis in Section~\ref{sec:adiabatic}.

Moreover, $A_1>0$ for an expanding universe, and the relation between the parametric time $\tau$ and the cosmic time $t$ thus reads
\beq
dt \ = \ e^{-\, \frac{(p+1) \left(A_1\,\tau+A_0\right)}{D-p-3}} \left[ \frac{\ell}{\rho(D-p-3)\cosh\left(\frac{\tau}{\rho}\right)}\right]^\frac{D-p-2}{D-p-3} \, d \tau
\eeq
so that the total time duration is
\beq
t_0 \ = \ e^{-\, \frac{(p+1)A_0}{D-p-3}} \left(\frac{\ell}{D-p-3}\right)^\alpha \int_{-\infty}^\infty d\tau  \ e^{-\, \alpha\, \frac{\tau}{\rho}\,\cos\Theta\,\sqrt{\frac{(p+1)}{(D-2)(D-p-2)}}}\left[ \frac{1}{\rho\,\cosh\left(\frac{\tau}{\rho}\right)}\right]^\alpha \ ,\label{ttaukp}
\eeq
where
\beq
\alpha \ = \ \frac{D-p-2}{D-p-3} \ .
\eeq

The initial singularity is approached as $\tau \to -\,\infty$, and the scale factor of the spatial coordinates vanishes there, but as $\tau \to +\infty$ one encounters an ``internal Big Crunch''. Indeed, the scale factor of the spatial coordinates diverges in the limit, but the size of the internal sphere vanishes again, for all choices of $\Theta$. Note that the complete evolution of this class of cosmologies occurs within a \emph{finite interval} of cosmic time, since the integral in eq.~\eqref{ttaukp} is finite.
Letting
\bea
\nu_\pm &=& \frac{D-p-2}{\rho(D-p-3)}\left[1 \ \pm \ \cos{\Theta}\, \sqrt{\frac{p+1}{(D-2)(D-p-2)}}\right] \ , \nonumber \\
\delta_\pm &=& \frac{1}{\rho(D-p-3)} \left[ 1 \ \pm \ \cos{\Theta} \sqrt{\frac{(p+1)(D-p-2)}{(D-2)}}\right] \ , \label{nudelta}
\eea
one can see that $\nu_\pm >0$ in all cases of interest, for which $D-p-2>1$, so that at early and late times
\bea
t &\sim& a \, e^{\nu_-\,\tau} \qquad  (\tau \to -\,\infty) \ , \nonumber \\
t &\sim& t_0  \ - \  b\, e^{-\,\nu_+\,\tau} \qquad  (\tau \to +\,\infty) \ ,
\eea
with $a$ and $b$ a pair of positive constants and $t_0$ the total span of cosmic time.

The limiting behaviors of the metric and the dilaton are captured by
\bea
ds^2 &\sim& - \ dt^2 \ + \ \left(\frac{t}{a}\right)^{2\widetilde{\alpha}_A^{\,-}}\, d\vec{x}^2 \ +  \ell^2 \, ds_{D-p-2,k'=1}^2 \ \left(\frac{t}{a}\right)^{2\widetilde{\alpha}_C^{\,-}} \ , \nonumber \\
e^\phi &\sim& e^{\phi_0} \ \left(\frac{t}{a} \right) ^{\widetilde{\alpha}_\phi^{\,-}}
\eea
at early cosmic times, and by
\bea
ds^2 &\sim& - \ dt^2 \ + \ \left(\frac{t_0\,-\,t}{b}\right)^{2\,\widetilde{\alpha}_A^+}  d\vec{x}^2 \ + \  \ell^2 \, ds_{D-p-2,k'=1}^2 \ \left(\frac{t_0\,-\,t}{b}\right)^{2\,\widetilde{\alpha}_C^+}\ , \nonumber \\
e^\phi &\sim& e^{\phi_0} \left(\frac{t_0\,-\,t}{b}\right)^{ \widetilde{\alpha}_\phi^+} 
\eea
at late cosmic times, where
\bea
\widetilde{\alpha}_A^\pm &=& \, \mp\ \frac{\cos {\Theta}}{\rho\,\nu_\pm} \ \sqrt{\frac{(D-p-2)}{(p+1)(D-2)}} \ ,\nonumber \\
\widetilde{\alpha}_C^\pm &=& \, \frac{\delta_\pm}{\nu_\pm} \ , \nonumber \\
\widetilde{\alpha}_\phi^\pm &=& \, \mp\  \frac{\sin{\Theta}}{2\,\rho\,\nu_\pm }\, \sqrt{\frac{(D-2)(D-p-2)}{(D-p-3)}} \ . \label{acphik'pos}
\eea

An expanding universe must have $\widetilde{\alpha}_A^- >0$, so that $-\,\frac{\pi}{2}<{\Theta}<\frac{\pi}{2}$. The evolution then starts from an initial singularity, but the scale factor for the spatial coordinates $x^i$ diverges after a finite interval of cosmic time. The scale factor of the internal space vanishes initially if $\widetilde{\alpha}_C^->0$, which is the case if $\cos\Theta<\sqrt{\frac{D-2}{(p+1)(D-p-2)}}$, is initially finite if $\cos\Theta=\sqrt{\frac{D-2}{(p+1)(D-p-2)}}$ and is initially infinite if $\cos\Theta>\sqrt{\frac{D-2}{(p+1)(D-p-2)}}$, but vanishes in all cases at the end of the evolution since, as we have seen, $\cos\Theta>0$ for an expanding universe, in view of eqs.~\eqref{acphik'pos} and~\eqref{nudelta}. These settings can thus give rise to ``dynamical compactifications'' from symmetric early stages, albeit in setups that are singular within the low-energy effective theory. 
\begin{figure}
    \centering
    \includegraphics[width=0.5\linewidth]{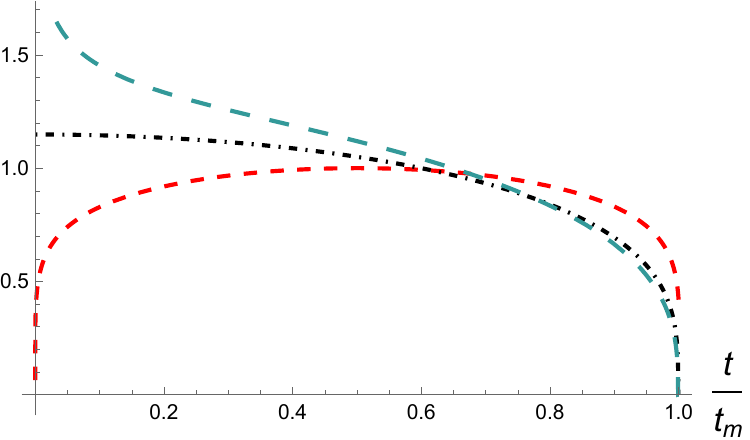}
    \caption{The radius of the internal sphere (for $D=10$, $p=2$) as a function of $\frac{t}{t_m}$, where $t$ denotes the cosmic time and $t_m$ its total span, for $\cos\Theta=0$ (red, dashed), $\cos\Theta=\frac{2}{3}$ (green, wide dashed) and $\cos\Theta=1$ (black, dash-dotted).}
      \label{fig:kprime_pos}
\end{figure}

One can verify that the exponents characterizing the asymptotic behaviors satisfy
\bea
&& (p+1)\widetilde{\alpha}_A^\pm \ + (D-p-2)\widetilde{\alpha}_C^\pm \ = \ 1 \ , \nonumber \\
&& p(p+1)\left(\widetilde{\alpha}_A^\pm\right)^2  \ +  \ 2(p+1)(D-p-2) \widetilde{\alpha}_A^\pm \,\widetilde{\alpha}_C^\pm \nonumber\\ &&+  (D-p-2)(D-p-3) \left(\widetilde{\alpha}_C^\pm\right)^2  -  \frac{4}{D-2}  \bigl(\widetilde{\alpha}_\phi^\pm\bigr)^2 \ = \ 0 \ , 
\label{constr_alphatilde}
\eea
which are again the conditions met in Section~\ref{sec:t0curv0}.
Therefore, at both ends, which are separated by a finite interval of the cosmic time $t$, the solutions approach the free Kasner-like behavior captured by eqs.~\eqref{free}, albeit with exponents that are generally different in the two cases. The net effect of the internal curvature is thus twofold: it confines the evolution to a finite interval of cosmic time, while also connecting pairs of different free-like exponents. 

For the special case of a string-inspired four-dimensional Universe with flat spatial slices ($D=10,p=2)$ with $(k,k')=(0,1)$
\beq
\nu_\pm \ = \  \frac{6}{5\,\rho} \left[ 1 \ \pm \ \frac{1}{4}\, \cos\Theta \right] \ , \qquad
\delta_\pm \ = \  \frac{1}{5\,\rho} \left[ 1 \ \pm \ \frac{3}{2}\, \cos\Theta \right] \ , 
\eeq
and consequently
\bea
\widetilde{\alpha}_A^\pm &=& \mp \ \frac{5}{12}\, \frac{\cos{\Theta}}{1 \ \pm \ \frac{1}{4}\, \cos{\Theta}}\ , \nonumber \\
\widetilde{\alpha}_C^\pm &=& \frac{1}{6}\,\frac{ 1 \ \pm \ \frac{3}{2}\, \cos{\Theta}}{1 \ \pm \ \frac{1}{4}\, \cos{\Theta}} \ , \nonumber \\
\widetilde{\alpha}_\phi^\pm &=& \mp \ \sqrt{\frac{5}{3}}\ \frac{\sin{\Theta}}{1 \ \pm \ \frac{1}{4}\, \cos{\Theta}}\ . \label{free_alphas}
\eea
\begin{figure}
    \centering
    \includegraphics[width=0.4\linewidth]{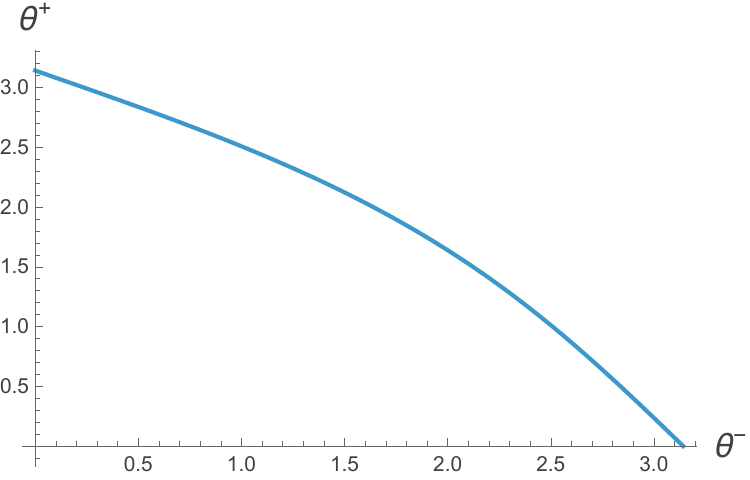}
    \caption{The relation between the free-Kasner angles corresponding to early and late times for the system with $k'=1$, in $D=10$ and for $p=2$ is a nearly linear one-to-one map.}
      \label{fig:thetamp}
\end{figure}

The limiting behaviors of these curved cosmologies at early and late times are thus captured by two distinct free Kasner solutions, with
\bea
\widetilde{\alpha}_A^\pm &=& \frac{1}{9}\ \Big(1 \ + \ 4 \,\cos\theta^\pm \Big) \ , \nonumber \\
\widetilde{\alpha}_C^\pm  &=& \frac{1}{9}\ \Big(1 \ - \ 2\, \cos\theta^\pm\Big) \ , \nonumber \\
\widetilde{\alpha}_\phi^\pm  &=& \frac{4}{3} \ \sin \theta^\pm \ , \label{param_theta21}
\eea
and the curved dynamics links the two angles $\theta^+$ and $\theta^-$ according to
\beq
\cos\,\theta^+ \ = \ - \ \frac{8 \ + \ 17\,\cos\,\theta^-}{17 \ + \ 8\,\cos\,\theta^-} \ .
\eeq

    \item If $k'=-1$ the internal space is a compact hyperbolic quotient, but eq.~\eqref{eq:XW_ham_constrt0} retains the same form. 
    
    If $\frac{1}{\rho^2}=0$, an option that was not available in the previous case, $A'$ and $\phi'$ must vanish and the solution for $X$ takes the form
    \beq
{X} \ = \  -  \ \log \left[(D-p-3)\,\left|\frac{\tau}{\ell}\right| \right] \ .
\eeq
Consequently one obtains
\bea
ds^2 &=& - \ d\tau^2 \  \left[ \frac{\ell}{(D-p-3)\left| \tau \right|}\right]^{2\,\frac{(D-p-2)}{(D-p-3)}} \ + \ d\vec{x}^2 \nonumber \\
&+& \ell^2 \, \left[ \frac{\ell}{(D-p-3)\left| \tau \right|} \right]^\frac{2}{D-p-3} ds_{D-p-2,k'=-1}^2 \ , \nonumber \\
\phi &=& \phi_0 \ , \label{kpnegrhoinfinite}
\eea
where the dependence on $A_0$ has been absorbed by rescaling $\ell$ and the $\vec{x}$ coordinates.
In terms of the cosmic time $t$, defined via
\beq
dt \ = \ \left[ \frac{\ell}{(D-p-3) \left|{\tau}\right|}\right]^\frac{D-p-2}{D-p-3} \, d \tau \ ,
\eeq
the metric finally becomes
\beq
ds^2 \ = \  - \ dt ^2 \ + \  d\vec{x}^2 \ + \  t^2 \, ds_{D-p-2,k'=-1}^2  \ , \label{milne}
\eeq
and one can let $0<t<\infty$. This is the type-IV solution in Table~\ref{tab:solsTkp11}.

The result is the direct product of a $(p+2)$-dimensional Minkowski space with a compact internal space obtained by quotienting a hyperbolic space whose size grows linearly with $t$. In the absence of quotienting, this solution would describe the direct product of a Milne universe with a Euclidean space along the $\vec{x}$ directions. For brevity, in the following we shall call Milne-like this type of behavior.
\begin{figure}
        \centering
        \includegraphics[width=0.5\linewidth]{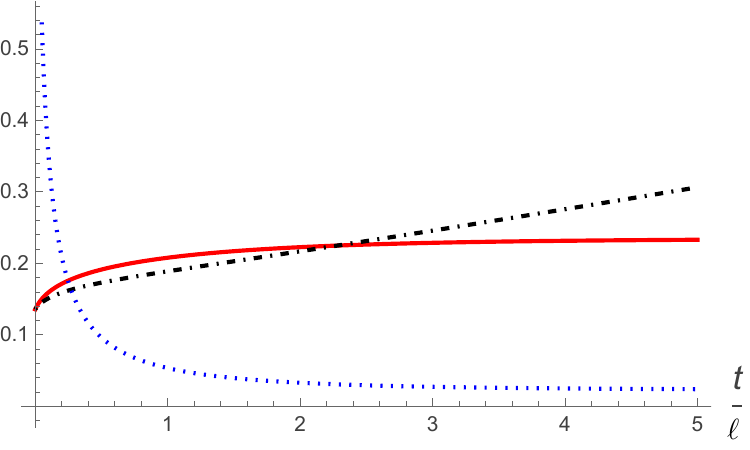}
                \caption{An example with $k'=-1$, a finite value of $\rho$ and a decreasing dilaton (blue, dotted) for $D=10$, $p=2$ as a function of $\frac{t}{\ell}$. Note that $e^A$ (red, solid) quickly approaches a constant value, while $e^C$ (black, dot-dashed) quickly approaches a linear behavior.}
        \label{fig:kprime_neg}
    \end{figure}
    
On the other hand, for finite values of $\rho$, the integration constants $A_1$ and $\phi_1$ can be parametrized as in eqs.~\eqref{Aphi}, and
    \beq
{X} \ = \  -  \ \log \left[\frac{(D-p-3)\,\rho}{\ell}\, \sinh \left|\frac{\tau}{\rho}\right| \right] \label{Xkpneg} \ ,
\eeq
so that the solution becomes
\bea
ds^2 &=& - \ d\tau^2 \ e^{-\, \frac{2(p+1) A_1\,\tau}{D-p-3}} \left[ \frac{\ell}{\rho (D-p-3)\sinh\left|\frac{\tau}{\rho}\right|}\right]^{2\,\frac{(D-p-2)}{(D-p-3)}} \ + \ e^{2 A_1\tau}\, d\vec{x}^2 \nonumber \\
&+& \ell^2 \, e^{-\, \frac{2(p+1) A_1\,\tau}{D-p-3}} \left[ \frac{\ell}{\rho (D-p-3)\sinh\left|\frac{\tau}{\rho}\right|}\right]^\frac{2}{D-p-3} ds_{D-p-2,k'=-1}^2 \ , \nonumber \\
\phi &=& \phi_1\,\tau \ + \ \phi_0 \ ,
\label{sinhtau} 
\eea
where  $A_1>0$ for an expanding universe.  This is the type-V solution in Table~\ref{tab:solsTkp11}.

The complete evolution is now captured working in the range $- \infty<\tau<0$, with $\tau \to - \infty$ corresponding to the origin of cosmic time. With this choice, one obtains a cosmological solution with \emph{finite past} and \emph{infinite future}. 
In detail, the cosmic time $t$ can now be defined starting from
\beq
t \ = \ e^{-\, \frac{(p+1)A_0}{D-p-3}}\,\rho \left[\frac{\ell}{(D-p-3)\rho}\right]^{\frac{(D-p-2)}{(D-p-3)}} \int_{-\infty}^{\frac{\tau}{\rho}} \ e^{-\, \frac{(p+1) A_1\,\rho\,u}{D-p-3}} \left(  \frac{1}{\sinh\left|{u}\right|}\right)^{\frac{(D-p-2)}{(D-p-3)}} \, d u \ ,
\eeq
where according to eqs.~\eqref{Aphi} the combination $A_1\,\rho$ is independent of $\rho$.
A singularity is now present in the finite past, and the limiting behavior as $\tau \to - \infty$ is power-like in cosmic time, as for $k'=1$, with exponents $\widetilde{\alpha}_A^{\,-}$, $\widetilde{\alpha}_C^{\,-}$ and $\widetilde{\alpha}_\phi^{\,-}$ that can be parametrized as in eqs.~\eqref{acphik'pos}, so that
\bea
ds^2 &=& - \ dt^2 \ + \ \left(\frac{t}{a}\right)^{2\widetilde{\alpha}_A^{\,-}}\, d\vec{x}^2 \ + \ \ell^2 \,  \left(\frac{t}{a}\right)^{2\widetilde{\alpha}_C^{\,-}} \ ds_{D-p-2,k'=-1}^2 \ , \nonumber \\
e^\phi &=& e^{\phi_0} \ \left(\frac{t}{a} \right) ^{\widetilde{\alpha}_\phi^{\,-}} \label{kpnegrhofinite}
\eea
while the universe has an infinite future evolution. After an interval of cosmic time of order $\rho$, $\phi$ approaches a constant value while the metric approaches 
\beq
ds^2 \ = \  - \ dt ^2 \ + \  d\vec{x}^2 \ + \  t^2 \, ds_{D-p-2,k'=-1}^2  \ , \label{milne2}
\eeq
so that in the large-$t$ limit these cosmologies evolve into a Milne-like universe, while the dilaton approaches a constant value.~\footnote{It would be a Milne universe if the internal space were not projected and thus genuinely noncompact. This type of solution will be found repeatedly in the following sections, where we continue to use this nomenclature.} We shall make this choice for the range of $\tau$ in the following but, for example, one could also work in the alternative range $0<\tau<\infty$, with $\tau = 0$ corresponding to the origin of cosmic time, and the resulting cosmology would be characterized by an infinite past and a finite future.
\end{enumerate}

One can address the stability of these solutions, starting from  $X$, which is fully determined by eq.~\eqref{eq_X} once $\rho$ and the origin of time are specified. The only non-trivial options involve deformations of $A_1$ and $\phi_1$, which are given in eqs.~\eqref{Aphi}. These are induced by variations of $\theta$, and our discussion in Section~\ref{sec:gen_stab}
 indicates that the system is stable away from the special choices for which one of the $\alpha's$ vanishes. These correspond to $\cos\Theta=0$, $\sin\Theta=0$, and to the additional values
\beq
\cos\Theta \ = \ \mp \ \sqrt{\frac{D-2}{(p+1)(D-p-2)}} 
\eeq
that are zeros of $\delta_\pm$. In fact, as we saw, $\nu_\pm$ never vanishes.

The solutions with $k \neq 0$ and $k'=0$ can be obtained from these by interchanging $A$ with $C$ and $p$ with $D-p-3$.

\subsection[\texorpdfstring{{\mdseries\textsc{Systems with $k \neq 0$ and $k' \neq 0$}}}{Systems with k ≠ 0 and k' ≠ 0}]
{{\mdseries\textsc{Systems with $k \neq 0$ and $k' \neq 0$}}}

In this case the harmonic-gauge equations become
 \bea
 && A'' \ = \ - \ \frac{k\,p}{\ell^2}\ e^{2(B-A)}  \ ,  \nonumber \\
&& C'' \ = \ - \ \frac{k'(D-p-3)}{\ell^2}\ e^{2(B-C)} \ ,  \nonumber  \\
  &&  \phi'' \ = 0  \ , \nonumber \\
&& p(p+1)\left(A'\right)^2 \,+\, 2(p+1)(D-p-2) A'C'\,+\, (D-p-2)(D-p-3)\left(C'\right)^2 \,-\, \frac{4\,(\phi')^2}{D-2}  \, \nonumber\\
 && + \ \frac{k\,p(p+1)}{\ell^2}\ e^{2(B-A)}\, + \, \frac{k'(D-p-3)(D-p-2)}{\ell^2}\ e^{2(B-C)} \,= \, 0  \ , \label{EqB_back_FnoT} 
  \eea
  so that $\phi'$ is determined by the Hamiltonian constraint, while the metric takes the form
\beq
ds^2 \ = \ - \ e^{2\left[(p+1)A+(D-p-2)C\right]} \,d\tau^2 \,+\, e^{2A} \,\ell^2\, ds_{p+1,k}^2 \,+\, e^{2C} \,\ell^2\, ds_{D-p-2,k'}^2 \ .
\eeq

The reduced system for $A$ and $C$ can be formulated more conveniently defining
  \bea
X &=& (p+1)A \ + \ (D-p-3) C \ , \nonumber \\
Y &=& p A \ + \ (D-p-2) C \ , 
  \eea
  and one is left with
  \bea
X'' &=& - \ \frac{k'(D-p-3)^2}{\ell^2}\, e^{2X} \  - \ \frac{k \,p(p+1)}{\ell^2}\, e^{2 Y} \ , \nonumber \\
Y'' &=& - \ \frac{k'(D-p-2)(D-p-3)}{\ell^2}\, e^{2X} \  - \ \frac{k \,p^2}{\ell^2}\, e^{2 Y} \ , \nonumber \\
&-& \frac{p(D-p-2)}{(D-2)} (X')^2  \,+\, \frac{2(D-p-2)(p+1)}{(D-2)}\, X'Y'\,-\, \frac{(p+1)(D-p-3)}{(D-2)} \,(Y')^2 \nonumber \\  &+& \frac{k\,p(p+1)}{\ell^2}\ e^{2 Y}\, + \, \frac{k'(D-p-3)(D-p-2)}{\ell^2}\ e^{2 X} \,= \, \frac{4}{(D-2)}\, (\phi')^2 \ . \label{eqskkp2}
  \eea

\subsubsection[\texorpdfstring{{\mdseries\textsc{An Adiabatic Approximation for the Solutions with $(k,k'=1)$}}}{An Adiabatic Approximation for the Solutions with (k,k'=1)}]
{{\mdseries\textsc{An Adiabatic Approximation for the Solutions with $(k,k'=1)$}}}\label{sec:adiabatic}

Some information on asymptotic regimes where $e^{2Y}$ is subdominant with respect to $e^{2X}$, or equivalently where $A \gg C$, can be deduced by relying on the analysis in Section~\ref{sec:kzerokpnotzero}. For $k'=1$, the preceding conditions can be satisfied in the past and in the future, depending on the value of $\Theta$. In the \emph{finite} past of these cosmologies, this demands that 
\beq
\cos\Theta < \sqrt{\frac{(p+1)}{(D-2)(D-p-2)}} \ \equiv \ \cos \Theta_0 \ , 
\eeq
while in their \emph{finite} future it demands that
\beq
\cos\Theta >  \ - \ \cos \Theta_0 \ . 
\eeq
On the other hand, the condition only applies to the \emph{finite} past of the cosmologies with $k'=-1$, but cannot hold in their \emph{infinite} future. 
\begin{figure}[ht]
\centering
\begin{tabular}{c}
\includegraphics[width=70mm]{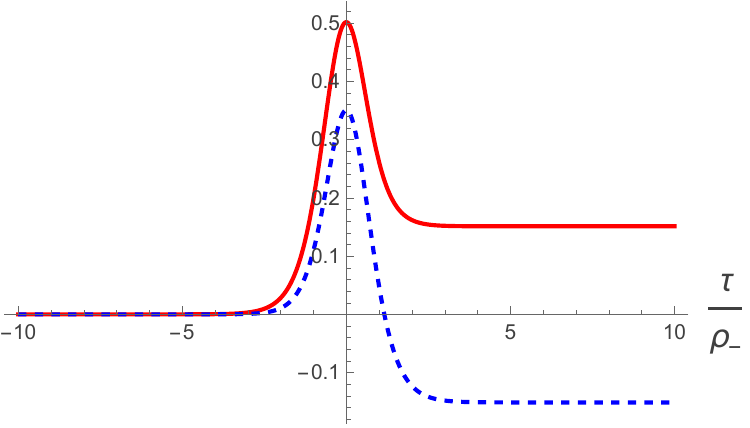} \\
\end{tabular}
 \caption{\small $\delta\,\rho$ (up to a positive overall factor) as a function of $\frac{\tau}{\rho_-}$ for $p=2$, $D=10$, $k=1$, and for $\cos\Theta_-=\frac{1}{2}\, \cos\Theta_0$  (red, solid) and for $\cos\Theta_-=-\, \frac{1}{2}\,\cos\Theta_0$ (blue, dashed). The actual magnitude of the effect is determined by $A_0$. }
 \label{fig:deltarho}
\end{figure}

Thus, adiabatic arguments can apply to the overall development for $k'=1$, when the total time duration $t_0$ is finite if $k=0$. Starting in the past with given values $(\rho_-,\Theta_-)$, if $A$ remains much larger than $C$, subdominant $k$ contributions introduce a mild time dependence in $\rho$ and $\Theta$, giving rise, in the future, to slightly altered values for them.
\begin{figure}[ht]
\centering
\begin{tabular}{ccc}
\includegraphics[width=65mm]{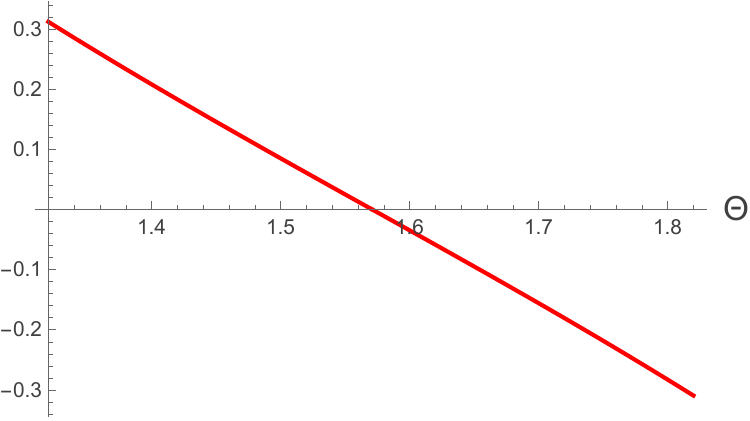} \quad  &
\includegraphics[width=65mm]{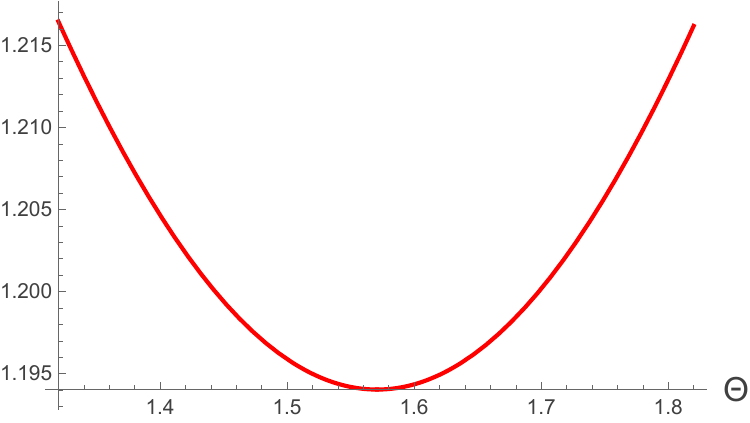}  \\
\end{tabular}
 \caption{\small The total variations of $\rho$ (left panel) and of $\Theta$ (right panel), up to overall normalizations determined by $A_0$, for $D=10$, $p=2$, $k=k'=1$, within the interval $(\pi+ \Theta_0, 2\pi - \Theta_0)$ for $\Theta$.}
\label{fig:deltatot}
\end{figure}
We now want to estimate these deformations and the resulting effect on $t_0$ in this type of cosmologies. To this end, we begin by noting that, in view of eqs.~\eqref{EqB_back_FnoT}, nonzero values of $k$ do not affect the evolution of $\phi$. Consequently, its zero mode, $\phi_1$, which depends on the combination $\frac{\sin\Theta}{\rho}$, should not change. This condition demands that
\beq
\delta\,\Theta \ = \ \tan\Theta\, \frac{\delta\,\rho}{\rho} \ ,
\eeq
while differentiating eq.~\eqref{eq_X} and making use of eqs.~\eqref{EqB_back_FnoT} in the result gives for $\rho$ the evolution equation
\beq
{\rho}' \ = \ k\, X'\,\frac{\rho^3\, p(p+1)}{\ell^2}\ e^{2X}\, e^{2(C-A)} \ . \label{rho_prime}
\eeq
This equation can be solved perturbatively, starting from an initial value $\rho_-$, by turning it into the integral equation
\beq
\delta\,\rho \ = \ k \int_{-\infty}^\tau \,d\tau \ \frac{\rho_-^3\, p(p+1)}{\ell^2} \, X_0'(\tau)\, e^{2X_0} \, e^{2(C_0-A_0)} \ ,
\eeq
where the $0$ subscript indicates that the different quantities are determined by the $k=0$ solution in eqs.~\eqref{coshtau}. In detail (see fig.~\ref{fig:deltarho}),
\beq
\delta\,\rho \ = \ - \ \frac{k\,\rho_-^3\,p(p+1)\,e^{-\, \frac{2(p+1)A_0}{D-p-3}}}{\ell^2} \left[\frac{\ell}{\rho_-(D-p-3)}\right]^{2\alpha}\int_{-\,\infty}^{x} dx' \ \frac{\tanh(x')}{\left[\cosh(x') \right]^{2\alpha}} \ e^{-\,\frac{2\,x'\,\cos\Theta_-}{(D-p-3)\cos\Theta_0}} \ ,
\eeq
where
\beq
\alpha \ = \ \frac{D-p-2}{D-p-3} \ ,
\eeq
so that the final value of $\rho$ is
\beq
\frac{\rho_+ \,-\, \rho_-}{\rho_-} \,=\, \frac{k\,p(p+1)\,e^{-\, \frac{2(p+1)A_0}{D-p-3}}}{(D-p-3)^{2\alpha}} \left(\frac{\ell}{\rho_-}\right)^{2\alpha-2}\int_{-\,\infty}^{\infty} dx' \ \frac{\tanh(x')}{\left[\cosh(x') \right]^{2\alpha}} \ e^{-\,\frac{2\,x'\,\cos\Theta_-}{(D-p-3)\cos\Theta_0}} \ ,
\eeq
and then
\beq
\Theta_+ \ = \ \Theta_- \ + \ \tan\Theta_-\, \frac{\rho_+\,-\,\rho_-}{\rho_-} \ .
\eeq
\begin{figure}[ht]
\centering
\begin{tabular}{ccc}
\includegraphics[width=70mm]{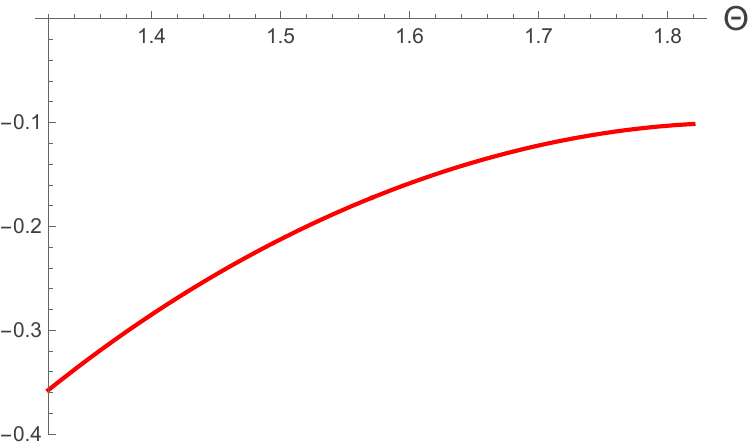}  \\
\end{tabular}
 \caption{\small The function $\delta\,t_0$ for $k=1$, up to a positive overall factor, as a function of $\Theta_-$, within the allowed range where $A \gg C$ initially. It is always \emph{negative}, which indicates a reduction of the cosmic time span for $k=1$ (and thus an increase for $k=-1$, since $\delta\,\rho$ is proportional to $k$). The actual magnitude of the effect is regulated by $A_0$. }
 \label{fig:deltat0}
\end{figure}
To first order in $\delta\,\rho$, the variation of the \emph{finite} total time span $t_0$ is then
\bea
\delta\,t_0 &=& e^{-\, \frac{(p+1)A_0}{D-p-3}}\,{\alpha} \left[\frac{\ell}{\rho_-(D-p-3)} \right]^\alpha \nonumber \\ &\times& \int_{-\,\infty}^\infty dx \  \frac{e^{-\,\alpha\,x\,\cos\Theta_-\cos\Theta_0}\left(x \, \frac{\cos\Theta_0}{\cos\Theta_-} + x\, \tanh x - 1\right)}{\left(\cosh x\right)^\alpha}\, \delta\rho(x) \ ,
\eea
and one finds numerically that the sign of $\delta\,t_0$ is opposite to the sign of $k$ (see fig.~\ref{fig:deltat0}). The time span is thus reduced for $k=1$ and increased for $k=-1$, due to the curvature of the spatial slices. Note that positive choices for $A_0$ can grant small values for both $\frac{\delta\,\rho}{\rho}$ and $\frac{\delta\,t_0}{t_0}$.

\subsubsection[\texorpdfstring{{\mdseries\textsc{Special Solutions with $k\neq 0$ and $k' \neq 0$}}}{Special Solutions with k ≠ 0 and k' ≠ 0}]
{{\mdseries\textsc{Special Solutions with $k\neq 0$ and $k' \neq 0$}}} \label{sec:kkpnoT}

The general setting with both $k$ and $k'$ is more complicated, but one can still find simple exact solutions where
\beq
Y \ - \ X \ = \ C \ - \ A \ = \ c \ ,
\eeq
with $c$ a constant, so that the system reduces to the single second-order equation
\beq
X'' \ = \ - \ k \,\frac{(D-p-3)(D-2)}{\ell^2} \, e^{2 X} \ , \label{eqXk}
\eeq
with
\beq
 k\, p \, e^{2c} \ = \ k'(D-p-3) \ . \label{kkp}
\eeq
Therefore, these solutions only exist if $k$ and $k'$ have identical signs and $0<p<D-2$, so that the spatial and internal slices can be curved. In this case, letting $k=k'$, the Hamiltonian constraint reduces to
\beq
\left(X'\right)^2 \ = \ \frac{1}{\rho^2} \ - \, k'\,\Delta^2\ e^{2 X}  \ , \label{hcrhokp}
\eeq
where
\beq
\Delta^2 \ = \ \frac{(D-p-3)(D-2)}{\ell^2} \ ,
\eeq
while
\beq
\phi \ = \ \pm \, \frac{\sqrt{D-1}}{2\,\rho}\,\tau \ + \ \phi_0 \ .
\eeq

One must now distinguish, as in the previous section, two cases.

\begin{enumerate}
\item If $k=k'=1$, the solution is
\beq
X \ = \ - \ \log\left[\Delta\,\rho\,\cosh\left(\frac{\tau}{\rho}\right)\right] \ , \label{solkkppos}
\eeq
where $- \infty< \tau<+\infty$, and then
\beq
A \ = \  \frac{X \ - \ c(D-p-3)}{(D-2)} \ , \qquad C \ = \ \frac{X \ + \ c(p+1)}{(D-2)} \ ,
\eeq
so that
\bea
ds^2 \!\!\!&=&\!\!\! \left(\frac{D-p-3}{p}\right)^{\frac{p+1}{D-2}}\Bigg[ \frac{- \ d\tau^2}{\left[ \Delta\,\rho\,\cosh\left(\frac{\tau}{\rho}\right)\right]^{\frac{2 (D-1)}{(D-2)}}} \nonumber \\ &+& \ell^2\,\frac{{p}\ ds^2_{p+1,1} \ + \ (D-p-3)\ ds^2_{D-p-2,1}}{(D-p-3) \left[ \Delta\,\rho\,\cosh\left(\frac{\tau}{\rho}\right)\right]^{\frac{2}{(D-2)}}} \Bigg] \ ,\nonumber \\
e^\phi &=& e^{\pm\, \frac{\tau\,\sqrt{D-1}}{2\,\rho} \ + \ \phi_0} \ . \label{specialkkppos}
\eea
By construction, these solutions are characterized by a constant ratio between the squared radii of the internal and spatial slices, equal to
\beq
e^{2c} \ = \ \frac{(D-p-3)}{p} \ ,
\eeq
in view of eq.~\eqref{kkp}, so that the two curvatures labeled by $k$ and $k'$ have comparable effects.
\begin{figure}[ht]
\centering
\begin{tabular}{ccc}
\includegraphics[width=70mm]{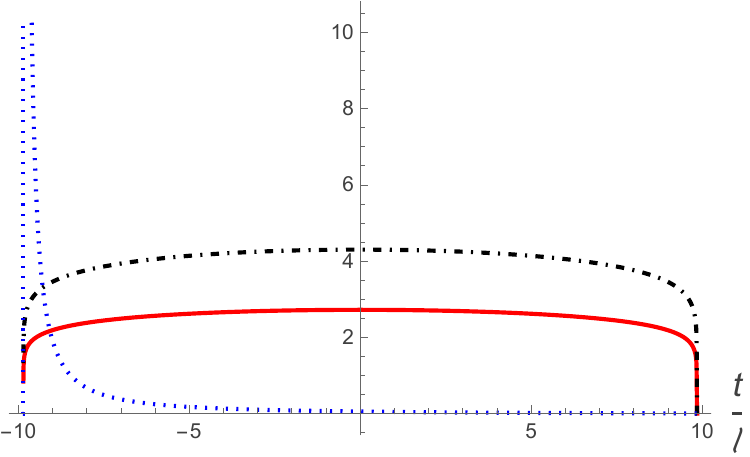} \quad  &
\includegraphics[width=70mm]{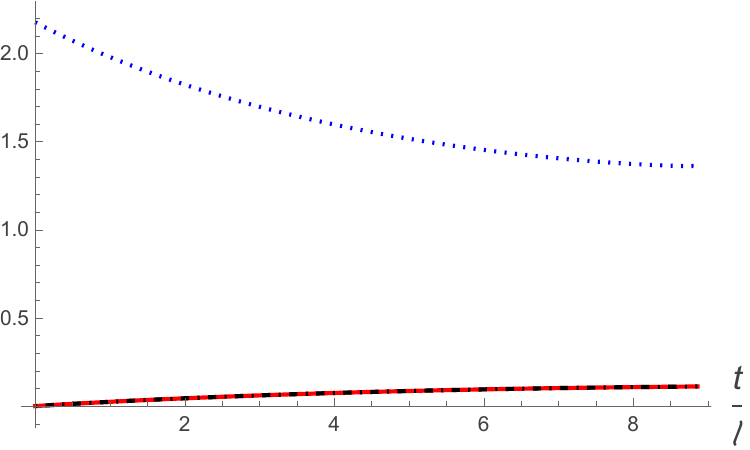}  \\
\end{tabular}
 \caption{\small Left panel: The solutions $e^A$ (red, solid), $e^C$ (black, dot-dashed) and $e^\phi$ (blue, dotted), in cosmic time for $D=10$, $p=2$, $k=k'=1$, within the interval $(-t_0<t<t_0)$. Right panel: how the combinations $(t-t_0)A'$ (red, solid), $(t-t_0)C'$ (black, dot-dashed) and $(t-t_0)\phi'$ (blue, dotted) approach their limiting values  $(\frac{1}{D-1}\simeq 0.11,\frac{1}{D-1}\simeq 0.11,\frac{(D-2)}{2\sqrt{D-1}}\simeq 1.33)$ as $t \to t_0^-$, for $D=10$, $p=2$. }
\label{fig:kkppos}
\end{figure}

These cosmologies have a finite past and a finite future span in cosmic time, and their complete evolution unfolds within an interval whose duration we now call
\beq
2\,t_0 \ = \ \left(\frac{D-p-3}{p}\right)^{\frac{p+1}{2(D-2)}} \int_{-\infty}^\infty \frac{d\tau}{\left[ \Delta\,\rho\,\cosh\left(\frac{\tau}{\rho}\right)\right]^{\frac{(D-1)}{(D-2)}}} \ .
\eeq

As $\tau \to \pm\,\infty$ the solution approaches
\bea
ds^2 \!\!\!&\sim&\!\!\! - \  e^{-\, \frac{2(D-1)\left|\tau\right|}{(D-2)\rho}}\ d\tau^2 \, + \, \ell^2\ e^{-\, \frac{2\left|\tau\right|}{(D-2)\rho}}\frac{\left( p\,ds^2_{p+1,1} \, + \, (D-p-3)\,ds^2_{D-p-2,1}\right)}{(D-p-3)}  \  , \nonumber \\
e^\phi &\sim& e^{\pm\, \frac{\tau\,\sqrt{D-1}}{2\,\rho}} \ ,
\eea
which becomes, in cosmic time
\bea
ds^2 \!\!\!&\sim&\!\!\! - \  dt^2 \, + \, \ell^2\ \left(1 \,-\, \frac{\left|t\right|}{t_0} \right)^\frac{2}{D-1} \frac{\left( p\,ds^2_{p+1,1} \, + \, (D-p-3)\,ds^2_{D-p-2,1}\right)}{(D-p-3)} \  , \label{specialkkpposcosmic} \\
e^\phi &=& \begin{cases}
     \left(1 \,-\, \frac{t}{t_0}\right)^{\mp\, \frac{(D-2)}{2\sqrt{D-1}}} \qquad\qquad t \to \ \ t_0 \ ; \nonumber \\
      \left(1 \,+\, \frac{t}{t_0}\right)^{\pm\, \frac{(D-2)}{2 \sqrt{D-1}}} \qquad\qquad t \to -\, t_0 \ . 
\end{cases} \ ,
\eea
for $t$ close to $\pm t_0$. 

\begin{figure}[ht]
\centering
\begin{tabular}{ccc}
\includegraphics[width=70mm]{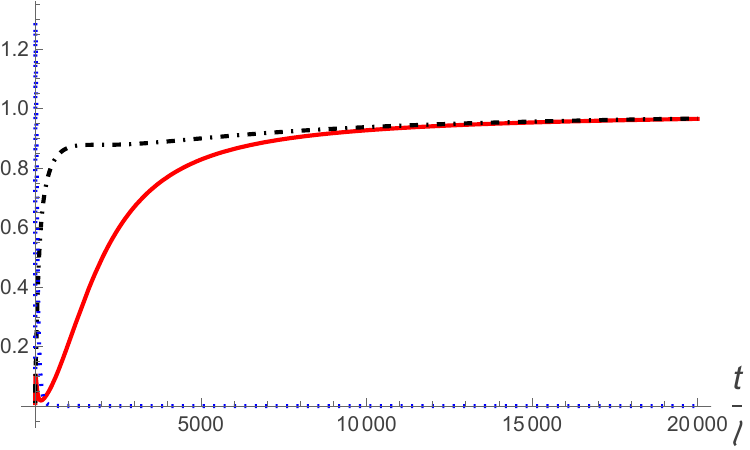} \quad  &
\includegraphics[width=70mm]{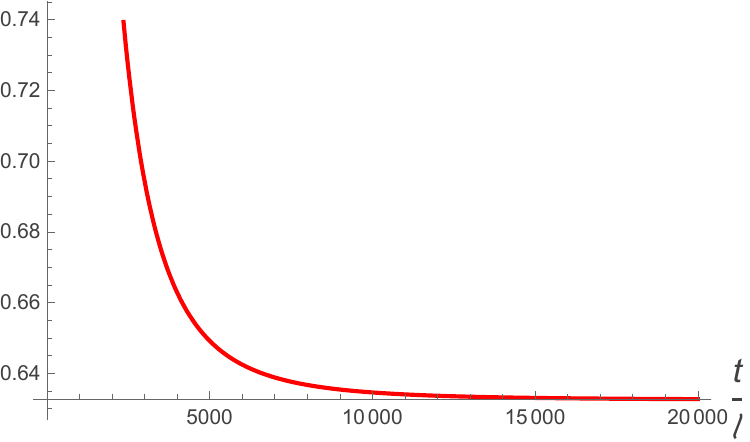}  \\
\end{tabular}
 \caption{\small The case $k=k'=-1$. Left panel: The limiting linear behavior of $t A'$ (red, solid) and $t C'$ (black, dot-dashed). Right panel: how $e^{A-C}$ approaches $\sqrt{\frac{p}{D-p-3}}=0.63$ for $D=10$, $p=2$. }
\label{fig:kkpminus}
\end{figure}
\item If $k=k'=-1$, the solution is
\beq
X \ = \ - \ \log\left[\Delta\,\rho\,\sinh\left(\frac{\left|\tau\right|}{\rho}\right)\right] \ , \label{solkkpneg}
\eeq
where $- \infty< \tau<0$, and then
\bea
ds^2 \!\!\!&=&\!\!\! - \  \frac{e^{\frac{2c(p+1)}{(D-2)}}\ d\tau^2}{\left[ \Delta\,\rho\,\sinh\left(\frac{\left|\tau\right|}{\rho}\right)\right]^{\frac{2 (D-1)}{(D-2)}}} \, + \, \ell^2\,\frac{e^{-\,\frac{2c(D-p-3)}{(D-2)}}\ ds^2_{p+1,-1} \, + \, e^{\frac{2c(p+1)}{(D-2)}}\ ds^2_{D-p-2,-1}}{\left[ \Delta\,\rho\,\sinh\left(\frac{\left|\tau\right|}{\rho}\right)\right]^{\frac{2}{(D-2)}}} \nonumber \\
e^\phi &=& e^{\pm\, \frac{\tau\,\sqrt{D-1}}{2\,\rho} \ + \ \phi_0} \ , \label{specialkkpneg}
\eea
and there is again a finite past but an infinite future, so that the cosmic time span is $0<t<+\infty$. In this case the limiting behavior at early times is captured by
\bea
ds^2 \!\!\!&\sim&\!\!\! - \  dt^2 \, + \, \ell^2\ \left|M\,t\right|^\frac{2}{D-1} \left[ p\,ds^2_{p+1,-1} \, + \, (D-p-3)\,ds^2_{D-p-2,-1}\right] \  , \nonumber \\
e^\phi &\sim& \left|M\,t\right|^{\pm\,\frac{1}{\sqrt{D-1}}} \ , \label{specialkkpnege}
\eea
where $M$ is a scale determined by $\rho$, which is a free Kasner behavior with two identical exponents for $A$ and $C$. On the other hand, at late times
\bea
ds^2 \!\!\!&\sim&\!\!\! - \  dt^2 \, + \,\frac{t^{2}}{(D-2)} \Big[ p \ ds^2_{p+1,-1} \, + \, (D-p-3) \ ds^2_{D-p-2,-1}\Big] \  , \nonumber \\
e^\phi &\sim& e^{\phi_0} \ , \label{milnelike}
\eea
so that one ends up with direct products of linearly expanding spatial and internal slices. In particular, in the isotropic case, when $p=D-2$, these cosmologies approach a Milne universe.
Note that eqs.~\eqref{milnelike} solve the system exactly in the $\rho \to \infty$ limit.
\end{enumerate}

\subsubsection{\sc The Stability Issue}

 In order to address the stability of these fine-tuned exact solutions, one can expand eqs.~\eqref{eqskkp2}, letting
 \beq
X \ =  X_0 \ + x \ , \qquad Y \ = \ X_0 \ + \ c \ + y \ ,
 \eeq
where $X_0$ is the background solution given in eq.~\eqref{solkkppos} for $k'=1$ and in eq.~\eqref{solkkpneg} for $k'=-1$. Linearization in $x$ and $y$ then gives the system
\beq
\left(\begin{array}{c} x'' \\ y'' \end{array} \right) \ + \ 2 k' \frac{(D-p-3)}{\ell^2}\  e^{2 X_0} {\cal M } \left(\begin{array}{c} x \\ y \end{array} \right)  \ = \ 0 \ ,
\eeq
where
 \beq
{\cal M} \ = \  \left( \begin{array}{cc} (D-p-3)  & (p+1) \\ (D-p-2) & p\end{array} \right)
 \eeq
 after using eqs.~\eqref{eqskkp2} and \eqref{eqXk}.
 
${\cal M}$ has the following eigenvalues and eigenvectors:
\bea
\lambda_1 &=& D-2 \ , \qquad \qquad v_1 \ = \ \left(\begin{array}{c} 1 \\ 1  \end{array}  \right) \nonumber \\
\lambda_0 &=& - \ 1 \ , \qquad \qquad v_2 \ = \ \left(\begin{array}{c} p+1 \\ -\left(D-p-2\right)  \end{array}  \right) \ . \label{v_array}
\eea
The first corresponds to time translations and deformations of the width $\rho$, the two parameters that determine $X_0$, while the second involves deformations affecting $X-Y$, which are captured by the second-order equation (with $k=k'$)
\beq
\xi'' \ - \ 2\,k'\,\frac{(D-p-3)}{\ell^2}\ e^{2X_0}\, \xi \ = \ 0 \ . \label{eqxi2}
\eeq
In addition, the Hamiltonian constraint~\eqref{hcrhokp} demands that, at all times,
\beq
\delta\,\phi'\ = \ 0 \ , 
\eeq
since $\phi'$ is purely determined by $\rho$ that is kept constant.
 Demanding that the deformation correspond to the non-trivial eigenvector $v_2$ in eq.~\eqref{v_array} links $\delta\,A$ and $\delta\,C$ to a single independent deformation $x$ according to
\beq
\delta\,A \ = \ \left(D-p-2\right) x \ , \qquad \delta\,C \ = \ - \ \left(p+1\right) x \ .
\eeq

The unperturbed solution $X_0$ is given in eq.~\eqref{solkkppos} for $k=k'=1$, which leads to
    \beq
x'' \ - \ \frac{2}{(D-2)\,\rho^2\,\cosh^2\left(\frac{\tau}{\rho}\right)}\, x \ = \ 0 \ , \label{eqxikkppos}
\eeq
where $-\infty<\tau<+ \infty$,
and in eq.~\eqref{solkkpneg} for $k=k'=-1$, which leads to
    \beq
x'' \ + \ \frac{2}{(D-2)\,\rho^2\,\sinh^2\left(\frac{\tau}{\rho}\right)}\, x \ = \ 0 \ , \label{eqxikkpneg}
\eeq
where $-\infty<\tau<0$.

These equations can be solved in terms of Legendre functions, but as $\tau \to \pm\,\infty$ in the first case, and as $\tau \to - \infty$ in the second, the solutions are at most linear. For finite values of $\rho$, the perturbations can thus overcome the backgrounds at early and late times for $k=k'=1$, and at early times for $k=k'=-1$. The former setup is thus \emph{unstable in the past and in the future}, while the latter is \emph{unstable in the past}. On the other hand, as $\tau \to 0^-$ eq.~\eqref{eqxikkpneg} reduces to
\beq
x'' \ + \ \frac{2}{(D-2)\,\tau^2}\ x \ = \ 0 \ , \label{eqxikkpneg0}
\eeq
which is solved by~\footnote{If $D=10$ the two indicial roots coincide, but the other independent solution, $(-\tau)^\frac{1}{2}\log(-\tau)$ is still sub-dominant.}
\beq
x \ = \  C_1 \, (-\tau)^{\frac{1 + \sqrt{\frac{D-10}{D-2}}}{2}} \ +\ C_0 \, (-\tau)^{\frac{1 - \sqrt{\frac{D-10}{D-2}}}{2}} \ ,
\eeq
and the two contributions are always subdominant with respect to $X_0$, which has a logarithmic singularity as $\tau \to 0^-$, so these solutions for $k=k'=-1$ are \emph{stable in the future} for finite values of $\rho$ if $D \leq 10$ and are \emph{unstable} if $D>10$. However, in the $\rho \to \infty$ limit, eq.~\eqref{eqxikkpneg0} describes the perturbations for all values of $\tau$, and as $\tau \to - \infty$ the solutions override the background, which is everywhere logarithmic in this limit. Therefore, the exact solution in eqs.~\eqref{milnelike} is \emph{unstable in the past}.

\section{\sc Cosmologies with Only Tension} \label{sec:tnot0curv0}
This case was already discussed at length in~\cite{ms21_nonsusy} for $D=10$~\footnote{We warn the reader that the present notation is slightly at variance with respect to that paper.}, so that we confine ourselves to complementing those results. 

In general, it is convenient to define, in the harmonic gauge, the three quantities
    \bea
    W &=& (p+1)A \ + \ (D-p-2) C \ + \ \frac{\gamma}{2}\, \phi  \, \nonumber \\
    V &=&  A  - \ C \ , \qquad K \ = \ \phi \ + \ \frac{\gamma(D-2)^2}{8} \ A 
    \label{Wdef2}
    \eea
but there is a special value of $\gamma$,
  \beq 
    \gamma_c \ = \ 4 \frac{\sqrt{D-1}}{D-2} \ , \label{gammac1}
\eeq 
where $W \,-\,\frac{\gamma}{2}\,K$ is proportional to $V$, so that three quantities are not independent. Let us begin by considering this case.

\subsection[\texorpdfstring{{\mdseries\textsc{Solutions with $\gamma = \gamma_c$}}}{Solutions with  gamma = gammac }]
{{\mdseries\textsc{Solutions with $\gamma = \gamma_c$}}}

In this case one can work in terms of $W$, $A$ and $C$, which satisfy
\bea
W'' &=& 0 \ , \nonumber \\
A'' &=& C'' \ = \ \frac{T}{D-2}\,e^{2W} \ , \label{eqs_ACW}
\eea
so that
    \beq
W \ = \ \beta\,\tau \ + \ W_0 \ , \label{Wtau}
    \eeq
while the Hamiltonian constraint reduces to
 \begin{align}
&p(p+1)\left(A'\right)^2 \,+\, 2(p+1)(D-p-2) A'C'\,+\, (D-p-2)(D-p-3)\left(C'\right)^2  \nonumber \\
&-\, \frac{(D-2)\,\left(W'- (p+1)A'-(D-p-2)C'\right)^2}{(D-1)} 
 \,- \, {T} \, e^{\, 2\,W}\,= \, 0 \ . \label{EqB_red_T}
 \end{align}

\subsubsection[\texorpdfstring{{\mdseries\textsc{The ${\beta = 0}$ Case\ (($0a$) solutions)}}}{The beta = 0 Case \ ((0a) solutions)}]
{{\mdseries\textsc{The ${\beta = 0}$ Case \ (($0a$) solutions)}}}     

    If $\beta=0$, the solution takes the form
    \bea
A &=& A_1\,\tau \ + \ A_0 \ + \ \frac{T\,e^{2 W_0}\,\tau^2}{2(D-2)} \ , \nonumber \\
C &=& C_1\,\tau \ + \ C_0 \ + \ \frac{T\,e^{2 W_0}\,\tau^2}{2(D-2)} \ , \nonumber \\
\phi &=& \phi_1\,\tau \ + \ \phi_0  \ - \ \frac{T\,e^{2 W_0}\,\sqrt{D-1}\tau^2}{4} \ ,
    \eea
    where 
    \beq
    W_0 \ = \ (p+1)A_0 \ + \ (D-p-2)C_0 \ + \ \frac{\gamma_c}{2}\, \phi_0 \ ,
    \eeq
while the Hamiltonian constraint reduces to
\beq
T\,e^{2\,W_0} \ = \ - \ \frac{(A_1-C_1)^2(p+1)(D-p-2)}{D-1} \ ,\label{neg_T2}
\eeq
and is only consistent if $T<0$. In this case
\beq
A_1 \ - \ C_1 \ = \ \pm  \sqrt{\frac{(D-1) \left|T\,e^{2\,W_0} \right|}{(p+1)(D-p-2)}} \ , 
\eeq
while
\beq
\phi_1 \ = \ - \ \frac{2}{\gamma_c} \left[(D-1) C_1 \ \pm \  \sqrt{\frac{(D-1)(p+1) \left|T\,e^{2\,W_0}\right|}{(D-p-2)}}  \right] \ ,
\eeq
so that the solutions
\bea
ds^2 &=& -\,  e^{2 W_0-\,\gamma_c\left(\phi_1\,\tau\,+\,\phi_0\right) \,-\, \frac{\left|T\,e^{2\,W_0}\right| (D-1)\,\tau^2}{(D-2) }} \, d\tau^2  + e^{-\, \frac{\left|T\,e^{2\,W_0}\right|\tau^2}{(D-2)}}\left[ e^{2 A_1\,\tau} d\vec{x}_{p+1}^2 + e^{2 C_1\,\tau} d\vec{y}_{D-p-2}^2 \right] \ , \nonumber \\
e^\phi &=& e^{\phi_1\,\tau\,+\,\phi_0 \, + \, \frac{\left|T\,e^{\,W_0}\right|\, \sqrt{D-1}\,\tau^2}{4}} \label{sol_gamma_c_beta0}
\eea
form a two-parameter family, which can be taken to be $W_0$ and $C_1$, after absorbing $A_0$ and $C_0$ by rescaling the $\vec{x}$ and $\vec{y}$ coordinates.

These cosmologies evolve within a finite interval of cosmic time of total duration $t_0^++t_0^-$, with
\beq
t_0^+ \ + \ t_0^- \ = \ \int_{-\infty}^{+\infty} d\tau \ e^{-\,\frac{\gamma_c}{2}\left(\phi_1\tau+\phi_0\right)} \ e^{-\, \frac{|T|e^{2W_0}(D-1) \tau^2}{2 (D-2)}} \ ,
\eeq
and the asymptotic behavior at early and late times is captured by
\bea
ds^2&\sim& - \ dt^2 \ + \ \left(t_0^\pm \ - \ |t|\right)^\frac{2}{D-1}\left(d\vec{x}^2 \ + \ d\vec{y}^2\right) \ , \nonumber \\
e^\phi &\sim&  \left(t_0^\pm \ - \ |t|\right)^{-\,\frac{2}{\gamma_c}} \ , \label{asymptbeta0}
\eea
so that there is strong coupling at both ends, as expected from the free-Kasner behavior of eqs.~\eqref{param_theta} with $\theta=-\frac{\pi}{2}$ close to its initial singularity.

\subsubsection[\texorpdfstring{{\mdseries\textsc{The ${\beta \neq 0}$ Case \ (($0b$) solutions)}}}{The beta ≠ 0 Case \ ((0b) solutions)}]
{{\mdseries\textsc{The ${\beta \neq 0}$ Case \ (($0b$) solutions)}}}
\begin{figure}[ht]
\centering
\begin{tabular}{ccc}
\includegraphics[width=70mm]{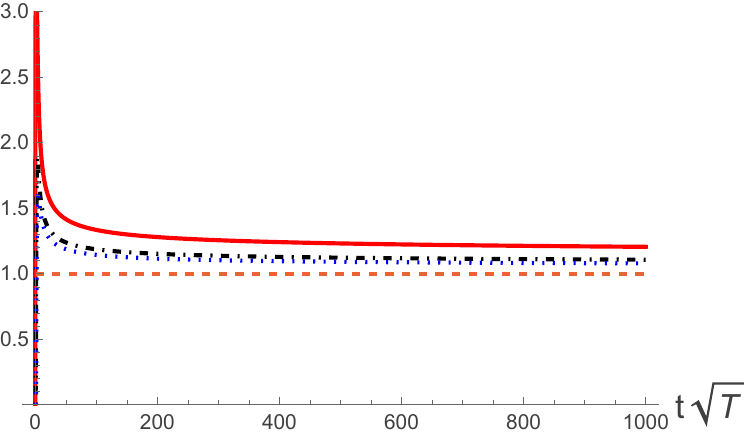} \quad  &
\includegraphics[width=70mm]{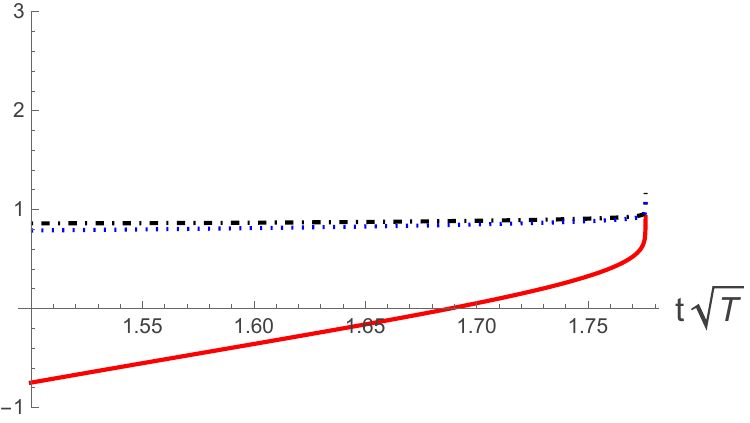}  \\
\end{tabular}
 \caption{\small Typical solutions with $\gamma=\gamma_c$, $D=10$, $p=2$. Left panel ($T>0$): The large-time behavior of $(D-1) u A'(u)$ (red, solid), $(D-1) u C'(u)$ (black, dot-dashed), $- \frac{\gamma_c}{2} \, u \phi'(u))$ (blue, dotted) and the asymptotic value 1 (orange, dashed). Right panel ($T<0$): the behavior of $(D-1)(v-v_0) A'(v)$ (red, solid), $(D-1) (v-v_0) C'(v)$  (black, dot-dashed), $- \frac{\gamma_c}{2} \, (v-v_0) \phi'(v))$  (blue, dotted) close to the final singularity.}
\label{fig:gammacriticalTposneg}
\end{figure}

For $\beta \neq 0$, up to a translation of $\tau$, eqs.~\eqref{EqB_back_F} are solved by
    \bea
A &=& \frac{T}{4 (D-2)\beta^2}\, e^{2 \,\beta\,\tau} \ + \ A_1\,\tau \ + \ A_0  \ , \nonumber \\
C &=& \frac{T}{4 (D-2)\beta^2}\, e^{2 \,\beta\,\tau} \ + \ C_1\,\tau \ + \ C_0 \ , \nonumber \\
\phi &=& - \ \frac{T\,\gamma_c \,(D-2)}{32\,\beta^2}\, e^{2 \,\beta\,\tau} \ + \ \phi_1\,\tau \ + \ \phi_0 \ , 
    \eea
where
\bea
&& (p+1)A_1 \ + \ (D-p-2)C_1 \ + \ \frac{\gamma_c}{2}\, \phi_1 \ = \ \beta \ , \nonumber \\
&& (p+1)A_0 \ + \ (D-p-2)C_0 \ + \ \frac{\gamma_c}{2}\, \phi_0 \ = \ 0 \ ,
\eea
due to the preceding condition on $W$.

The Hamiltonian constraint then reduces to
\bea
\left[(p+1)A_1 \,+\,(D-p-2) C_1\right]^2 \,=\, (p+1) A_1^2\,+\, (D-p-2)C_1^2 \,+\, \frac{4\,(\phi_1)^2}{D-2}    \ , \label{ham_c02}
\eea
which has exactly the form obtained for the free Kasner solutions of Section~\ref{sec:t0curv0}.
Indeed, the independent choices of $A_1$, $C_1$ and $\phi_1$ can again be parametrized in terms of an angle $\theta$, since the three exponents 
\beq
\alpha_A \ = \ \frac{A_1}{\mu} \ , \qquad \alpha_C \ = \ \frac{C_1}{\mu} \ , \qquad \alpha_\phi \ = \ \frac{\phi_1}{\mu} \ , \qquad 
\eeq
and
\beq
\mu \ = \ (p+1) A_1 \ + \ (D-p-2) C_1 \ ,
\eeq
grant that
\beq
(p+1) \alpha_A \ + \ (D-p-2) \alpha_C \ = \ 1 \ .
\eeq
One thus recovers the first of eqs.~\eqref{free}, while the Hamiltonian constraint~\eqref{ham_c02} can similarly be turned into the second one. However, here the range of $\theta$ is to be restricted to the region $\sin\theta \neq -1$, so that 
\beq
W'(\theta) \ \equiv \ \beta(\theta) \ = \ \mu \left(1 \ + \ \sin\theta\right) 
\eeq
does not vanish.

Letting 
\beq
t \ = \ \frac{1}{\mu}\ e^{\mu\tau}
\eeq
the complete solution finally reads
\bea
ds^2 &=& - \ dt^2\, e^{\frac{T(D-1)}{2 (D-2)\beta^2(\theta)}\, \left(\mu t\right)^{2(1+\sin\theta)}\,-\, \gamma_c\,\phi_0} \ + \ \left(\mu t\right)^{2\alpha_A(\theta)}\, e^{\frac{T}{2 (D-2)\beta^2(\theta)}\, \left(\mu t\right)^{2(1+\sin\theta)}\,+\, 2\,A_0} d\vec{x}^2 \nonumber \\ &+& \left(\mu t\right)^{2\alpha_C(\theta)}\, e^{\frac{T}{2 (D-2)\beta^2(\theta)}\, \left(\mu t\right)^{2(1+\sin\theta)}\,+\, 2\,C_0} d \vec{y}^2 \ , \nonumber \\
e^\phi &=& \left(\mu t\right)^{\frac{2}{\gamma_c}\sin\theta}\ e^{-\,\frac{T\,\gamma_c \,(D-2)}{32\,\beta^2(\theta)}\, \left(\mu t\right)^{2(1+\sin\theta)}\,+\, \phi_0} \ , \label{gammac_init}
\eea
where $0<t<\infty$ if $\mu>0$ and $- \infty<t<0$ if $\mu<0$. Up to time reversal, one can assume that $\mu>0$, which we do in the following, and then $t$ plays the role of cosmic time near the initial singularity. Moreover, $A_0$ and $C_0$ can be absorbed by rescaling $\vec{x}$ and $\vec{y}$, so the solution depends only on the scale $\mu$ and on the two parameters $\theta$ and $\phi_0$. 

Irrespective of the sign of $T$, all these cosmologies have a \emph{finite} past, and as $t \to 0$ they approach 
\bea
ds^2 &\sim& - \ dt^2 \ + \ \left(\mu t\right)^{2\alpha_A(\theta)}\, d\vec{x}^2 \nonumber \ +\  \left(\mu t\right)^{2\alpha_C(\theta)}\, d \vec{y}^2 \ , \nonumber \\
e^\phi &\sim& \left(\mu t\right)^{\frac{2}{\gamma_c}\,\sin\theta} \ , \label{gammac_init2}
\eea
which is one of the free Kasner solutions of Section~\ref{sec:t0curv0}.

When $T>0$, these cosmologies have an \emph{infinite} future, and their large-time behavior is captured by
\bea
ds^2 &\sim& - du^2 \ + u^\frac{2}{D-1} \left(d\vec{x}^2 \ + \ d\vec{y}^2\right) \ , \nonumber \\
e^\phi &\sim& u^{-\, \frac{2}{\gamma_c}} \ , \label{criticalTpos}
\eea
as $u$, the proper time, becomes very large. Note that this final behavior is universal and independent of $\theta$. 

On the other hand, when $T<0$ the system has a \emph{finite} future and undergoes a big crunch at
\beq
v_0(\theta) \ = \ \int_0^\infty dt \ e^{-\, {\frac{ |T|(D-1)}{4 (D-2)\beta^2(\theta)}\, \left(\mu t\right)^{2(1+\sin\theta)}}\,-\, \frac{1}{2}\, \gamma_c\,\phi_0} \ .
\eeq
The limiting behavior as $t\to+\infty$ is captured by
\bea
ds^2 &\sim& - dv^2 \ + \left(v_0(\theta) \ - \ v\right)^\frac{2}{D-1} \left(d\vec{x}^2 \ + \ d\vec{y}^2\right) \ , \nonumber \\
e^\phi &\sim&  \left(v_0(\theta) \ - \ v\right)^{-\, \frac{2}{\gamma_c}} \ , \label{criticalTneg}
\eea
where now the proper time $v$ is close to $v_0(\theta)$.
This behavior occurs at strong coupling, and is characterized by a universal free-Kasner exponent, independent of $\theta$, which is identical to what we found in eqs.~\eqref{asymptbeta0}, although the total time span $v_0(\theta)$ depends on $\theta$ and also on $|T|$. The typical features of solutions of this type (for $D=10, p=2$) are illustrated in fig.~\ref{fig:gammacriticalTposneg}.

In summary, the past limiting behavior of these backgrounds is Kasner-like and independent of the sign of $T$. The exponents that characterize the future asymptotics, which is isotropic, are universal and independent of both $\theta$ and the sign of $T$, and are captured by eqs.~\eqref{param_theta} with $\theta\,=\,-\,\frac{\pi}{2}$. However, for $T>0$ the limiting behavior is approached at weak coupling in the infinite future, as $u\to\infty$, so that it is directly captured by a free Kasner solution of Section~\ref{sec:t0curv0}, while for $T<0$ it is approached at strong coupling in the finite future, as $v \to v_0^-$, so that the correspondence is with the time reversal of the previous free Kasner solution. However, as we have stressed, $v_0$ is not universal, but depends on both $\theta$ and $|T|$.

For $T>0$ the future asymptotics is fixed, so there is no future stability issue, but small early perturbations are governed by small variations $\delta\alpha_A$, $\delta\alpha_C$ and $\delta\alpha_\phi$ induced by a small deformation $\delta\theta$. Therefore, the solutions are \emph{stable in the past} away from the special values of $\theta$ where one of these exponents vanishes, and this result applies to tensions of both signs. 
On the other hand, for $T<0$ perturbations affecting $v_0$ introduce instabilities close to the final big crunch.

\subsection[\texorpdfstring{{\mdseries\textsc{Solutions with $\gamma \neq \gamma_c$}}}{Solutions with  gamma ≠  gammac }]
{{\mdseries\textsc{Solutions with $\gamma \neq \gamma_c$}}} \label{sec:gamma_not_gammac}

For $\gamma \neq \gamma_c$ one can work with three variables $V$, $K$ and$W$, which are independent, and one can invert eqs.~\eqref{Wdef2} to obtain
\bea
A &=& -{16}\ \frac{W  \ - \ \frac{\gamma}{2}\, K \ + \ (D - p - 2) V}{(D-2)^2 \left(\gamma^2 \ - \ \gamma_c^2\right)} \ , \nonumber \\
C &=& A \ - \ V   \ , \nonumber \\
\phi &=& K \ -  \ \frac{\gamma(D-2)^2}{8}\, A \ , 
\eea
while the second-order equations become
 \bea
 V '' &=&  0  \nonumber \ ,  \\
K'' &=& 0  \ , \nonumber \\
W'' &=&   - \ \frac{D-2}{16}\left(\gamma^2 -\gamma_c^2\right)T\ e^{2W} \ , \label{ACphieqsn}
  \eea
The first two of eqs.~\eqref{ACphieqsn} can be integrated to give
\beq
V \ = \ v_1\,\tau \ + \ v_0 \ , \qquad K \ = \ K_1\,\tau \ + \ K_0 \ .
\eeq
while the Hamiltonian constraint becomes the energy-conservation condition for $W$
\beq
(W')^2 \ +\  \frac{(D-2)\bigl(\gamma^2-\gamma_c^2\bigr)}{16} \,T\,e^{2W}  \ = \ {\cal E} \ .  \label{en_W}
\eeq

It is now convenient to trade $K_1$ for
\beq
\phi_1 \ = \ \frac{2\,\gamma(D-p-2) v_1 \ - \ \gamma_c^2\, K_1}{\gamma^2-\gamma_c^2} \ , 
\eeq
the contribution to $\phi$ linear in $\tau$, thus obtaining for ${\cal E}$ the convenient expression
\beq
{\cal E} \ = \ \frac{\left(\gamma^2-\gamma_c^2\right)^2 \phi_1^2}{4\,\gamma_c^2} \ -  \   \frac{(D-p-2)(p+1) \,v_1^2}{(D-2)\gamma_c^2} \left(\gamma^2-\gamma_c^2 \right) \ . \label{ham_constr}
\eeq
Defining also
\beq
\phi_0 \ = \ \frac{2\,\gamma(D-p-2) v_0 \ - \ \gamma_c^2\, K_0}{\gamma^2-\gamma_c^2} \ ,
\eeq
the solutions finally take the form~\footnote{Note that here $v_1$ is redefined by an overall factor of two compared to~\cite{ms21_nonsusy}.}
\bea
ds^2 \!\! &=&\!\! - \ e^{\omega\left(D-1\right) W \,-\,\gamma\left(\phi_1\,\tau+\phi_0\right)}\ d\tau^2 \ \nonumber \\ \!\!&+&\!\! e^{\omega\, W\,+\, 2\left(v_1\tau+v_0\right)\,-\,\frac{\gamma}{D-1}\left(\phi_1\,\tau+\phi_0\right)}\left[ e^{-\,2 \frac{\left(p+1\right)\left(v_1\,\tau+v_0\right)}{D-1}} d\vec{x}^2 \ + \ e^{-\,2 \frac{\left(D-p-2\right)\left(v_1\,\tau+v_0\right)}{D-1}}d\vec{y}^2\right] , \nonumber \\
e^\phi\!\!&=&\!\!  e^{\phi_1\,\tau+\phi_0} \ e^{- \ \frac{(D-2)^2\,\gamma\,\omega\,W}{16}}  \ .
\eea
where 
\beq
\omega \ = \ \frac{32}{(D-2)^2\left(\gamma_c^2 \ - \ \gamma^2\right)} \ = \ \frac{2}{D-1} \ \frac{1}{1 \ - \ \left(\frac{\gamma}{\gamma_c}\right)^2}\label{omega} \ .
\eeq
For later convenience, we also define
\beq
\Xi \ = \ 16(p+1) + \gamma^2 (D-2)^2(D-p-3) \ . \label{Xi}
\eeq

In analogy with~\cite{ms21_nonsusy}, it is now convenient to define
\beq
\Delta^2 \ = \ \frac{(D-2)\bigl| T\left(\gamma_c^2-\gamma^2\right)\bigr|}{16} \ , \label{Deltagamma}
\eeq
so that the Hamiltonian constraint takes the form
\beq
(W')^2 \ = \ \epsilon_1\,\epsilon_2 \,\Delta^2 \,e^{2W} \ + \ \epsilon_3\left|{\cal E}\right| \ , \label{cons_W}
\eeq
where
\beq
\epsilon_1 \ = \ \sign\left(\gamma_c^2 \ - \ \gamma^2 \right) \ , \qquad \epsilon_2 \ = \ \sign\left( T \right) \ , \qquad \epsilon_3 \ = \ \sign\left({\cal E} \right) \ .
\eeq

Naively, there are eight sign choices in eq.~\eqref{cons_W}, but there are some restrictions and some additions:
\begin{itemize}
\item if $\epsilon_1\,\epsilon_2 <0$, then $\epsilon_3>0$;
\item if $\epsilon_1>0$, eq.~\eqref{ham_constr} demands $\epsilon_3>0$;
\item if $\epsilon_1=-1$, there are two branches of solutions labeled by $\epsilon'=\pm 1$;
\item furthermore, ${\cal E}$ can be zero if $\epsilon_1 \epsilon_2>0$, and we label these cases with $\epsilon_3=0$.
\end{itemize}
\begin{table}[hbt!]
 \begin{center}
  \scalebox{0.78}{
\begin{tabular}{ ||c||c|c|c|| } 
 \hline\hline
  name & $\epsilon_1$ & $\epsilon_2$ & $\epsilon_3$ \\ [0.5ex] 
  \hline\hline
 $(1a)$ & $+$ & $-$ &  $+$ \\ 
 [0.5ex] 
  \hline
 $(1b, \epsilon'=1)$ & $-$ & $+$ &  $+$  \\ 
 [0.5ex] \hline
  $(1b, \epsilon'=-1)$ & $-$ & $+$ &  $+$ \\ 
 [0.5ex] \hline
  $(2a)$ &  $+$ & $+$ &  $0$ \\ 
 [0.5ex] 
  \hline
  $(2b)$ &  $+$ & $+$ &  $+$\\ 
 [0.5ex] 
  \hline
   $(2c,\epsilon'=1)$ & $-$ & $-$ &  $-$ \\ 
 [0.5ex] 
  \hline
     $(2c,\epsilon'=-1)$ & $-$ & $-$ &  $-$ \\ 
 [0.5ex] 
  \hline
     $(2d,\epsilon'=1)$ & $-$ & $-$ &  $0$ \\ 
 [0.5ex] \hline
      $(2d,\epsilon'=-1)$ & $-$ & $-$ &  $0$ \\ 
 [0.5ex] 
  \hline
   $(2e,\epsilon'=1)$ & $-$ & $-$ &  $+$ \\ 
 [0.5ex]  \hline
   $(2e,\epsilon'=-1)$ & $-$ & $-$ &  $+$ \\ 
 [0.5ex] 
 \hline\hline
\end{tabular}}
 \end{center}
 \vskip 12pt 
 \caption{\small The eleven families of solutions for $\gamma \neq \gamma_c$. The cases with $\epsilon_3=0$ have ${\cal E}=0$, while there are two branches of solutions in all cases with $\epsilon_1=-1$, or equivalently $\gamma>\gamma_c$. The asymptotic behaviors of these solutions in the past and in the future are summarized in Table~\ref{tab:solsTkp12}.}\vskip 12pt
 \label{tab:signs}
 \end{table}
 
\begin{figure}[ht]
\centering
\begin{tabular}{ccc}
\includegraphics[width=65mm]{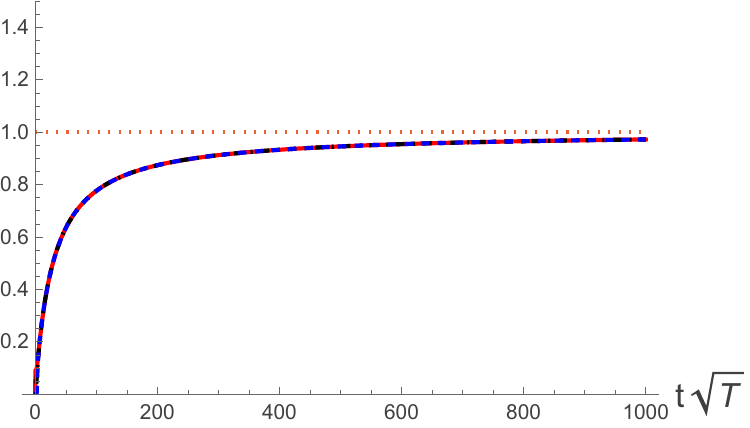} \quad  &
\includegraphics[width=65mm]{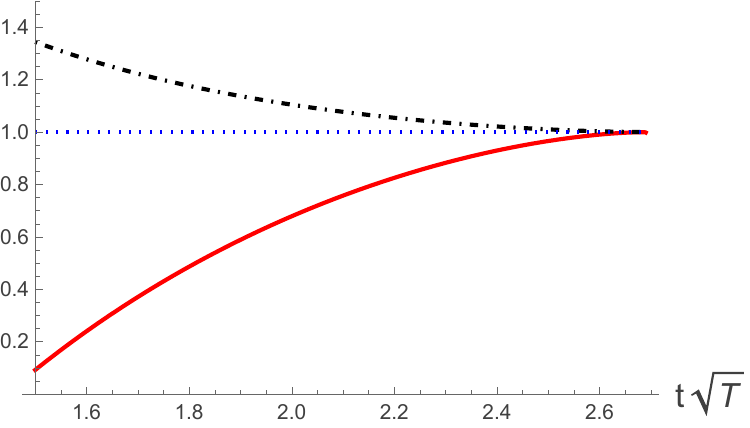} \\
\end{tabular}
 \caption{\small Typical solutions with $\gamma< \gamma_c$, $D=10$, $p=2$. Left panel ($T>0$): $\frac{(D-2)^2 \gamma^2}{16}\, u A'(u)$ (red, solid), $\frac{(D-2)^2 \gamma^2}{16}\, u C'(u)$ (black, dot-dashed), $- \frac{\gamma}{2} \, u \phi'(u))$ (blue, dotted) are practically superposed, since the Lucchin-Matarrese limit is also an exact solution, and approach one (orange, dotted) for large cosmic times, in agreement with eqs.~\eqref{lm}. Right panel ($T<0$): the behavior of $(p+1)(v-v_0) A'(v)+(D-p-2)(v-v_0) C'(v)$  (red, solid), compared with the free Kasner value 1  (blue, dotted) and $\frac{(v-v_0)}{\beta_\phi^+(\eta)}\,\phi'(v)$ (black,dot-dashed), which approach one close to the final singularity.}
\label{fig:gammasubcriticalTposneg}
\end{figure}

The independent families of solutions are as follows  (see Table~\ref{tab:signs}):
    \begin{enumerate}
        \item[(1a)] If $\gamma< \gamma_c$ and $T<0$, the independent solutions of the Hamiltonian constraint~\eqref{ham_constr} can be parametrized as
    \bea
\phi_1 &=& \frac{2\,\gamma_c}{\rho\left(\gamma_c^2 - \gamma^2\right)}\, \cos\eta \ , \nonumber \\ 
v_1 &=& \frac{\gamma_c}{\rho} \, \sqrt{\frac{(D-2)}{(p+1)(D-p-2)\left(\gamma_c^2 - \gamma^2\right)}}\, \sin \eta \ , \label{phi1vn}
\eea
while
 \beq
    W \ = \ - \ \log\left[\Delta \, \rho \,\cosh\left(\frac{\tau}{\rho}\right)\right] \ .
    \eeq
        \item[(1b, $\epsilon'=\pm 1$)] if $\gamma>\gamma_c$ and $T>0$, the independent solutions of the Hamiltonian constraint~\eqref{ham_constr} can be parametrized as
\bea
\phi_1 &=& \frac{2\,\epsilon'\,\gamma_c}{\rho\left(\gamma^2 - \gamma_c^2\right)}\, \cosh\zeta \ , \nonumber \\ 
v_1 &=& \frac{\gamma_c}{\rho} \, \sqrt{\frac{(D-2)}{(p+1)(D-p-2)\left(\gamma^2 - \gamma_c^2\right)}}\, \sinh \zeta \ , \label{phi1v1zetatnegn}
\eea
where $\epsilon'=\pm 1$, while
 \beq
    W \ = \ - \ \log\left[\Delta \, \rho \,\cosh\left(\frac{\tau}{\rho}\right)\right] \ .
    \eeq
\item[(2a)] If $\gamma<\gamma_c$, $T>0$, and ${\cal E}=0$, the only option is $\phi_1=0$ and $v_1=0$, while
     \beq
      W \ = \ - \ \log\left(\Delta\,\tau\right) \ .  \label{Wlog}
     \eeq
     and $0< \tau< \infty$.

    \item[(2b)] If $\gamma<\gamma_c$, $T>0$, and ${\cal E}>0$, the solutions can still be parametrized as in eqs.~\eqref{phi1vn}, but now
   \beq
    W \ = \ - \ \log\left[\Delta \, \rho \,\sinh\left(\frac{\left|\tau\right|}{\rho}\right)\right] \ , \label{Wsinh}
    \eeq
    and $- \infty< \tau< 0$.

    \item[(2c)] If $\gamma > \gamma_c$, $T<0$, and ${\cal E} <0$, the solutions can be parametrized as
    \bea
\phi_1 &=& \frac{2\,\,\gamma_c}{\rho\left(\gamma^2 - \gamma_c^2\right)}\, \sinh\zeta \ , \nonumber \\ 
v_1 &=& \epsilon'\frac{\gamma_c}{\rho} \, \sqrt{\frac{(D-2)}{(p+1)(D-p-2)\left(\gamma^2 - \gamma_c^2\right)}}\, \cosh \zeta \ , \label{phi1v1zetatnegn2}
\eea
with $\epsilon'=\pm 1$, and
    \beq 
    W=-\log\Big[\Delta\rho\,\sin\left({\frac{\tau}{\rho}}\right)\Big] \label{Wsin} \ .
    \eeq 

         \item[(2d)] If $\gamma>\gamma_c$, $T<0$, and ${\cal E}=0$, $W$ is  as in eq.~\eqref{Wlog}, but there are two one-parameter families of solutions, with
\beq
v_1 \ = \ \frac{\epsilon'}{2}\, \phi_1\, \sqrt{\frac{(D-2)}{(D-p-2)(p+1)}\,\left(\gamma^2 \,-\, \gamma_c^2\right)} \ , \label{v1propphi1}
\eeq
and $\epsilon'= \pm 1$.

  \item[(2e)] If $\gamma>\gamma_c$, $T<0$, and ${\cal E}>0$, $W$ is still given in eq.~\eqref{Wsinh}, while the parametrization is given in eqs.~\eqref{phi1v1zetatnegn}.
\end{enumerate}
\begin{figure}[ht]
\centering
\begin{tabular}{ccc}
\includegraphics[width=65mm]{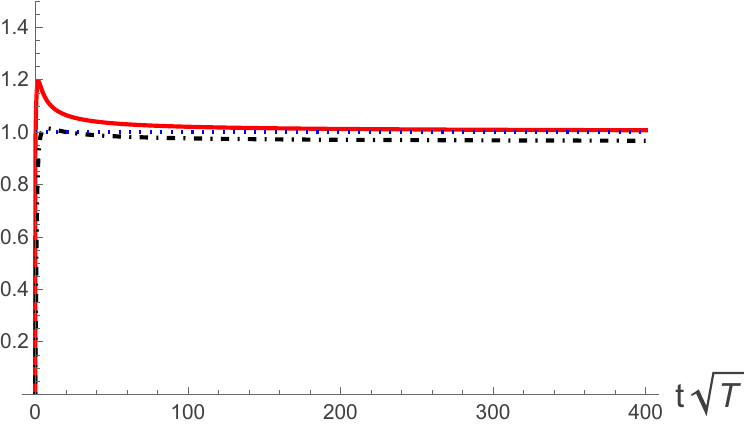} \quad  &
\includegraphics[width=65mm]{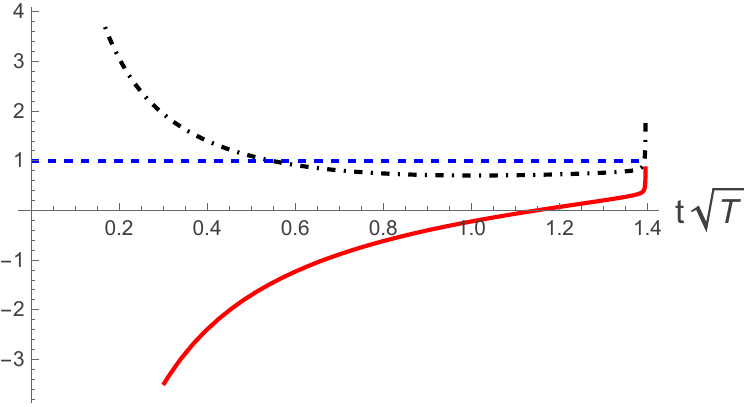} \\
\end{tabular}
 \caption{\small Typical solutions with $\gamma> \gamma_c$, $D=10$, $p=2$. Left panel ($T>0$): $(p+1) t A'(t)+(D-p-2)t C'(t)$ (red, solid), $\frac{1}{\beta_\phi^+(\zeta)}\, t \phi'(t)$ (black, dot-dashed), and their limiting value one (blue, dotted) are readily superposed. Right panel ($T<0$): the behavior of $(p+1)(v-v_0) A'(v)+(D-p-2)(v-v_0) C'(v)$  (red, solid), compared with the free Kasner value 1  (blue, dotted) and $\frac{(v-v_0)}\beta_\phi^+(\eta)$ (black,dot-dashed), which approach one close to the final singularity.}
\label{fig:gammasupercriticalTposneg}
\end{figure}

More details on these types of solutions can be found in Appendix~\ref{app:tension_only}.

\section{\sc Cosmologies with Tension and One Curvature}\label{sec:Tensionkp}

When the tension $T$ and, say, the internal curvature $k'$ are simultaneously present, a change of variables can simplify the analysis. In the harmonic gauge, one can work with the combinations
\bea
X&=&(p+1)A\ + \ (D-p-3)C \ , \nonumber  \\
W&=&(p+1)A\ +\ (D-p-2)C \ + \ \frac{\gamma}{2}\, \phi \ , \nonumber  \\
K&=&\phi\ + \ \gamma \ \frac{(D-2)^2}{8}\, A \ , \label{XWK}
\eea
from which one can recover the original functions of eqs.~\eqref{Eqs_back} as
\bea \label{eq:XW_ABCphin}
A &=& \frac{16 (D-p-2)}{\Xi}\, X  \ - \ \frac{16 (D-p-3)}{\Xi}\, W \ + \ \frac{8(D-p-3)\gamma}{\Xi}\, K \ , \nonumber \\
B &=& \frac{(D-2)^2(D-p-2)\gamma^2}{\Xi} \,X \ + \ \frac{16 (p+1)}{\Xi}\, W \ - \ \frac{8(p+1)\gamma}{\Xi} \,K \ , \nonumber \\
C &=& \frac{\left[(D-2)^2 \gamma^2 - 16 (p+1)\right]}{\Xi} \,X \ + \ \frac{16(p+1)}{\Xi} \,W \ - \ \frac{8(p+1)\gamma }{\Xi} \,K \ , \nonumber \\
\phi &=& -\frac{2(D-2)^2(D-p-2)\gamma}{\Xi} \,X \ + \ \frac{2(D-2)^2 (D-p-3)\gamma }{\Xi} \,W \ + \ \frac{16(p+1)}{\Xi} \,K \ ,
\eea
where we have defined
\beq
\Xi \ = \ 16(p+1)\ + \ (D-2)^2(D-p-3)\gamma^2 \ .
\eeq
In terms of $X$, $W$ and $K$ the equations become
\bea \label{eq:tadpole-curvature_system}
X''&=&- \ \frac{k'\left(D-p-3\right)^2}{\ell^2}\ e^{2X}\ + \ T \ e^{2W} \ , \nonumber  \\
W''&=&- \ \frac{k'\left(D-p-2\right)\left(D-p-3\right)}{\ell^2}\ e^{2X}\ - \ \frac{D-2}{16}\left(\gamma^2 -\gamma_c^2\right)T\ e^{2W} \ ,  \nonumber \\
K''&=&0 \ ,
\eea
where the ``critical'' value of $\gamma$ is
\beq
\gamma_c\ =\ \frac{4\sqrt{D-1}}{D-2} \ , \label{gammac1_2}
\eeq 
while the Hamiltonian constraint becomes
\bea \label{eq:XW_ham_constrkp2}
0 &=& \frac{k'\left(D-p-3\right)\left(D-p-2\right)}{\ell^2}\,e^{2X} \ - \ T e^{2W} \ - \ \frac{1}{\Xi}\Big[ 16(D-2)(D-p-3) (W')^2   \nonumber \\
&-&  32(D-2)(D-p-2)W'X'-(D-p-2) (D-2)^2 \left(\gamma^2-\gamma_c^2\right)(X')^2 \nonumber \\
&+& \frac{64 (p+1)}{D-2}(K')^2\Big] \ .
\eea

The system~\eqref{eq:tadpole-curvature_system} implies that $K'$ is a constant, which in its turn determines the ``energy'' in the Hamiltonian constraint. However, the resulting system is still complicated, since the two nonlinear equations couple $X$ and $W$.

\subsection[\texorpdfstring{{\mdseries\textsc{Special Exact Solutions with $W=X+c$}}}{Special Solutions with W=X+c}]
{{\mdseries\textsc{Special Exact Solutions with $W=X+c$}}}

The system of eqs.~\eqref{eq:tadpole-curvature_system} and~\eqref{eq:XW_ham_constrkp2} is not exactly solvable in general, but it determines a simple class of exact cosmologies. These can be obtained by demanding that
\beq
W \ = \ X \ + \ c \ ,
\eeq
where $c$ is a constant, a condition that is equivalent to
\beq
\gamma\,\phi \ + \ 2\,C \ = \ 2\,c \ .
\eeq
The system~\eqref{eq:tadpole-curvature_system} then yields the consistency condition
\beq
\left[1 \ + \ \frac{(D-2)}{16}\left( \gamma^2 \ - \ \gamma_c^2\right) \right] T\,e^{2c} \ = \ - \ \frac{k'(D-p-3)}{\ell^2} \ , \label{e2c}
\eeq
which can be recast in the more convenient form
\beq
\frac{(D-2)}{16}\left( \gamma^2 \ - \ \gamma_0^2\right) T\,e^{2c} \ = \ - \ \frac{k'(D-p-3)}{\ell^2} \ , \label{e2cn}
\eeq
where
\beq
\gamma_0 \ = \ \frac{4}{D-2} \ . \label{gamma0}
\eeq
Therefore, the combination of $\left( \gamma^2 \ - \ \gamma_0^2\right) T$ determines the consistent sign choice of $k'$ for these solutions. There are three possible cases: 
\begin{enumerate}
\item for $\left( \gamma_0^2 \ - \ \gamma^2\right)T>0$, these special solutions exist if $k'=1$;
\item for $\left( \gamma_0^2 \ - \ \gamma^2\right)T<0$, these special solutions exist if $k'=-1$.
\item Finally, if $\gamma = \gamma_0$, there are no solutions of this type unless $k'=0$, but then one is led back to the cases of Section~\ref{sec:tnot0curv0}.
\end{enumerate}
These conditions can be summarized by stating that non-trivial solutions of this type only exist if
\beq
\left(\gamma_0^2 \ - \ \gamma^2\right) k'\,T \ > \ 0 \ , \label{ineqkpT}
\eeq
and in general the equation for $X$ reduces to
\beq
X'' \ = \ - \ \frac{k'(D-p-3)}{\ell^2}\, \left\{ D-p-3 \ + \ \frac{(D-2)}{\left[ \left(\frac{\gamma}{\gamma_0}\right)^2 \ - \ 1 \right]}\right\}\, e^{2X} \ . \label{eqX2}
\eeq
The expression within curly brackets is negative for $\gamma< \gamma_0$ and positive for $\gamma>\gamma_0$, so that taking the inequality~\eqref{ineqkpT} into account, one can cast eq.~\eqref{eqX2} in the form
\beq
X'' \ = \ \sign(T)\,\Delta^2 \, e^{2X} \ , 
\eeq
where
\beq
\Delta^2 \ = \ - \ 
\sign(T)\ \frac{k'(D-p-3)}{\ell^2}\, \left\{ D-p-3 \ + \ \frac{(D-2)}{\left[ \left(\frac{\gamma}{\gamma_0}\right)^2 \ - \ 1 \right]}\right\} \ , \label{eqDelta}
\eeq
is always positive.

The Hamiltonian constraint then leads to
\beq
\left(X'\right)^2  \ = \ \sign(T) \ \Delta^2 \, e^{2X} \ + \ \frac{1}{\rho^2} \ ,
\eeq
where
\beq
\frac{1}{\rho^2} \ = \ \frac{64 (p+1) \left(K'\right)^2 }{(D-2) \left[ (D-2)^2(D-p-2)\gamma^2 \ + \ 16\,p \right]} \ = \ \frac{\left(K'\right)^2}{\sigma^2} \ ,
\eeq
so that one can work with
\beq
K'\ = \ \epsilon \, \frac{\sigma}{\rho} \ , \label{kprimesigma}
\eeq
with $\epsilon=\pm 1$, and then
\beq
K \ = \ \frac{\epsilon \, \sigma\,\tau}{\rho} \ + \ K_0 \ .
\eeq
The general solution is therefore determined as
\bea \label{eq:XW_ABCphin2}
A &=& \frac{16}{\Xi}\, X  \ + \  \frac{8(D-p-3)}{\Xi}\left[ \gamma\left( \frac{\epsilon \, \sigma\,\tau}{\rho}\,+\,K_0\right) \,-\,2\,c\right]\ , \nonumber \\
B &=& \left[1 \ +\ \frac{\gamma^2(D-2)^2}{\Xi} \right]X  \ - \ \frac{8(p+1)}{\Xi}\left[ \gamma\left( \frac{\epsilon \, \sigma\,\tau}{\rho}\,+\,K_0\right) \,-\,2\,c\right]\ , \nonumber \\
C &=& \frac{\gamma^2 (D-2)^2}{\Xi} \,X  \ - \ \frac{8(p+1)}{\Xi}\left[ \gamma\left( \frac{\epsilon \, \sigma\,\tau}{\rho}\,+\,K_0\right) \,-\,2\,c\right]  \ ,  \\
\phi &=& -\,\frac{2\,\gamma(D-2)^2}{\Xi} \,X \ + \  \frac{1}{\Xi} \left[16(p+1)\left( \frac{\epsilon \, \sigma\,\tau}{\rho}\,+\,K_0\right) \ + \ 2(D-2)^2 (D-p-3)\gamma\,c\right]\ , \nonumber
\eea
and one must now distinguish three cases.

\subsection[\texorpdfstring{{\mdseries\textsc{$T>0, \, \rho = \infty$}}}{T>0, rho=infty}]
{{\mdseries\textsc{$T>0, \, \rho = \infty$}}}

In this case the solution is
\beq
K \ = \ K_0 \ , \qquad X \ = \ - \ \log\left[ \Delta \left|\tau\right| \right] \ ,
\eeq
so that the background is
\bea
ds^2 \!\!&=&\!\! \left( \Delta \left|\tau\right| \right)^{-\,\frac{32}{\Xi}} \, d\vec{x}^2 \nonumber \\
\!\!&+&\!\! \left[\frac{- \ d\tau^2}{\left( \Delta \left|\tau\right| \right)^2} \ + \ {\ell^2 \ ds_{D-p-2,k'}^2} \right]\! \left( \Delta \left|\tau\right| \right)^{\frac{-\,2 \gamma^2(D-2)^2}{\Xi}} \,e^{\frac{16(p+1)}{\Xi}\left( 2 c \,-\, \gamma K_0\right)}\ , \nonumber \\
e^\phi \!\!&=&\!\! \ \left( \Delta \left|\tau\right| \right)^{\frac{2\,\gamma(D-2)^2}{\Xi}}e^ {\frac{16(p+1) K_0}{\Xi} \ + \ \frac{2(D-2)^2 (D-p-3)\gamma\,c}{\Xi}} \ . \label{tposrhoinf}
\eea

One can recast these expressions in cosmic time, but the $\gamma=0$ case must be treated separately. If $\gamma=0$, one obtains
\bea
ds^2 &=& - \ dt^2\ + \ e^{2t \sqrt{\frac{T}{(p+1)(D-2)}}} d\vec{x}^2 \ + \ \frac{(D-2)(D-p-3)}{T} \  ds_{D-p-2,k'=1}^2 \ , \nonumber \\
e^\phi &=& e^{K_0}  \ , \label{gamma_0_t2}
\eea
which describes the direct product of a de Sitter spacetime with an internal sphere of constant radius.
 On the other hand, if $\gamma \neq 0$ the preceding results imply that the link between cosmic time $t$ and parametric time $\tau$ is
 \beq
\Delta\,t \ = \ \frac{\Xi}{\gamma^2(D-2)^2} \left|\Delta\,\tau\right|^{-\, \frac{(D-2)^2\gamma^2}{\Xi}} \, \,e^{\frac{8(p+1)}{\Xi}\left( 2 c \,-\, \gamma K_0\right)}\ ,
 \eeq
 so that, up to rescalings,
 \bea
ds^2 \ &=& - \ dt^2 \ + \ \left(\Delta\,t\right)^{2\,\frac{\gamma_0^2}{\gamma^2}} d\vec{x}^2 \ + \
\frac{\gamma^4(D-2)^4}{\Xi^2}\, \left(\Delta\,t\right)^2 \, \ell^2 \, ds_{D-p-2,k'}^2 \ , \nonumber \\
e^\phi &=& \left(\frac{\gamma^2(D-2)^2 \Delta\,t}{\Xi} \right)^{\,-\,\frac{2}{\gamma}} \ \,e^{\frac{2c}{\gamma}} \ , \label{exactkpminus}
 \eea
 where $\Delta\,\ell$ is given in eq.~\eqref{eqDelta}. This solution combines a Lucchin-Matarrese (LM) spacetime with a linearly expanding internal space, which has $k'=1$ for $\gamma<\gamma_0$ and $k'=-1$ for $\gamma>\gamma_0$.
     \begin{figure}[ht]
\centering
\begin{tabular}{ccc}
\includegraphics[width=70mm]{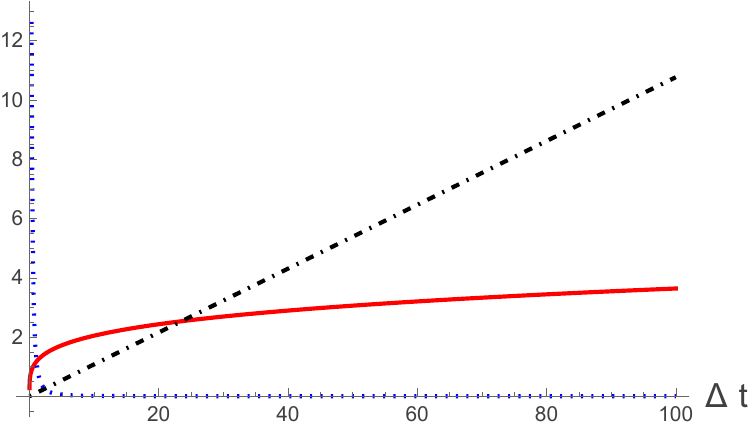} \quad  &
\includegraphics[width=70mm]{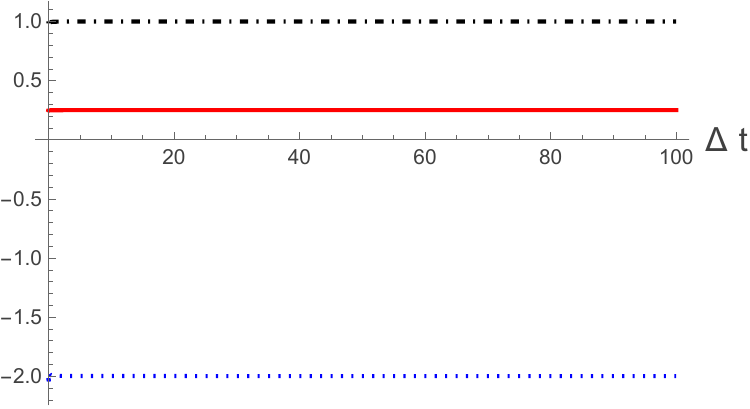} \\
\end{tabular}
 \caption{\small Left panel: $e^A$ (red, solid), $e^C$ (black, dot-dashed) and $e^\phi$ (blue dotted, essentially superposed to the horizontal axis) for the solution~\eqref{exactkpminus} with $k'=-1$ and $\gamma=1$ obtained numerically starting from fine-tuned initial conditions. Right panel: $t A'$ (red, solid), $t C'$ (black, dot-dashed) and $t \phi'$ (blue dotted) for the same solution.}
\label{fig:exactplusminus}
\end{figure}

 There are no values of $\gamma$ for which $\Delta\,\ell=1$, as would be the case for a Milne universe. The expansion is accelerated for $\gamma<\gamma_0$, and thus for $k'=1$ in view of inequality~\eqref{ineqkpT}, and is decelerated for $\gamma>\gamma_0$, and thus for $k'=-1$, again in view of inequality~\eqref{ineqkpT}. Note how the late-time behavior of these cosmologies is discontinuous at $\gamma=0$.
 
\subsection[\texorpdfstring{{\mdseries\textsc{$T>0, \, \rho < \infty$}}}{T>0, rho=infty}]
{{\mdseries\textsc{$T>0, \, \rho < \infty$}}}

If $T>0$, so that, in view of eq.~\eqref{ineqkpT}, $k'(\gamma_0^2-\gamma^2)>0$, and $\rho$ is finite, up to a translation of $\tau$ the solution for $X$ is
\beq
X \ = \ - \ \log\left[ \Delta\,\rho\, \sinh\left( \frac{\left|\tau\right|}{\rho} \right) \right] \ , \label{solposT}
\eeq
with $- \infty<\tau<0$, so that, after rescaling the $\vec{x}$ coordinates,
\bea
ds^2 \!\!&=&\!\! \left[ \Delta\,\rho\, \sinh\left( \frac{\left|\tau\right|}{\rho} \right) \right]^{-\,\frac{32}{\Xi}} \ e^{\frac{16\epsilon(D-p-3)\gamma\, \sigma\,\tau}{\rho \,\Xi}}\, d\vec{x}^2 \nonumber \\
\!\!&+&\!\! \left[\frac{- \ d\tau^2}{\left[ \Delta\,\rho\, \sinh\left( \frac{\left|\tau\right|}{\rho} \right)\right]^2} \ + \ {\ell^2 \ ds_{D-p-2,k'}^2} \right]\! \left[ \Delta\,\rho\, \sinh\left( \frac{\left|\tau\right|}{\rho} \right) \right]^{\frac{-\,2 \gamma^2(D-2)^2}{\Xi}} \!\!\!\!\! e^{\frac{16(p+1)}{\rho\,\Xi}\left[ 2 c \rho - \gamma\left(\epsilon\sigma\tau+\rho K_0\right)\right]}, \nonumber \\
e^\phi \!\!&=&\!\! \ \left[ \Delta\,\rho\, \sinh\left( \frac{\left|\tau\right|}{\rho} \right) \right]^{\frac{2\,\gamma(D-2)^2}{\Xi}}e^ {\frac{16(p+1)}{\Xi} \left( \frac{\epsilon \, \sigma\,\tau}{\rho}\,+\,K_0\right) \ + \ \frac{2(D-2)^2 (D-p-3)\gamma\,c}{\Xi}} \ . \label{Tkppos_gammaless}
\eea

Special solutions of this type exist with $\gamma=0$ and $k'=1$. In this case, the preceding expressions reduce to
\bea
ds^2 \!\!&=&\!\! \left[ \Delta\,\rho\, \sinh\left( \frac{\left|\tau\right|}{\rho} \right) \right]^{-\,\frac{2}{p+1}} \ d\vec{x}^2 \ +\  \left[\frac{- \ d\tau^2}{\left[ \Delta\,\rho\, \sinh\left( \frac{\left|\tau\right|}{\rho} \right)\right]^2} \ + \ {\ell^2 \ ds_{D-p-2,k'=1}^2} \right]  e^{2 c} \nonumber \\
e^\phi \!\!&=&\!\! e^{\frac{\epsilon\,\tau}{2}\,\sqrt{\frac{p(D-2)}{(p+1)}}\,+\, K_0} \ , \label{gamma_0_tau}
\eea
and can be exactly recast in cosmic time as
\bea
ds^2 &=& - \ dt^2\ + \ \left[\sinh\left(t\,\sqrt{\frac{T(p+1)}{(D-2)}}\right)\right]^\frac{2}{p+1} d\vec{x}^2 \ + \ \frac{(D-2)(D-p-3)}{T} \  ds_{D-p-2,k'=1}^2 \ , \nonumber \\
e^\phi &=& e^{K_0} \left[\tanh\left(\frac{t}{2}\,\sqrt{\frac{T(p+1)}{(D-2)}}\right) \right]^{\frac{\epsilon}{2} \sqrt{\frac{p(D-2)}{(p+1)}}} \ , \label{gamma_0_t}
\eea
where $0<t<\infty$. These special solutions are actually independent of $\rho$ and describe the direct product of expanding spacetimes and internal spheres whose constant radii are determined by the tension $T$. They interpolate between early Kasner-like epochs near $t=0$ (at weak coupling if $\epsilon=1$, and at strong coupling if $\epsilon=-1$) and final de Sitter-like expansions with bounded couplings. The two options reflect the $\phi \to -\, \phi$ symmetry of the action~\eqref{eqs4} for $\gamma=0$. The comparison with the exact solutions of eqs.~\eqref{Tnokpgammazeroc}, where the internal space is flat, reveals that their dominant early-time behaviors agree if $\sin\eta = -\, \sqrt{\frac{(D-p-2)}{(p+1)(D-2)}}$, while the late-time behaviors are different for all values of $\eta$.

For $\gamma \neq 0$, we could not obtain simple expressions as above in cosmic time, but one can still explore in detail the limiting early and late behaviors, as in previous sections. To begin with, at early times $(\tau \to -\ \infty)$,
\beq
X \ \sim \ W \ \sim \ - \ \frac{\left|\tau\right|}{\rho} \ ,
\eeq
and using also eq.~\eqref{kprimesigma}, one can conclude that
\bea \label{eq:XW_ABCphi2}
A &\sim& \frac{8}{\rho\,\Xi} \Big[ - \,2 \,\left|\tau\right|\ + \ (D-p-3)\,\gamma\,\epsilon\,\sigma\, \tau\Big]   \ , \nonumber \\
B &\sim& \frac{1}{\rho\,\Xi} \Big\{ - \left[(D-2)^2(D-p-2)\gamma^2 \ + \ 16(p+1)  \right] \left|\tau\right| \ -\ 8 (p+1) \,\gamma\,\epsilon\,\sigma\, \tau \Big\} \ , \nonumber \\
C &\sim& \frac{1}{\rho\,\Xi} \Big[ - \, (D-2)^2 \gamma^2\,\left|\tau\right| \ -\ 8 (p+1) \,\gamma\,\epsilon\,\sigma\, \tau \Big]  \ , \nonumber \\
\phi &\sim& \frac{1}{\rho\,\Xi}\left[ {2(D-2)^2\gamma} \,\left|\tau\right| \ + \ {16(p+1)}\,\epsilon\,\sigma\, \tau \right] \ . \label{6.25}
\eea
As $\tau \to - \ \infty$, $B \to - \infty$ for both choices of $\epsilon$, and so all these solutions have a singularity in the finite past, where the limiting behavior is free Kasner-like, with exponents
\bea
\alpha_A^- &=& \frac{16 \ + \ 8 (D-p-3)\,\gamma\,\epsilon\,\sigma}{  \left[(D-2)^2(D-p-2)\gamma^2 \ + \ 16(p+1)  \right] \ - \ 8 (p+1) \,\gamma\,\epsilon\,\sigma} \ , \nonumber \\
\alpha_C^- &=& \frac{(D-2)^2 \gamma^2 \  - \ 8 (p+1) \,\gamma\,\epsilon\,\sigma }{\left[(D-2)^2(D-p-2)\gamma^2 \ + \ 16(p+1)  \right] \ - \ 8 (p+1) \,\gamma\,\epsilon\,\sigma}   \ , \nonumber \\
\alpha_\phi^- &=&  \frac{ {-\,2(D-2)^2\gamma}  \ + \ {16(p+1)}\,\epsilon\,\sigma }{\left[(D-2)^2(D-p-2)\gamma^2 \ + \ 16(p+1)  \right] \ - \ 8 (p+1) \,\gamma\,\epsilon\,\sigma}\ . \label{alphaskpone}
\eea
For a cosmology that expands near the initial singularity, one must have $\alpha_A^->0$, which is always the case for $\epsilon=+1$, but only if 
\beq
\gamma^2 \, <  \,\frac{- \,8 (D-p-3)\,+\,8 \sqrt{ 4(p+1)(D-2)(D-p-2) \ + \ p^2(D-p-3)^2}}{(D-2)^2(D-p-2)(D-p-3)}   
\eeq
for $\epsilon=-1$. In particular, for $\gamma=0$ one recovers the exponents of eqs.~\eqref{gamma_0_t}, and the results for the metric are independent of $\epsilon'$.
In the complementary region, one obtains ``bounce'' cosmologies, where the spatial slices first contract and then eventually expand.
Moreover, if $\gamma \neq 0$, the scale factor $\alpha_C^-$ of the internal coordinates is always negative for $\epsilon=1$, so the internal space initially contracts from a divergent size, while it is manifestly positive for $\epsilon=-1$, so the internal space initially expands from a vanishing size. Finally, the string coupling vanishes at the initial singularity for $\epsilon=1$ and diverges for $\epsilon =-1$.

The limiting behavior at late times is captured by the exact solutions obtained for infinite $\rho$ in eqs.~\eqref{gamma_0_t2} and \eqref{exactkpminus}.

 \subsection[\texorpdfstring{{\mdseries\textsc{$T<0$}}}{T<0}]
{{\mdseries\textsc{$T<0$}}}

If $T<0$, eq.~\eqref{ineqkpT} implies that $k'=-1$ for $\gamma<\gamma_0$ and $k'=1$ for $\gamma>\gamma_0$. In both cases, the solution for $X$ is
\beq
X \ = \ - \ \log\left[ \Delta\,\rho\, \cosh\left( \frac{\tau}{\rho} \right) \right] \ , \label{solnegT}
\eeq
and $-\, \infty<\tau<+\,\infty$, so that
\bea
ds^2 \!\!&=&\!\! \left[ \Delta\,\rho\, \cosh\left( \frac{\tau}{\rho} \right) \right]^{-\,\frac{32}{\Xi}} \ e^{\frac{16\epsilon(D-p-3)\gamma\, \sigma\,\tau}{\rho \,\Xi}}\, d\vec{x}^2 \nonumber \\
\!\!&+&\!\! \left[\frac{- \ d\tau^2}{\left[ \Delta\,\rho\, \cosh\left( \frac{\tau}{\rho} \right)\right]^2} \ + \ {\ell^2 \ ds_{D-p-2,k'}^2} \right]\! \left[ \Delta\,\rho\, \cosh\left( \frac{\tau}{\rho} \right) \right]^{\frac{-\,2 \gamma^2(D-2)^2}{\Xi}} \!\!\!\!\! e^{\frac{16(p+1)}{\rho\,\Xi}\left[ 2 c \rho - \gamma\left(\epsilon\sigma\tau+\rho K_0\right)\right]}, \nonumber \\
e^\phi \!\!&=&\!\! \ \left[ \Delta\,\rho\, \cosh\left( \frac{\tau}{\rho} \right) \right]^{\frac{2\,\gamma(D-2)^2}{\Xi}}e^ {\frac{16(p+1)}{\Xi} \left( \frac{\epsilon \, \sigma\,\tau}{\rho}\,+\,K_0\right) \ + \ \frac{2(D-2)^2 (D-p-3)\gamma\,c}{\Xi}} \ . \label{neg_T}
\eea
If $\gamma=0$, the background of eq.~\eqref{neg_T} reduces to
\bea
ds^2 \!\!&=&\!\! \left[ \Delta\,\rho\, \cosh\left( \frac{\tau}{\rho} \right) \right]^{-\,\frac{2}{p+1}} \ d\vec{x}^2 \ +\  \left[\frac{- \ d\tau^2}{\left[ \Delta\,\rho\, \cosh\left( \frac{\tau}{\rho} \right)\right]^2} \ + \ {\ell^2 \ ds_{D-p-2,k'=-1}^2} \right]  e^{2 c} \nonumber \\
e^\phi \!\!&=&\!\! e^{\frac{\epsilon\,\tau}{2}\,\sqrt{\frac{p(D-2)}{(p+1)}}\,+\,K_0} \ , 
\eea
and there is again a simple cosmic-time presentation, which reads
\bea
ds^2 &=& - \ dt^2\ + \ \left[\sin\left(t\,\sqrt{\frac{\left|T\right|}{(D-2)}}\right)\right]^{\frac{2}{p+1}} d\vec{x}^2 \ + \ \frac{(D-2)(D-p-3)}{|T|}\  ds_{D-p-2,k'=-1}^2 \ , \nonumber \\
e^\phi &=& e^{K_0} \left[\tan\left(\frac{t}{2}\,\sqrt{\frac{|T|}{(D-2)}}\right) \right]^{\frac{\epsilon}{2} \sqrt{\frac{p(D-2)}{(p+1)}}} \ . \label{metricgamma0-kp}
\eea

Note that the $\rho $-dependence disappeared in these expressions after turning to cosmic time.
These cosmologies expand and then contract within the finite interval $t \in \left[ 0,{\pi}\,\sqrt{\frac{(D-2)}{|T|}}\right]$ of cosmic time, with early weak coupling and late strong coupling for $\epsilon=1$, and vice versa for $\epsilon=-1$, while the internal manifold is a static (quotiented) hyperboloid. Note that the early-time behavior of this solution recovers the $k'=0$ result of eqs.~\eqref{limiting_gamma0_early} for a special choice of $\eta$, such that
\beq
\sin\eta \ = \ \sqrt{\frac{(D-p-2)}{(p+1)(D-2)}} \ , 
\eeq
which also eliminates the contribution of ${\cal O}(t)$ from $C$. The opposite value of $\sin \eta$ has the same effect at the other end of the time evolution: the negative curvature present in this example connects the two limiting behaviors.

For general values of $\gamma$, these solutions are still characterized by a finite past and a finite future. Their limiting behaviors at the initial and final singularities, which are separated by a finite span of proper time, are free Kasner-like, with exponents
\bea
\alpha_A^\pm &=& \frac{16 \ \pm \ 8 (D-p-3)\,\gamma\,\epsilon\,\sigma}{  \left[(D-2)^2(D-p-2)\gamma^2 \ + \ 16(p+1)  \right] \ \mp \ 8 (p+1) \,\gamma\,\epsilon\,\sigma} \ , \nonumber \\
\alpha_C^\pm &=& \frac{(D-2)^2 \gamma^2 \  \mp \ 8 (p+1) \,\gamma\,\epsilon\,\sigma }{\left[(D-2)^2(D-p-2)\gamma^2 \ + \ 16(p+1)  \right] \ \mp \ 8 (p+1) \,\gamma\,\epsilon\,\sigma}   \ , \nonumber \\
\alpha_\phi^\pm &=&  \frac{ {-\,2(D-2)^2\gamma}  \ \pm \ {16(p+1)}\,\epsilon\,\sigma }{\left[(D-2)^2(D-p-2)\gamma^2 \ + \ 16(p+1)  \right] \ \mp \ 8 (p+1) \,\gamma\,\epsilon\,\sigma}\ . \label{alphaskpone2}
\eea
As $\gamma \to 0$, these expressions recover the asymptotics emerging from eqs.~\eqref{metricgamma0-kp}.

\subsection{\sc The Stability Issue} \label{sec:staility_sec6}

 The special exact solutions of the system described in this section are based on fine-tuned initial conditions consistent with $W - X = c$. Therefore, it is important to perform a stability analysis aimed at determining whether or not there exist perturbations that make this difference grow unbounded. On the other hand, perturbations that do not affect the difference $W-X$ can be ignored, as they are mere modifications of the two parameters (width and center) that characterize the background cosmologies, which we denote collectively by $X_0$ in the subsequent analysis. 
 
 In order to address the stability issue, one can expand eqs.~\eqref{eq:tadpole-curvature_system}, letting
 \beq
X \ =  X_0 \ + x \ , \qquad W \ = \ X_0 \ + \ c \ + w \ , \label{XWxw}
 \eeq
where $X_0$ is given in eq.~\eqref{solposT} for $T>0$ and in eq.~\eqref{solnegT} for $T<0$. Linearization in $x$ and $w$ then gives the system
\beq
\left(\begin{array}{c} x'' \\ w'' \end{array} \right) \ = \ - \ 2 k'\, \frac{(D-p-3)}{\ell^2}\  e^{2 X_0} {\cal M } \left(\begin{array}{c} x \\ w \end{array} \right) \ ,
\eeq
where
 \beq
{\cal M} \ = \  \left( \begin{array}{cc} (D-p-3)  & \frac{16}{(D-2)\left( \gamma^2 \,-\,\gamma_0^2\right)} \\ (D-p-2) &  \ \ -\ 1 \ + \ \frac{16}{(D-2)\left( \gamma^2 \,-\,\gamma_0^2\right)} \end{array} \right)
 \eeq
 after using eqs.~\eqref{eq:tadpole-curvature_system} and \eqref{e2c}. One can see that
 \bea
\mathrm{Tr}\left({\cal M}\right) &=& D-p-4   \ + \ \frac{16}{(D-2)\left( \gamma^2 \,-\,\gamma_0^2\right)} \ , \nonumber \\
\det\left({\cal M}\right) &=& - \ (D-p-3)   \ - \ \frac{16}{(D-2)\left( \gamma^2 \,-\,\gamma_0^2\right)} \ , \label{tracedet}
 \eea
and that ${\cal M}$ has the manifest eigenvector $\left(\begin{array}{c} 1 \\ 1 \end{array}\right)$. However, this only affects the width and center of the solution, and thus is irrelevant for the present stability analysis. The corresponding eigenvalue
 \beq
\lambda_1\ = \ (D-p-3)   \ + \ \frac{16}{(D-2)\left( \gamma^2 \,-\,\gamma_0^2\right)} \ , \label{eqlamda1}
 \eeq
is positive when $\gamma>\gamma_0$ and negative when $\gamma<\gamma_0$. The other eigenvalue,
\beq
\lambda_0 \ = \ - \ 1 \ ,
\eeq
is determined by the trace in eqs.~\eqref{tracedet}, and is always negative, and the corresponding eigenvector $\xi(\tau)$ satisfies
\beq
\xi'' \ - \ 2\,k'\,\frac{(D-p-3)}{\ell^2}\ e^{2X_0}\, \xi \ = \ 0 \ . \label{eqxi}
\eeq

In addition, the Hamiltonian constraint~\eqref{eq:XW_ham_constr} demands that $\delta\,K'=0$, 
or, in view of eqs.~\eqref{eq:XW_ABCphin}
\beq
\delta\,\phi \ + \ \frac{\gamma}{8}\,\left(D\,-\,2\right)^2 \, \delta\,A \ = \ 0 \ ,
\eeq
after selecting the relevant eigenvector with eigenvalue $\lambda_0=-1$, for which the small perturbation $w$ and $x$ in eq.~\eqref{XWxw} are related according to
\beq
w \ = \ - \ \frac{(D-p-2)(D-2)}{16} \left(\gamma^2 \,-\, \gamma_0^2\right) x \ .
\eeq

The preceding conditions imply that the effect of this perturbation of $A$, $C$ and $\phi$ is described by
\beq
\delta\,A \ = \ \frac{D-p-2}{D-2}\, x \ , \qquad \delta\,C \ = \ - \ \frac{p}{D-2}\, x \, \qquad \delta\,\phi \ = \ - \ \frac{\gamma}{8}\,(D-2)(D-p-2)\, x \ .
\eeq
For $T>0$, $X_0$ is given in eq.~\eqref{solposT}, and in view of eqs.~\eqref{eqDelta} and \eqref{eqlamda1}, one can link $\Delta$ and $\lambda_1$, so that when formulated for $x$ eq.~\eqref{eqxi} becomes
    \beq
x'' \ + \ \frac{2}{\lambda_1\,\rho^2\,\sinh^2\left(\frac{\tau}{\rho}\right)}\, x \ = \ 0 \ , \label{eqxi21}
\eeq
where $-\infty<\tau<0$. The general solution reads
\beq
x \ = \ C_1\, P_{-\,\frac{1}{2}+\nu}\left[-\coth\left(\frac{\tau}{\rho}\right) \right] \ + \  C_0\, Q_{-\,\frac{1}{2}+\nu}\left[-\coth\left(\frac{\tau}{\rho}\right) \right] \ ,
\eeq
where $P$ and $Q$ are Legendre functions of order ${-\,\frac{1}{2}+\nu}$, with
\beq
\nu^2\ = \ \frac{1}{4}\left(1 \ - \ \frac{8}{\lambda_1}\right) \ .
\eeq
The two types of Legendre functions behave at most linearly in $\tau$ for $\tau$ large and negative, where these perturbations can overcome the background, so that these solutions are \emph{unstable in the past} under these spatially constant perturbations. However, the perturbations behave as $\left|\tau\right|^{\frac{1}{2} \pm \nu}$ as $\tau \to 0^-$, so that \emph{future instabilities} are present when $\frac{1}{2} - \nu$ is negative. This is the case when $\lambda_1<0$, and thus when $\gamma<\gamma_0$ and $k'=1$. This result concerns, in particular, the special solution with $\gamma=0$ of eqs.~\eqref{gamma_0_tau} and \eqref{gamma_0_t}. On the other hand, the solutions with $k'=-1$ and $\gamma>\gamma_0$ are \emph{stable in the future}. Still, in the $\rho \to \infty$ limit the perturbations behave as
\beq
x \ = \ C_1\, \left|\tau\right|^{\frac{1}{2}+\nu} \ + \  C_0\, \left|\tau\right|^{\frac{1}{2}-\nu}  \ ,
\eeq
for arbitrary negative values of $\tau$, and thus in the past override the background, which is logarithmic in this case. The future behavior is as for finite values of $\rho$, but in the $\rho \to \infty$ limit the solution in eqs.~\eqref{tposrhoinf} is \emph{always unstable in the past}, irrespective of the value of $\gamma$.

For $T<0$, $X_0$ is given in eq.~\eqref{solnegT}, and in view of eqs.~\eqref{eqDelta} and \eqref{eqlamda1}, one can link $\Delta$ and $\lambda_1$, and eq.~\eqref{eqxi} becomes for $x$
    \beq
x'' \ - \ \frac{2}{\lambda_1\,\rho^2\,\cosh^2\left(\frac{\tau}{\rho}\right)}\, x \ = \ 0 \ , \label{eqxi212}
\eeq
where $-\infty<\tau<\infty$. The general solution reads
\beq
x \ = \ C_1\, P_{-\,\frac{1}{2}+\nu}\left[\tanh\left(\frac{\tau}{\rho}\right) \right] \ + \  C_0\, Q_{-\,\frac{1}{2}+\nu}\left[\tanh\left(\frac{\tau}{\rho}\right) \right] \ ,
\eeq
where $P$ and $Q$ are Legendre functions of order ${-\,\frac{1}{2}+\nu}$, with
\beq
\nu^2\ = \ \frac{1}{4}\left(1 \ - \ \frac{8}{\lambda_1}\right) \ .
\eeq
The two types of Legendre functions behave at most linearly in $\tau$ for $|\tau|$ large, where these perturbations can overcome the background, so that instabilities are present in this case, both in the past and in the future. 

\subsection[\texorpdfstring{{\mdseries\textsc{Special Exact Solutions with Tension and Only $k$}}}{Special Exact Solutions with Tension and Only k}]
{{\mdseries\textsc{Special Exact Solutions with Tension and Only $k$}}} \label{sec:Tensionk}

There is a special cosmology of this type, which rests on the reduced system in eqs.~\eqref{D8_sol} and \eqref{EqB_red_d2}, with $\gamma=0$. The solution at stake can be conveniently studied in the gauge $B=0$, and if $\phi'=0$, letting
\beq
\sigma \ = \ \sqrt{\frac{|T|}{(D-1)(D-2)}} \ ,
\eeq
one finds
\begin{itemize}
\item If $T<0$, the Hamiltonian constraint demands that $k=-1$, and then
\bea
ds^2 &=& - \,dt^2 \ + \ \frac{\sin^2\left( \sigma \, t\right)}{\sigma^2} \ ds_{D-1,k=-1}^2 \ , \nonumber \\
e^\phi &=& e^{\phi_0} \ ,
\eea
which is a $D$-dimensional AdS space in a hyperbolic slicing, so that the zeroes of the trigonometric factor are coordinate singularities;
\item If $T>0$ and $k=1$ 
\bea
ds^2 &=& - \,dt^2 \ + \  \frac{\cosh^2\left( \sigma \, t\right)}{\sigma^2}\ ds_{D-1,k=1}^2 \ , \nonumber \\
e^\phi &=& e^{\phi_0} \ ,
\eea
which is a $D$-dimensional dS space in a spherical slicing;
\item If $T>0$ and $k=0$,
\bea
ds^2 &=& - \,dt^2 \ + \ e^{2\,t\,\sigma}\, d\vec{x}^2 \ , \nonumber \\
e^\phi &=& e^{\phi_0} \ ,
\eea
which is the familiar flat-slicing presentation of (half of) $D$-dimensional dS space;
\item If $T>0$ and $k=-1$,
\bea
ds^2 &=& - \, dt^2 \ + \  \frac{\sinh^2\left( \sigma \, t\right)}{\sigma^2}\  ds_{D-1,k=-1}^2 \ , \nonumber \\
e^\phi &=& e^{\phi_0} \ ,
\eea
which is a $D$-dimensional dS space in a hyperbolic slicing.
\end{itemize}

In the general case, the convenient change of variables in the harmonic gauge rests on the combinations
\bea
W&=&(p+1)A\ +\ (D-p-2)C \ + \ \frac{\gamma}{2}\, \phi \ , \nonumber  \\
Y&=&p A\ + \ (D-p-2)C \ , \nonumber  \\
L&=&\phi\ + \ \gamma \ \frac{(D-2)^2}{8}\, C \ , \label{WYL}
\eea
from which one can recover the original functions of eqs.~\eqref{Eqs_back} as
\bea \label{eq:YW_ABCphin}
A &=& \frac{\left[(D-2)^2 \gamma^2 - 16 (D-p-2)\right]}{\widetilde{\Xi}} \,Y \ + \ \frac{16(D-p-2)}{\widetilde{\Xi}} \,W \ - \ \frac{8(D-p-2)\gamma }{\widetilde{\Xi}} \,L \ , \nonumber \\
B &=& \frac{(D-2)^2(p+1)\gamma^2}{\widetilde{\Xi}} \,Y \ + \ \frac{16 (p+1)}{\widetilde{\Xi}}\, W \ - \ \frac{8(D-p-2)\gamma}{\widetilde{\Xi}} \,L \ , \nonumber \\
C &=& \frac{16 (p+1)}{\widetilde{\Xi}}\, Y  \ - \ \frac{16 p}{\widetilde{\Xi}}\, W \ + \ \frac{8 \,p \,\gamma}{\widetilde{\Xi}}\, L \ , \nonumber \\
\phi &=& -\frac{2(D-2)^2(p+1)\gamma}{\widetilde{\Xi}} \,Y \ + \ \frac{2(D-2)^2 p\,\gamma }{\widetilde{\Xi}} \,W \ + \ \frac{16(D-p-2)}{\widetilde{\Xi}} \,L \ ,
\eea
where we have defined
\beq
\widetilde{\Xi} \ = \ 16(D-p-2)\ + \ (D-2)^2\,p\,\gamma^2 \ .
\eeq
In terms of $Y$, $W$ and $L$ the equations become
\bea \label{eq:tadpole-curvature_system_k}
Y''&=&- \ \frac{k\,p^2}{\ell^2}\ e^{2Y}\ + \ T \ e^{2W} \ , \nonumber  \\
W''&=&- \ \frac{k\,p(p+1)}{\ell^2}\ e^{2Y}\ - \ \frac{D-2}{16}\left(\gamma^2 -\gamma_c^2\right)T\ e^{2W} \ ,  \nonumber \\
L''&=&0 \ ,
\eea
where the ``critical'' value of $\gamma$ is as above, while the Hamiltonian constraint becomes
\bea \label{eq:XW_ham_constr}
0 &=& \frac{k\,p(p+1)}{\ell^2}\,e^{2Y} \ - \ T e^{2W} \ - \ \frac{1}{\widetilde{\Xi}}\Big[ 16(D-2)\,p\, (W')^2   \nonumber \\
&-&  32(D-2)(p+1)W'Y'-(p+1) (D-2)^2 \left(\gamma^2-\gamma_c^2\right)(Y')^2 \nonumber \\
&+& \frac{64 (D-p-2)}{D-2}(L')^2\Big] \ .
\eea
These solutions can be obtained from those of Section~\ref{sec:Tensionkp} interchanging $k'$ with $k$, $p$ with $D-p-3$ and $A$ with $C$.

\section[\texorpdfstring{{\mdseries\textsc{Solutions with Tension and Both $k$ and $k'$}}}{Special Solutions with Tension and Both k and k'}]
{{\mdseries\textsc{Solutions with Tension and Both $k$ and $k'$}}}  \label{sec:Tkkp}

For all these solutions $0<p<D-3$, a restriction allowing for curved spatial and internal slices.  Once more, the treatment is different if $\gamma=0$.

 \subsection[\texorpdfstring{{\mdseries\textsc{Solutions with $\gamma=0$}}}{Solutions with gamma=0}]
{{\mdseries\textsc{Solutions with $\gamma=0$}}} 
For $\gamma=0$, one can conveniently work with $A$, $C$ and $\phi$, referring to eqs.~\eqref{Eqs_back_F} and \eqref{EqB_back_F}, which involve three different exponents. Special solutions exist with constant $A$ and $C$, and in this case the system reduces, in the cosmic gauge, to
 \bea
 && 0 \ = \  \frac{T}{(D-2)} \ - \ \frac{k\,p}{\ell^2}\ e^{-2A}  \label{Eqs_back_F2} \ ,  \\
&& 0 \ = \ \frac{T}{(D-2)} \ - \ \frac{k'(D-p-3)}{\ell^2}\ e^{-2C} \ ,  \nonumber  \\
  &&  \phi'' \ = \ 0  \ , \nonumber \\
&&  \frac{4\,(\phi')^2}{D-2}  \,=\,  - \,{T} \, + \, \frac{k\,p(p+1)}{\ell^2}\ e^{-2A}\, + \, \frac{k'(D-p-3)(D-p-2)}{\ell^2}\ e^{-2C}  \ , \label{EqB_back_F2} 
  \eea

The first two equations then imply that $k$ and $k'$ must have the same sign as $T$, and when used in the Hamiltonian constraint they reduce it to
\beq
(\phi')^2 \ = \ \frac{T}{4} \ ,
\eeq
so that this solution only exists if $T$, $k$ and $k'$ are all positive. In detail, one then finds
\bea
ds^2 &=& - \ dt^2 \ + \ \frac{D-2}{T} \left[ p \, ds_{p+1,k=1}^2 \ + \ (D-p-3) ds_{D-p-2,k'=1}^2 \right] \ , \nonumber \\
\phi &=& \phi_0 \ + \ \frac{\epsilon}{2}\, \sqrt{T}\, t \ , \label{kkpTgamma0}
\eea
where $\epsilon\,=\,\pm \, 1$, so that the metric is static, while  the dilaton evolves linearly in cosmic time. The string coupling increases for $\epsilon=1$ and decreases for $\epsilon=-1$.

Moreover, the solutions in Section~\ref{sec:Tensionk} afford some extensions to spacetimes of dimension $p+2$, which are still characterized by constant values of $C$. Letting
\beq
\sigma_p \ = \ \sqrt{\frac{|T|}{(p+1)(D-2)}} \ ,
\eeq
for $T<0$ the more general backgrounds read
\bea
ds^2 &=& - \,dt^2 \ + \  \frac{\sin^2\left( \sigma_{p} \, t\right)}{\sigma^2_p} \ ds_{p+1,k=-1}^2 \ + \  \frac{1}{\sigma^2_{D-p-4}}\ ds_{D-p-2,k'=-1}^2 \ , \nonumber \\
e^\phi &=& e^{\phi_0} \ , \label{adsm1m1}
\eea
where $0<t<\frac{\pi}{\sigma_p}$ and where the two endpoints host coordinate singularities. These are direct products of $AdS_{p+2}$ and time-independent internal hyperbolic spaces of dimension $D-p-2$.  Similarly, for $T>0$ one can obtain direct products of $dS_{p+2}$ and internal spheres of dimension $D-p-2$, in three ways as before:
\begin{itemize}
\item If $T>0$ and $k=1$ 
\bea
ds^2 &=& -\, dt^2  \ + \  \frac{\cosh^2\left( \sigma_{p} \, t\right)}{\sigma^2_{p}}\  ds_{p+1,k=1}^2 \ +\  \frac{1}{\sigma^2_{D-p-4}}\  ds_{D-p-2,k'=1}^2 \ , \nonumber \\
e^\phi &=& e^{\phi_0} \ , \label{ds11}
\eea
where $-\infty<t<+\infty$ and the spacetime portion is $dS_{p+2}$ in a spherical slicing;
\item If $T>0$ and $k=0$,
\bea
ds^2 &=& - \, dt^2 \ + \ e^{2\,t\,\sigma_{p}}\, d\vec{x}^2 \ + \  \frac{1}{\sigma^2_{D-p-4}}\  ds_{D-p-2,k'=1}^2\ , \nonumber \\
e^\phi &=& e^{\phi_0} \ , \label{ds01}
\eea
where $-\infty<t<+\infty$ and the spacetime portion is $dS_{p+2}$ in a flat slicing, which was already discussed in eqs.~\eqref{gamma_0_t2};
\item If $T>0$ and $k=-1$,
\bea
ds^2 &=&- \,dt^2 + \frac{\sinh^2\left( \sigma_{p} \, t\right)}{\sigma^2_{p}}\   ds_{p+1,k=-1}^2 \ +  \ \frac{1}{\sigma^2_{D-p-4}}\ ds_{D-p-2,k'=1}^2 \ , \nonumber \\
e^\phi &=& e^{\phi_0} \ , \label{dsm11}
\eea
where $0<t<+\infty$ and the spacetime portion is $dS_{p+2}$ in a hyperbolic slicing. We listed together these solutions, since they are different parametrizations of the same vacua although, strictly speaking, the second would belong to Section~\ref{sec:Tensionkp} since it has $k=0$.
\end{itemize}

 \subsection[\texorpdfstring{{\mdseries\textsc{Special Solutions with $\gamma\neq 0$}}}{Special Solutions with gamma≠0}]
{{\mdseries\textsc{Special Solutions with $\gamma\neq 0$}}} 

If $\gamma \neq 0$, one can define the three variables
\bea
W &=& (p+1)A \ +\ (D-p-2) C \ + \ \frac{\gamma}{2}\, \phi \ , \nonumber \\
X &=& (p+1)A \ + \ (D-p-3) C \ , \nonumber \\
Y &=& p A \ + \ (D-p-2) C \ ,
\eea
which determine $A$, $C$ and $\phi$ according to
\bea
A &=& \frac{(D-p-2)}{(D-2)}\, X \ - \ \frac{(D-p-3)}{(D-2)}\, Y \ , \nonumber \\
C &=& - \ \frac{p}{(D-2)} \, X \ + \ \frac{(p+1)}{(D-2)}\, Y \ , \nonumber \\
\phi &=& \frac{2}{\gamma} \left[ W \ - \ \frac{(D-p-2)}{(D-2)}\, X \ - \ \frac{(p+1)}{(D-2)}\, Y \right] \ .
\eea
The system thus reduces to
\bea
X'' &=& T\,e^{2W} \, - \, \frac{k'(D-p-3)^2}{\ell^2}\, e^{2X} \, - \, \frac{k\,p(p+1)}{\ell^2}\, e^{2Y} \ , \nonumber \\
Y'' &=& T\,e^{2W}  \, - \, \frac{k'(D-p-2)(D-p-3)}{\ell^2}\, e^{2X} \, - \, \frac{k\,p^2}{\ell^2}\, e^{2Y} \ ,  \label{eqs_kkpT}\\
W'' &=& - \ \frac{T(D-2)}{16} \left(\gamma^2 \,-\,\gamma_c^2 \right)\, e^{2W}  \, - \, \frac{k'(D-p-2)(D-p-3)}{\ell^2}\,e^{2X}\, - \, \frac{k\,p(p+1)}{\ell^2}\,e^{2 Y} \ , \nonumber
\eea
while the Hamiltonian constraint becomes
\bea
\!\!\!&&\!\!\! 32 \, W' \left[(D-p-2) X'+(p+1) Y'\right] \  -\  (D-2) (D-p-2) \left(X'\right)^2 \left[\gamma_0^2 (D-p-2)+\gamma ^2 p\right]\nonumber \\
\!\!\!&&\!\!\! -\  (D-2) (p+1) \left(Y'\right)^2 \left[\gamma ^2 (D-p-3)+(p+1)\gamma_0^2\right] \ - \  16 (D-2) \left(W'\right)^2 \nonumber \\ 
\!\!\!&&\!\!\!+ \  2 (p+1)(D-2) (D-p-2)  \left(\gamma ^2-\gamma_0^2\right)  X' Y' \nonumber \\
\!\!\!&&\!\!\!+ \ (D-2)^2 \,\gamma^2\left[\frac{k'(D-p-2) (D-p-3) \ e^{2 X}}{\ell^2}\ + \ \frac{k\, p (p+1) \ e^{2 Y}}{\ell^2}\ -\ T\ e^{2 W} \right]= 0  \,.
\eea

Special solutions exist when the three variables differ by constants, so that
\bea
Y \ = \ X \ + \ y \ , \qquad W \ = \ X \ + \ w \ , \label{YWyw}
\eea
with $y$ and $w$ two constants. In this case the system reduces to
\bea
X'' &=& e^{2X} \left[ - \, \frac{k'(D-p-3)^2}{\ell^2} \, + \, T\,e^{2w} \, - \, \frac{k\,p(p+1)}{\ell^2}\, e^{2y} \right] \ , \nonumber \\
X'' &=& e^{2X} \left[ - \, \frac{k'(D-p-2)(D-p-3)}{\ell^2} \,+\, T\,e^{2w} \, - \, \frac{k\,p^2}{\ell^2}\, e^{2y} \right] \ , \label{sys_YWyw} \\
X'' &=& e^{2X} \left[ - \, \frac{k'(D-p-2)(D-p-3)}{\ell^2} \,- \, \frac{T(D-2)}{16} \left(\gamma^2 \,-\,\gamma_c^2 \right)\, e^{2w} \, - \, \frac{k\,p(p+1)}{\ell^2}\,e^{2 y}  \right] \ , \nonumber
\eea
These equations imply that
\beq
\frac{k' (D-p-3)}{\ell^2} \ = \  \frac{k \,p}{\ell^2}\, e^{2y} \ = \ T\,e^{2w} \frac{(D-2)\left(\gamma_0^2 \ - \ \gamma^2\right)}{16} \ . \label{cond_YWyw}
\eeq
Therefore, $k$ and $k'$ must have the same sign as $T\left(\gamma_0^2 \ - \ \gamma^2\right)$ and, moreover, $\gamma \neq \gamma_0$. For that special value of $\gamma$, both $k$ and $k'$ must vanish, and so one is led back to the case with tension only. Otherwise the second-order equations can be turned into
\beq
X'' \ = \ - \frac{k(D-p-3)}{\ell^2}\left[D-1 \ - \ \frac{\gamma^2 \,-\,\gamma_c^2}{\gamma^2\,-\,\gamma_0^2} \right] e^{2X} \ = \ \frac{k}{\ell^2}\,\frac{(D-p-3)(D-2) \gamma^2}{\gamma_0^2 \,-\,\gamma^2}\, e^{2X} \ ,
\eeq
while the Hamiltonian constraint implies that ${\cal E}=0$ in this case. Letting
\beq
\epsilon\,\Delta^2 \ = \ \frac{k}{\ell^2}\,\frac{(D-p-3)(D-2) \gamma^2}{\gamma_0^2 \,-\,\gamma^2} \ ,
\eeq
where
\beq
\epsilon \ = \ k \, \sign\left(\gamma_0^2 \ - \ \gamma^2\right) \ = \ \sign(T)\ ,
\eeq
the solution only exists for $\epsilon=1$, and thus for $k=k'=1$ if $\gamma<\gamma_0$ and for $k=k'=-1$ for $\gamma>\gamma_0$, so that $T>0$ in both cases. The solution for $X$, 
\beq
X \ = \ - \ \log\left(\Delta\left|\tau\right|\right) \ , \label{XYW_back}
\eeq
with $- \infty<\tau<0$ implies that
\bea
A &=& \frac{1}{D-2}\, X \ - \ \frac{D-p-3}{D-2}\, y \ , \nonumber \\
B &=& \frac{D-1}{D-2} \, X \ + \ \frac{p+1}{D-2}\, y \ , \nonumber \\
C &=& \frac{1}{D-2} \, X \ + \ \frac{p+1}{D-2}\, y \ , \nonumber \\
\phi &=& \frac{2}{\gamma} \left[  - \ \frac{1}{D-2}\, X \ + \ w \ - \ \frac{p+1}{D-2}\, y \right] \ ,
\eea
and these expressions determine the background
\bea
ds^2 &=& \left(\frac{D-p-3}{p}\right)^\frac{p+1}{D-2} \left\{- \ \left(\Delta\left|\tau\right|\right)^{-\,\frac{2(D-1)}{D-2}}d\tau^2\right. \nonumber \\ &+& \left.\frac{\ell^2 \left(\Delta\left|\tau\right|\right)^{-\,\frac{2}{D-2}}}{D-p-3}\left[p\,ds_{p+1,k}^2 \,+\,(D-p-3)\,ds_{D-p-2,k}^2  \right] \right\} \ , \nonumber \\
e^\phi &=&  \left(\Delta\left|\tau\right|\right)^{\frac{2}{\gamma(D-2)}} \left(\frac{D-p-3}{p}\right)^{-\,\frac{p+1}{\gamma(D-2)}}\left[\frac{T \,k\,\ell^2\left(D-2\right)\left(\gamma_0^2\,-\,\gamma^2\right)}{16(D-p-3)} \right]^{-\,\frac{1}{\gamma}} \ . \label{sol_gammanot0kkpT}
\eea

In cosmic time
\bea
ds^2 &=& - \ dt^2 \ +\ \frac{\gamma^2\, t^2}{(D-2)\left|\gamma_0^2\,-\,\gamma^2\right|} \left( p\,ds_{p+1,k}^2 \,+\,(D-p-3)\,ds_{D-p-2,k}^2 \right) \ , \nonumber \\
e^\phi &=& \left[ \frac{\gamma^2\,t^2\,T}{16}\right]^{-\,\frac{1}{\gamma}} \ , \label{sol_gammanot0kkpT_CT}
\eea
so that there are spheres with linearly growing radii when $\gamma<\gamma_0$ and direct products of Milne-like universes and linearly growing (quotiented) hyperboloids when $\gamma>\gamma_0$. Note the sharp difference between these solutions and those obtained for $\gamma=0$.

 \subsection[\texorpdfstring{{\mdseries\textsc{The Stability Issue}}}{The Stability Issue}]
{{\mdseries\textsc{The Stability Issue}}}  \label{sec:stabilitykkpT}

Given the special nature of the exact solutions that we just discussed, it is important to address their stability under perturbations of the initial conditions, along the lines of what we did in Section~\ref{sec:Tensionkp}.

Let us begin by considering the solutions with $\gamma=0$ of eqs.~\eqref{kkpTgamma0}. To this end, we linearize eqs.~\eqref{Eqs_back_F} and \eqref{EqB_back_F} around them, letting
\beq
A \ = \ A_0(t) \ + \ \delta\,a \ , \qquad C \ = \ C_0 \ + \ \delta\,c \ , \quad \phi \ = \ \phi_0 \ + \ \delta\,\phi \ ,
\eeq
which leads to the system
\bea
&& \delta\,\ddot{a} \ + \ 2(p+1) \dot{A}_0 \,\delta\,\dot{a}\ = \ \frac{2 k p}{\ell^2}\, e^{-2 A_0}\,\delta\,a \ , \nonumber \\
&& \delta\,\ddot{c} \ + \ (p+1) \dot{A}_0 \,\delta\,\dot{c}\ = \ \frac{2 k' (D-p-3)}{\ell^2}\, e^{-2 C_0}\,\delta\,c \ , \nonumber \\ && \delta\,\ddot{\phi} \ + \ (p+1) \left(\dot{A}_0 \,\delta\,\dot{\phi}\,+\, \delta\,\dot{a}\, \dot{\phi}_0\right) \ = \ 0 \ , \label{pert71}
\eea
while the linearized Hamiltonian constraint reads
\bea
&& p(p+1) \dot{A}_0\,\delta\,\dot{a} \, + \, (p+1)(D-p-2) \dot{A}_0\,\delta\,\dot{c} \,-\, \frac{k p(p+1)}{\ell^2}\, e^{-2A_0}\,\delta\,a \nonumber \\  && - \, \frac{4}{D-2}\ \dot{\phi}_0\, \delta\,\dot{\phi} \ - \ \frac{k'(D-p-2)(D-p-3)}{\ell^2}\, e^{-2C_0}\,\delta\,c \ = \ 0 \ .
\eea

We can now analyze the different cases above.
\begin{enumerate}
    \item For the solution in eqs.~\eqref{kkpTgamma0}, after using the background equations, the first two of eqs.~\eqref{pert71} reduce to
    \bea
&& \delta\,\ddot{a} \ = \ \frac{2 T}{D-2}\,\delta\,a \ , \nonumber \\
&& \delta\,\ddot{c} \ = \ \frac{2 T}{D-2}\,\delta\,c \ ,
\eea
while the linearized Hamiltonian constraint gives
\beq
\delta\,\dot{\phi} \ = \ \frac{t}{4\,\dot{\phi}_0}\Big[(p+1) \delta\,a \ + \ (D-p-2)\delta\,c \Big] \ .
\eeq
The perturbed scale factors $\delta\,a$ and $\delta\,c$ have modes that grow exponentially in the past or in the future, and consequently the same is true for $\delta\, \phi$. One can thus conclude that the dynamical system is \emph{unstable in the past and in the future}.
\item For the solution of eqs.~\eqref{adsm1m1} $\phi_0$ is constant, and it suffices to consider the equation for $\delta\,\phi$, which decouples from the rest and implies that
\beq
\delta\,\phi \ =  \ C_\phi \ \Big[ \sin\left(\sigma_p t\right)\Big]^{-\,(p+1)} \ ,
\eeq
where $C_\phi$ is a constant.
This expression diverges at the endpoints, and so one can conclude that the dynamical system is \emph{unstable in the past and in the future}.
\item For the solution of eqs.~\eqref{ds11}, $\delta\,\dot{\phi}$ can be obtained as above, but now one finds a bounded expression,
\beq
\delta\,\phi \ =  \ C_\phi \ \Big[ \cosh\left(\sigma_p t\right)\Big]^{-\,(p+1)} \ ,
\eeq
so that the stability of the dynamical system must be ascertained from the equation for $\delta\,a$, which reads
\beq
\delta\,\ddot{a}\,+\,2(p+1) \sigma_p\, \tanh(\sigma_p t)\,\delta\,\dot{a} \ = \ \frac{2 k p\,\sigma_p^2}{\ell^2\,\cosh^2(\sigma_p t)}\ \delta\,a \ ,
\eeq
and from the equation for $\delta\,c$, which has a similar form. In the asymptotic regions $t \to \pm \infty$ this reduces to
\beq
\delta\,\ddot{a}\,\pm \,2(p+1) \sigma_p\,\delta\,\dot{a} \ = \ 0 \ ,
\eeq
so that no growing modes exist at the two ends and the dynamical system is \emph{stable}. Note that this is the coordinate system that provides a global foliation for de Sitter space.
\item For the solution of eqs.~\eqref{ds01}, one finds
\beq
\delta\,\dot{\phi} \ = \ C_\phi \, e^{- \,(p+1)\sigma_p t} \ ,
\eeq
and so the dynamical system is \emph{unstable in the past and stable in the future}, which is confirmed by the other equations.
\item Finally, for the solution of eqs.~\eqref{dsm11}, one finds
\beq
\delta\,\phi \ =  \ C_\phi \ \Big[ \sinh\left(\sigma_p t\right)\Big]^{-\,(p+1)} \ ,
\eeq
and so the dynamical system is \emph{unstable in the past and stable in the future}, which is confirmed by the other equations.
\end{enumerate}

For $\gamma \neq 0$ there are three perturbations, $\delta\,x$, $\delta\,y$ and $\delta\,w$, and linearizing eqs.~\eqref{eqs_kkpT} around the solution determined by eqs.~\eqref{YWyw}, \eqref{cond_YWyw} and \eqref{XYW_back} yields the system
\beq
\ddot{\xi} \ = \ \frac{2}{\tau^2}\, {\cal M} \, \xi \ ,
\eeq
where
\beq
\xi \ = \ \left(\begin{array}{c} \delta\,{x} \\ \delta\,{y} \\ \delta\,{w} \end{array} \right) 
\eeq
and
\beq
{\cal M} \ = \ \frac{1}{D-2} \left(\begin{array}{ccc} \left[1 \ - \ \left(\frac{\gamma_0}{\gamma} \right)^2 \right]
\left(D-p-3\right) & \left[1 \ - \ \left(\frac{\gamma_0}{\gamma} \right)^2 \right]\left(p+1\right) & \left(\frac{\gamma_0}{\gamma} \right)^2 (D-2) \\ &  & \\ \left[1 \ - \ \left(\frac{\gamma_0}{\gamma} \right)^2 \right]
\left(D-p-2\right) & \left[1 \ - \ \left(\frac{\gamma_0}{\gamma} \right)^2 \right] p & \left(\frac{\gamma_0}{\gamma} \right)^2 (D-2) \\ &  & \\ \left[1 \ - \ \left(\frac{\gamma_0}{\gamma} \right)^2 \right]
\left(D-p-2\right) & \left[1 \ - \ \left(\frac{\gamma_0}{\gamma} \right)^2 \right](p+1) & \left[ \left(\frac{\gamma_c}{\gamma} \right)^2 \ - \ 1 \right]  \end{array}   \right) \ .
\eeq
The matrix has the three eigenvectors,
\beq
\xi_1 \, = \, \left(\begin{array}{c} 1 \\ 1 \\ 1 \end{array} \right) \, , \quad \xi_2 \, = \, \left(\begin{array}{c} 1 \\ - \, 1 \\ 0 \end{array} \right) \, , \quad \xi_3 \,=\, \left(\begin{array}{c} 1 \\ 0 \\ - 1 \end{array} \right) \ , 
\eeq
with eigenvalues
\beq
\lambda_1 \,=\, 1 \ , \qquad \lambda_2\,=\,\lambda_3 \,=\, \frac{1}{D-2} \left[\left(\frac{\gamma_0}{\gamma} \right)^2 \ - \ 1 \right] \ .
\eeq

As in Section~\ref{sec:staility_sec6}, the first eigenvector simply affects the center of the solution, displacing it from $\tau=0$, while the others are genuine perturbations, so one is led to consider the Euler-type equation
\beq
\ddot{\xi}_\lambda \ = \ \frac{2\,\lambda}{\tau^2} \, {\xi}_\lambda \ ,
\eeq
with $\lambda=\lambda_2=\lambda_3$, whose general solution is
\beq
\xi_\lambda \ = \ C_1 \, \left|\tau\right|^{\alpha_+} \ + \ C_0 \, \left|\tau\right|^{\alpha_-} \ ,
\eeq
with
\beq
\alpha_\pm \ = \ \frac{1}{2}\left[ 1 \ \pm \ \sqrt{1 \ + \ 8\, \lambda_{2}}\right] \ .
\eeq
Comparing this result with the growth of the background solution of eq.~\eqref{XYW_back}, $-\,\log\left(\Delta\left|\tau\right|\right)$, as the universe expands, shows that perturbations overcoming it exist if $\gamma<\gamma_0$. Therefore, the system is~\emph{unstable in the future} if the spatial and internal sections have positive curvatures, since $\alpha_-<0$ in that case, while \emph{stable in the future} under these spatially constant perturbations in the complementary range $\gamma>\gamma_0$, where the spatial and internal sections have negative curvature. However, perturbations always override the background as $\tau \to - \infty$, so this system is \emph{always unstable in the past}. 

 \section{\sc Approximate Solutions} \label{sec:approx_sol}

 The general system of eqs.~\eqref{Eqs_back_F} and~\eqref{EqB_back_F} depends on the three parameters $(T,k,k')$. In the preceding section, we described the complete solutions that arise when two of them vanish. In addition, we found special classes of solutions that arise where two of the three contributions, or all of them, are proportional, but we could not find general solutions for the complete system. In this section, we characterize when exact results obtained in the presence of one parameter only among $(T,k,k')$ can approximate more general settings within one asymptotic region, by demanding that the other contributions be sub-dominant there.

\subsection[\texorpdfstring{{\mdseries\textsc{The $(T,k,k')=(0,0,1)$ Case}}}{The (T,t,k')=(0,0,1) Case}]
{{\mdseries\textsc{The $(T,k,k')=(0,0,1)$ Case}}} \label{sec:kppos_rest0}

 We begin by taking as our starting point the exact solutions with $(T,k,k')=(0,0,1)$ of eqs.~\eqref{coshtau}, which depend on the three parameters $(\rho,\Theta,\phi_0)$. 
\begin{enumerate}
 \item The $k$ contribution is subdominant if $e^{-A} \ll e^{-C}$. Letting
    \beq
     \cos\Theta_0 = \sqrt{\frac{p+1}{(D-2)(D-p-2)}} \  , \eeq
     one finds that:
 \begin{itemize}
     \item at early times $(\tau\to- \infty)$ this occurs if
     \beq \cos\Theta \ < \ \cos\Theta_0 \eeq
or equivalently if
     \beq
\Theta_0 \ < \ \Theta \ < 2\,\pi \ - \ \Theta_0  \ ; \label{ineqearlyk}
     \eeq
     \item At late times $(\tau\to +\infty)$ this occurs if
     $$ \cos\Theta \ > \ - \ \cos\Theta_0\ \  , $$
      or equivalently if
     \beq
-\ \pi\ + \ \Theta_0 \ < \ \Theta < \ \pi \ - \ \Theta_0 \ . \label{ineqlatek}
     \eeq
 \end{itemize}

 \item The $T$ contribution is subdominant if $e^{\gamma\,\phi} \ll  e^{-2 C}$.
 
 It is now convenient to define
\bea
\sin\Theta_1 &\equiv& {4}\,\sqrt{\frac{(D-2)}{\Xi (D-p-2)}} \ ,  \quad \sin\beta \ = \ \sin\Theta_1\,\sqrt{\frac{(p+1)(D-p-2)}{(D-2)}} \  ,  \nonumber \\
\cos\beta &=& \frac{\gamma}{4}\,\sin\Theta_1\, \sqrt{(D-2)(D-p-2)(D-p-3)}   \ , 
\eea
where $\Xi$ is defined in eq.~\eqref{Xi}, so that $\sin\Theta_1<1$. 
\begin{itemize}
    \item at early times $(\tau\to- \infty)$ the tadpole is sub-dominant if
    \beq
\sin \left({\Theta} \,-\,\beta\right) \ > \ - \ \sin\Theta_1 \ , \label{ineqsin}
    \eeq
    or equivalently if
    \beq
\beta \ - \ \Theta_1 \ < \ {\Theta} \ < \ \pi \ + \ \beta\ + \ \Theta_1 \ ; \label{ineqearlyT}
    \eeq
 \item at late times $(\tau\to +\infty)$  the tadpole is sub-dominant if
 \beq
\sin \left({\Theta} \,-\,\beta\right) \ < \sin\Theta_1 \ ,   \label{ineqsin2}
    \eeq
    or equivalently if
    \beq
\beta \ - \ \pi \ - \ \Theta_1 \ < \ {\Theta} \ < \ \beta \ + \ \Theta_1 \ . \label{ineqlateT}
    \eeq
\end{itemize}
  \item The $k$ and $T$ contributions are both subdominant if $e^{-A} \ll e^{-C}$ and $e^{\gamma\,\phi} \ll  e^{-2 C}$:
  \begin{itemize}
     \item at early times $(\tau\to- \infty)$ this occurs if both inequalities~\eqref{ineqearlyk} and \eqref{ineqearlyT} hold;
     \item at late times $(\tau\to + \infty)$ this occurs if both inequalities~\eqref{ineqlatek} and \eqref{ineqlateT} hold.
 \end{itemize}
\end{enumerate}
\subsection[\texorpdfstring{{\mdseries\textsc{The $(T,k,k')=(0,0,-1)$ Case}}}{The (T,t,k')=(0,0,-1) Case}]
{{\mdseries\textsc{The $(T,k,k')=(0,0,-1)$ Case}}} \label{sec:kpneg_rest0}
We now turn to the solutions with $(T,k,k')=(0,0,-1)$ of eqs.~\eqref{sinhtau}, which depend on the three parameters $(\rho,\Theta,\phi_0)$. For brevity, we refer to the definitions given for $k'=1$.
\begin{enumerate}
 \item The $k$ contribution is subdominant if $e^{-A} \ll e^{-C}$. 
 \begin{itemize}
     \item at early times $(\tau\to- \infty)$ this occurs if
     \beq \cos\Theta \ < \ \cos\Theta_0 \eeq
or equivalently if
     \beq
\Theta_0 \ < \ \Theta \ < 2\,\pi \ - \ \Theta_0  \ ; \label{ineqearlykn}
     \eeq
     \item At late times $(\tau\to 0^-)$  this never occurs.
 \end{itemize}

 \item The $T$ contribution is subdominant if $e^{\gamma\,\phi} \ll  e^{-2 C}$.

\begin{itemize}
    \item at early times $(\tau\to- \infty)$ the tadpole is sub-dominant if
    \beq
\sin \left({\Theta} \,-\,\beta\right) \ > \ - \ \sin\Theta_1 \ ,
    \eeq
    or equivalently if
    \beq
\beta \ - \ \Theta_1 \ < \ {\Theta} \ < \ \pi \ + \ \beta\ + \ \Theta_1 \ ; \label{ineqearlyTn}
    \eeq
 \item at late times $(\tau\to 0^-)$ this never occurs.
\end{itemize}
  \item The $k$ and $T$ contributions are both subdominant if $e^{-A} \ll e^{-C}$ and $e^{\gamma\,\phi} \ll  e^{-2 C}$.
  \begin{itemize}
     \item At early times $(\tau\to- \infty)$ this occurs if both inequalities~\eqref{ineqearlykn} and \eqref{ineqearlyTn} hold;
     \item At late times $(\tau\to 0^-)$ this never occurs.
 \end{itemize}
\end{enumerate}

\subsection{\sc Tension-Dominated Solutions} \label{sec:larger_tadpole} 

We can now describe the conditions under which the tadpole potential dominates over one or both curvature contributions. These conditions demand that
\beq
e^{\gamma\,\phi \,+\,2\,C} \ \gg \ 1 \ ,
\eeq
so that the contribution associated with $k'$ can be neglected, or that
\beq
e^{\gamma\,\phi \,+\,2\,A} \ \gg \ 1 \ ,
\eeq
so that the contribution associated with $k$ can be neglected, or both.
We follow the order of Section~\ref{sec:tnot0curv0}, and in particular the nomenclature of Section~\ref{sec:gamma_not_gammac} in cases with $\gamma \neq \gamma_c$. The interested reader can find several useful details in Table~\ref{tab:solsTkp12}.

\begin{itemize}
    \item[(0a) ] If $\gamma=\gamma_c$ and $\beta=0$, consistency demands that $T<0$. The curvature contributions associated to both $k'$ and $k$ are always subdominant in the past and future asymptotic regions, where the solution~\eqref{sol_gamma_c_beta0} approaches eq.~\eqref{asymptbeta0}.
    \item[(0b) ] If $\gamma=\gamma_c$ and $\beta \neq 0$, the solutions in eqs.~\eqref{gammac_init} depend on the angle $\theta$ in eqs.~\eqref{param_theta}, and it is convenient to let
    \bea
    \sin\theta_0 &=& \frac{1}{\sqrt{\frac{(D-2)(p+1)}{(D-p-2)}  + \ (D-1)^2}} \ , \qquad \sin\widetilde{\theta}_0 \ = \ \frac{1}{\sqrt{\frac{(D-2)(D-p-2)}{(p+1)}  + \ (D-1)^2}} \ , \nonumber \\ 
    \cos\sigma &=& (D-1)\,\sin\theta_0 \ , \qquad \cos\widetilde{\sigma} \ = \  (D-1)\,\sin\widetilde{\theta} _0 \ , \nonumber \\  \sin\sigma &=& \sqrt{\frac{(D-2)(p+1)}{(D-p-2)}}\, \sin\theta_0 \ , \qquad \sin\widetilde{\sigma} \ = \  \sqrt{\frac{(D-2)(D-p-2)}{(p+1)}}\, \sin\widetilde{\theta}_0 \ .
    \eea
    \begin{itemize}
    \item  at early times the tadpole dominates over the curvature $k'$ if $\gamma_c\,\alpha_\phi+2 \alpha_C<0$, and using eqs.~\eqref{param_theta} leads to the condition
    \beq
\sin\left(\theta \,-\,\sigma\right) \ < \ - \ \sin\theta_0 \ , 
    \eeq
    or equivalently to
    \beq
\pi \,+\, \theta_0 \,+\, \sigma \ < \ \theta  \ < \ 2\,\pi \,-\, \theta_0 \,+\, \sigma \ . \label{ineqgammaeq}
    \eeq
    \item  at late times the curvature $k'$ is never dominant for $T>0$, while it is always dominant for $T<0$.
    \item  at early times the tadpole dominates over the curvature $k$ if 
    $\gamma_c\,\alpha_\phi+2 \alpha_A<0$, and using eqs.~\eqref{param_theta} leads to the condition
    \beq
\sin\left(\theta \,+\,\widetilde{\sigma}\right) \ < \ - \ \sin\widetilde{\theta}_0 \ , 
    \eeq
    or equivalently to
    \beq
\pi \,+\, \widetilde{\theta}_0 \,-\, \widetilde{\sigma} \ < \ \theta  \ < \ 2\,\pi \,-\, \widetilde{\theta}_0 \,-\, \widetilde{\sigma} \ . \label{ineqgammaeq2}
    \eeq
        \item  at late times the curvature $k$ is never dominant for $T>0$, while it is always dominant for $T<0$.
    \end{itemize}
    \begin{figure}[ht]
\centering
\begin{tabular}{ccc}
\includegraphics[width=70mm]{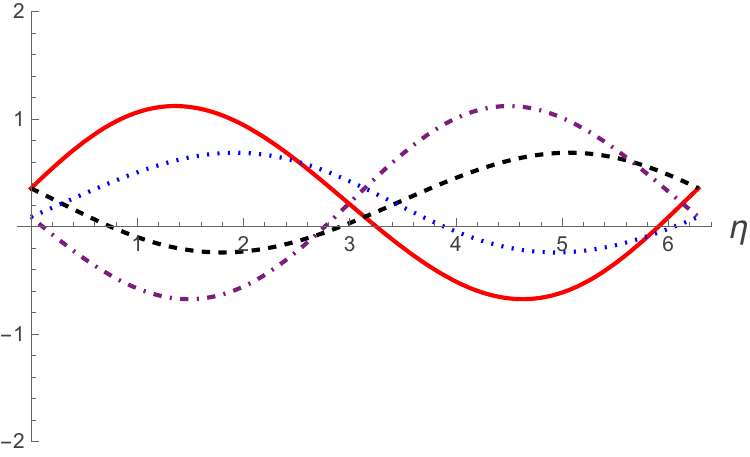} \quad  &
\includegraphics[width=70mm]{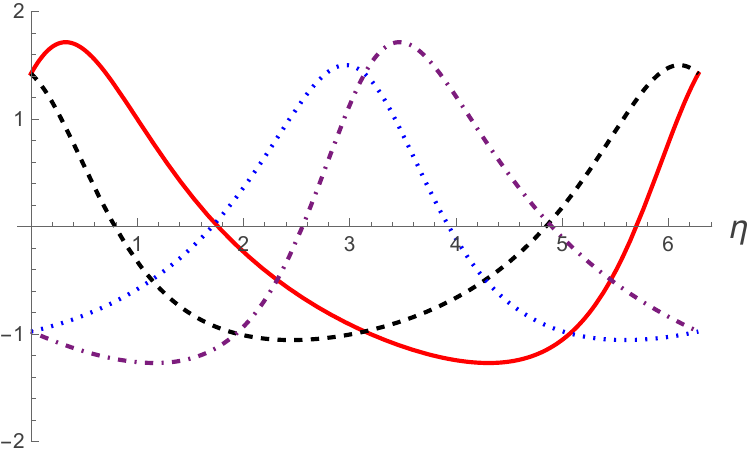} \\
\end{tabular}
 \caption{\small Typical (1a) inequalities with $\gamma< \gamma_c$ (with $D=10$, $p=2$) at early and late times, with the parametrization of eqs.~\eqref{phi1vn}.  Left panel: $\gamma=0.1$; right panel: $\gamma=0.9$. The curves correspond to $\gamma\,\beta_\phi^- + 2 \beta_A^-$ (red, solid), $\gamma\,\beta_\phi^- + 2 \beta_C^-$ (black, dashed), $\gamma\,\beta_\phi^+ + 2 \beta_A^+$ (magenta, dot-dashed),$\gamma\,\beta_\phi^+ + 2 \beta_C^+$ ( (blue, dotted).  The $\beta$ are the Kasner exponents in Appendix~\ref{app:tension_only}, and in the allowed regions the functions are negative.}
\label{fig:onea}
\end{figure}
    \item[(1a) ] If $\gamma<\gamma_c$, $T<0$ and ${\cal E}>0$, the solution is given in eqs.~\eqref{solutiongammamore1} and spans a finite range of cosmic time. The conditions for tadpole dominance over $k'$ at early and late times,
    \beq
    \gamma \,\beta_\phi^\pm \ + \ 2\, \beta_C^\pm \ < \ 0 \ ,
    \eeq
    where the $\beta$'s are defined in eqs.~\eqref{betaeta},
    are transcendental inequalities for the $\eta$ parameter, whose solutions can be studied graphically, as in fig.~\ref{fig:onea}. These conditions simplify for $\gamma=0$, and become
\beq
\sin\eta \ > \ \sqrt{\frac{D-p-2}{(p+1)(D-2)}}
\eeq
    at early times, and
\beq
\sin\eta \ < \ - \ \sqrt{\frac{D-p-2}{(p+1)(D-2)}}
\eeq
    at late times. 
    
    Letting 
    \beq
    \sin\eta_0 \ = \ \sqrt{\frac{D-p-2}{(p+1)(D-2)}} \ ,
    \eeq
    the two preceding inequalities become
    \beq
\eta_0 \ < \ \eta \ < \ \pi \ - \ \eta_0 \ ,  
    \eeq
    and 
    \beq
\pi \,+\, \eta_0 \ < \ \eta \ < \ 2\,\pi \ - \ \eta_0 \ . 
    \eeq
    The conditions for tadpole dominance over $k$ at early and late times,
    \beq
    \gamma \,\beta_\phi^\pm \ + \ 2\, \beta_A^\pm \ < \ 0 \ ,
    \eeq
    are also transcendental inequalities for the $\eta$ parameter, whose solutions can be studied graphically, as in fig.~\ref{fig:onea}. These conditions simplify for $\gamma=0$, and become
\beq
\sin\eta \ < \ - \ \sqrt{\frac{p+1}{(D-p-2)(D-2)}} 
\eeq
at early times, and
\beq
\sin\eta \ > \  \sqrt{\frac{p+1}{(D-p-2)(D-2)}} 
\eeq
at late times.
    \begin{figure}[ht]
\centering
\begin{tabular}{ccc}
\includegraphics[width=70mm]{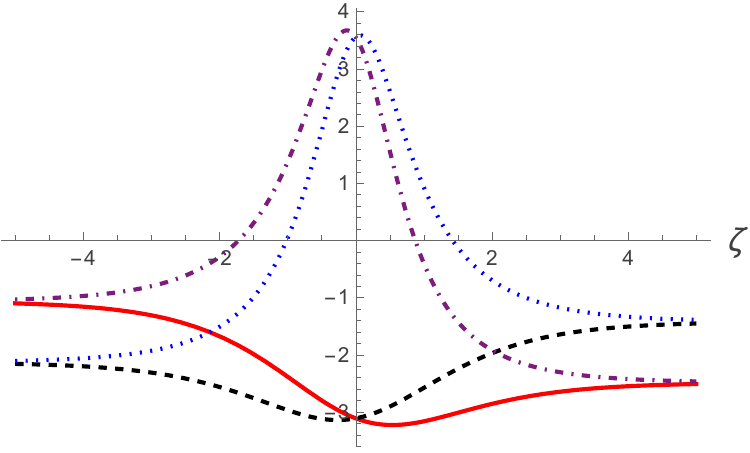} \quad  &
\includegraphics[width=70mm]{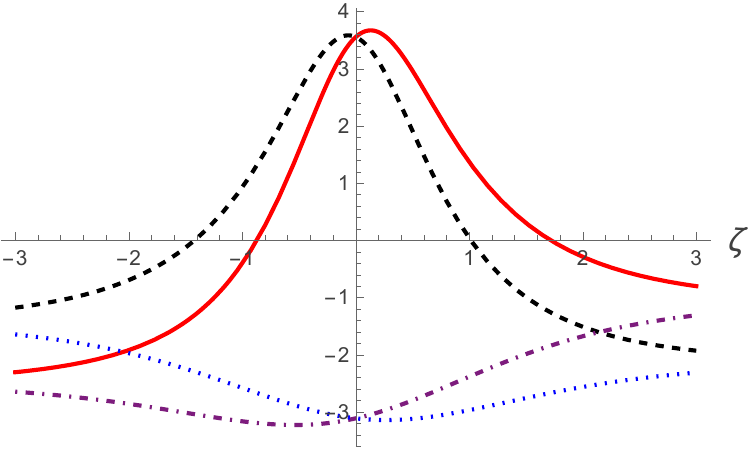} \\
\end{tabular}
 \caption{\small Typical (1b) inequalities with $\gamma> \gamma_c$ (with $D=10$, $p=2$, $\gamma=\frac{5}{2}$) at early and late times, with the parametrization of eqs.~\eqref{phi1v1zetatnegn}.  Left panel: $\epsilon'=1$; right panel: $\epsilon'=-1$. The curves correspond to $\gamma\,\beta_\phi^- + 2 \beta_A^-$ (red, solid), $\gamma\,\beta_\phi^- + 2 \beta_C^-$ (black, dashed), $\gamma\,\beta_\phi^+ + 2 \beta_A^+$ (magenta, dot-dashed),$\gamma\,\beta_\phi^+ + 2 \beta_C^+$ (blue, dotted). The $\beta$ are the Kasner exponents in Appendix~\ref{app:tension_only}.}
\label{fig:oneb}
\end{figure}
    \item[(1b, $\epsilon'$) ] If $\gamma>\gamma_c$, $T>0$ and ${\cal E}>0$, the solution is given in eqs.~\eqref{solutiongammamore1}, with the parametrization of eqs.~\eqref{phi1v1zetatnegn}.
    
    For $\epsilon'=1$ there is infinite past and finite future, and the conditions for tadpole dominance are (with the $\beta's$ defined in eqs.~\eqref{betaplusminus1b}):
    \begin{itemize}
    \item at early times:  $\gamma \,\beta_\phi^- \ + \ 2\, \beta_A^- \ >  \ 0$ for dominance over $k$ and $\gamma \,\beta_\phi^- \ + \ 2\, \beta_C^- \ >  \ 0$ for dominance over $k'$;
    \item at late times:  $\gamma \,\beta_\phi^+ \ + \ 2\, \beta_A^+ \ <  \ 0$ for dominance over $k$ and  $\gamma \,\beta_\phi^+ \ + \ 2\, \beta_C^+ \ <  \ 0$ for dominance over $k'$.
    \end{itemize}
    These conditions can be studied graphically as in fig.~\ref{fig:oneb}, and are never satisfied in the large-$\gamma$ limit.

     For $\epsilon'=-1$ there is finite past and infinite future, and the conditions for tadpole dominance are (with the $\beta's$ defined in eqs.~\eqref{betaplusminus1b}):
    \begin{itemize}
    \item at early times:  $\gamma \,\beta_\phi^- \ + \ 2\, \beta_A^- \ <  \ 0$ for dominance over $k$ and $\gamma \,\beta_\phi^- \ + \ 2\, \beta_C^- \ <  \ 0$ for dominance over $k'$;
    \item at late times:  $\gamma \,\beta_\phi^+ \ + \ 2\, \beta_A^+ \ >  \ 0$ for dominance over $k$ and $\gamma \,\beta_\phi^+ \ + \ 2\, \beta_C^+ \ >  \ 0$ for dominance over $k'$.
    \end{itemize}
    These conditions are never satisfied in the large-$\gamma$ limit.

      \item[(2a) ] If $\gamma<\gamma_c$, $T>0$ and ${\cal E}=0$, the solution is given in eqs.~\eqref{gammalessTpE0}.
      \begin{itemize}
      \item For $0\leq \gamma<\gamma_0$ the dominance conditions over $(k,k')$ do not hold at early times but do hold at late times;
      \item For $\gamma=\gamma_0$ the conditions never hold;
      \item For $\gamma_0< \gamma<\gamma_c$ the conditions do not hold at late times but do hold at early times.
      \end{itemize}

    \item[(2b) ] If $\gamma<\gamma_c$, $T>0$ and ${\cal E}>0$, the solution is given in eqs.~\eqref{Tnokpgammaless}, with the parametrization of eqs.~\eqref{betaeta}. 
    \begin{itemize}
    \item At early times the condition for tadpole dominance over $k$ is $\gamma\,\beta_\phi^- \,+\, 2 \beta_A^- <0$, while the condition for tadpole dominance over $k'$ is  $\gamma\,\beta_\phi^- \,+\, 2 \beta_C^- <0$. For $\gamma=0$ these conditions reduce to
    \beq
\sin\eta \ > \  \sqrt{\frac{p+1}{(D-p-2)(D-2)}} 
\eeq
and to
    \beq
\sin\eta \ < \ - \  \sqrt{\frac{D-p-2}{(p+1)(D-2)}} \ . \label{sinetaless}
\eeq
In general, the solutions can be studied graphically, and examples are provided by the (red, solid) and (black, dashed) curves in fig.~\ref{fig:onea}.
    \item At late times the behavior is like in case (2a), so that the conditions hold for $\gamma<\gamma_0$ and do not hold for $\gamma \geq \gamma_0$.
    \end{itemize}
    
    \item[(2c, $\epsilon'$) ] If $\gamma>\gamma_c$, $T<0$ and ${\cal E}<0$, the solution is given in eqs.~\eqref{neg_E_epsilonpos}, and there is a finite span of cosmic time. Tadpole dominance over $k$ and $k'$ holds at both ends in this case.
    
\item[(2d, $\epsilon'$) ] If $\gamma>\gamma_c$, $T<0$ and ${\cal E}=0$, the solution is given in eqs.~\eqref{neg_E_epsilonpos_lim}, with the parametrization~\eqref{betasmorethan}, and the tadpole always dominates at early times. On the other hand, at late times tadpole dominance over $k$ never holds for $\phi_1<0$ but always holds for $\phi_1>0$.

    \item[(2e, $\epsilon'$) ] If $\gamma>\gamma_c$, $T<0$ and ${\cal E}>0$, the solution is given in eqs.~\eqref{Tnokpgammaless}.
    The tadpole-dominance conditions over $k$ (over $k'$) are then, with the $\beta$'s of eqs.~\eqref{alphasgammamore} and the parametrization~\eqref{phi1v1zetatnegn},
    \begin{itemize}
    \item $\gamma\,\beta_\phi^- \,+\, \beta_A^- > 0$ ($\gamma\,\beta_\phi^- \,+\, \beta_C^- > 0$) if $\epsilon'=1$;
    \item $\gamma\,\beta_\phi^- \,+\, \beta_A^- < 0$ ($\gamma\,\beta_\phi^- \,+\, \beta_C^- < 0$) if $\epsilon'=-1$.
    \end{itemize}
    In general, the solutions can be studied graphically, and examples are provided by the (red, solid) and (black, dashed) curves in fig.~\ref{fig:oneb}. From the figure one can see that no solutions exist for $\epsilon'=1$, while they do exist for $\epsilon'=-1$.
    
    The late-time behavior is captured by eqs.~\eqref{gammalessTpE0}, or by eqs.~\eqref{lm} in cosmic time, but now $\omega<0$, and so the late-time behavior translates into $t \to 0$. The tadpole dominance conditions hold identically in the limit, since $\gamma>\gamma_0$ for these solutions.
 \end{itemize}   

 \section{\sc Scaling Solutions in Cosmic Time}  \label{sec:log_asympt}

In this section we derive consistency conditions granting that $A$, $C$ and $\phi$ have asymptotic logarithmic behaviors in cosmic time, and explore several solutions of this type.

Our starting point is provided by the system of eqs.~\eqref{Eqs_back_C} and~\eqref{EqB_back_C}, and by
\beq
A \ \sim \ \alpha_A \, \log(Mt) \ , \qquad C \ \sim \ \alpha_C \, \log(Mt) \ , \qquad \phi \ \sim \ \alpha_\phi \, \log(Mt) \ .  \label{asympt_sing3}
\eeq
We begin by addressing the behavior close to a singularity that is encountered at a finite value of the cosmic time $t$, say $t=0$. Referring to Appendix~\ref{app:tension_only}, these considerations apply to singularities that occur in the finite past (cases II; III; 0a; 0b; 1a; 1b, $\epsilon'=-1$; 2b; 2e, $\epsilon'=-1$, but also in the finite future (cases 0a; 0b, $T<0$; 1a; 1b,  $\epsilon'=1$; 2d, $\epsilon'$, $\phi_1>0$ in Tables~\ref{tab:solsTkp11} and \ref{tab:solsTkp12}). 

Eqs.~\eqref{asympt_sing3} turn the second-order equations into
\bea
\frac{\alpha_A}{t^2} \left[ -\,1 +(p+1)\alpha_A +(D-p-2)\alpha_C\right ] &=& \frac{T}{(D-2)}\, (Mt)^{\gamma\,\alpha_\phi}  -  \frac{k\,p}{\ell^2}\, (Mt)^{-\,2\alpha_A} , \nonumber \\
\frac{\alpha_C}{t^2} \left[ -\,1 +(p+1)\alpha_A +(D-p-2)\alpha_C\right ] &=& \frac{T}{(D-2)}\, (Mt)^{\gamma\,\alpha_\phi}  -  \frac{k'(D-p-3)}{\ell^2}\, (Mt)^{-\,2\alpha_C} \,, \nonumber \\
\frac{\alpha_\phi}{t^2} \left[ -\,1 +(p+1)\alpha_A +(D-p-2)\alpha_C\right ] &=& - \ \frac{T\,\gamma\,(D-2)}{8}\ (Mt)^{\gamma\,\alpha_\phi}\ ,  \label{sys_alpha}
\eea
and if all source terms due to tension and curvatures are subdominant in the limit, which is the case if
\beq
\gamma\,\alpha_\phi \ > \ - \ 2 \ , \qquad \alpha_A < 1  \ , \qquad \alpha_C < 1 \ , \label{free_dominance}
\eeq
the system admits a limiting free Kasner solution. To begin with, one readily recovers the familiar condition
\beq
(p+1)\alpha_A \ + \ (D-p-2)\alpha_C \ = \ 1 \ , \label{eqlin_0}
\eeq
while the Hamiltonian constraint~\eqref{EqB_back_C} reduces to
 \begin{align}
(p+1) \alpha_A^2 \ + \ (D-p-2) \alpha_C^2 \ + \ \frac{4}{D-2}\, \alpha_\phi^2 \ = \ 1 \ . \label{EqB_red02}
 \end{align}
As we have seen, the solutions of eqs.~\eqref{eqlin_0} and \eqref{EqB_red02} can be parametrized as in eqs.~\eqref{param_theta}, so that the inequalities in eq.~\eqref{free_dominance} translate into the conditions
 \beq
\frac{\gamma}{\gamma_c}\, \sin\theta \ > \ - \ 1   \ , \qquad \cos\theta \ < \  \sqrt{\frac{(D-2)(p+1)}{(D-p-2)}} \ , \qquad \cos\theta \ > \ - \ \sqrt{\frac{(D-2)(D-p-2)}{(p+1)}} \ . \label{conditions_asympt}
 \eeq
 The first condition is identically satisfied if $\gamma \leq \gamma_c$, but it can also be satisfied if $\gamma>\gamma_c$, within a range of values for $\theta$. This result is manifest for $\gamma< \gamma_c$, while for $\gamma=\gamma_c$ we saw that the solution only exists in the past for $\theta \neq - \,\frac{\pi}{2}$. In fact, all exact solutions with tension only of Section~\ref{sec:tnot0curv0} satisfy identically the first of these conditions, independently of the value of $\gamma$, consistent with an early finite-time Kasner behavior. This also occurs for $\gamma>\gamma_c$, since the parametrization~\eqref{betaplusminus1b} of the $\widehat{\beta}$ exponents restricts $\widehat{\beta}_\phi$ to positive values. These results are summarized in Table~\ref{tab:solsTkp12}. On the other hand, the second and third conditions are identically satisfied. The cases (0a; 0b, $T<0$), where $\gamma=\gamma_c$ and $\theta=-\,\frac{\pi}{2}$ in the finite future, are marginal, but the Kasner behavior emerges taking logarithmic corrections into account, as in the stability analysis in~\cite{dms_review}. 

 One can repeat the analysis if the singularity lies at infinite proper time, in the past or in the future. In these cases the free Kasner behavior is recovered if the tension and curvature contributions fall down faster than $\frac{1}{t^2}$ as $t \to \infty$. The preceding inequalities then become
  \beq
\frac{\gamma}{\gamma_c}\, \sin\theta \ < \ - \ 1   \ , \qquad \cos\theta \ > \  \sqrt{\frac{(D-2)(p+1)}{(D-p-2)}} \ , \qquad \cos\theta \ < \ - \ \sqrt{\frac{(D-2)(D-p-2)}{(p+1)}} \ . \label{conditions_asympt2}
 \eeq
 While the first inequality can be satisfied if $\gamma>\gamma_c$, as occurs in the past for cases (1b,  $\epsilon'=1$) and (2e, $\epsilon'=1$) and in the future for cases (1b, $\epsilon'=-1$) and (2d, $\epsilon'=\pm 1$, $\phi_1 \leq 0$) of Table~\ref{tab:solsTkp12}), the others cannot be satisfied. The case (0b, $T>0$, where $\gamma=\gamma_c$ and $\theta=-\,\frac{\pi}{2}$, is marginal, but the Kasner behavior emerges again taking logarithmic corrections into account, as in the stability analysis in~\cite{dms_review}. 
 
 If the system has a finite duration in cosmic time, this asymptotics does not exist. This is typically the case for positive curvatures, while for negative curvatures one obtains Milne-like universes, where $e^A \sim e^C \sim t$, but in this case the curvature terms are not negligible. The LM solution, to which we now turn, is precisely of this type. 

 The second type of asymptotics that we have met is found indeed in the LM attractor, which continues to play a role, under certain conditions, even if curvatures are present. 
 In this case $\gamma \neq 0$, and it is convenient to consider the more general Ansatz 
 \beq
A \ = \ \alpha_A \, \log(M_A t) \ , \qquad C \ = \ \alpha_C \, \log(M_C t) \ , \qquad \phi \ =  \ - \ \frac{2}{\gamma}\, \log(M_\phi t) \ .  \label{asympt_sing2}
\eeq
 involving the three scales $M_A$, $M_C$ and $M_\phi$, or equivalently allowing relative shifts.
 Eqs.~\eqref{sys_alpha} then become
  \bea
\frac{\alpha_A}{t^2} \left[ -\,1 +(p+1)\alpha_A +(D-p-2)\alpha_C\right ] &=& \frac{T}{(D-2) (M_\phi t)^2} \ - \  \frac{k\,p}{\ell^2}\, (M_A t)^{-\,2\alpha_A} , \nonumber \\
\frac{\alpha_C}{t^2} \left[ -\,1 +(p+1)\alpha_A +(D-p-2)\alpha_C\right ] &=& \frac{T}{(D-2) (M_\phi t)^2}  \ - \  \frac{k'(D-p-3)}{\ell^2}\, (M_C t)^{-\,2\alpha_C} , \nonumber \\
\frac{1}{t^2} \left[ -\,1 +(p+1)\alpha_A +(D-p-2)\alpha_C\right ] &=&  \frac{T\,\gamma^2 \,(D-2)}{16 (M_\phi t)^2}\  , \label{sys_alpha2}
\eea
while the Hamiltonian constraint reduces to
\bea
&& \frac{1}{t^2}\left[p(p+1) \alpha_A^2\,+\,2(p+1)(D-p-2)\alpha_A\,\alpha_C \,+\,(D-p-2)(D-p-3) \alpha_C^2 \,-\, \frac{4\,\alpha_\phi^2}{D-2} \right] \nonumber \\
&& = \ \frac{T}{\left(M_\phi\,t\right)^2}\, - \, \frac{k\,p(p+1)}{\ell^2 \left(M_A t\right)^{2\alpha_A}}\,-\, \frac{k'(D-p-2)(D-p-3)}{\ell^2 \left(M_C t\right)^{2 \alpha_C}} \ .
\eea
Substituting the last of eqs.~\eqref{sys_alpha2} in the first two gives
\bea
\frac{T(D-2)}{16(M_\phi t)^2} \left(\alpha_A\,\gamma^2\,-\,\gamma_0^2\right) &=& - \  \frac{k\,p}{\ell^2}\, (M_A t)^{-\,2\alpha_A} \ , \nonumber \\
\frac{T(D-2)}{16(M_\phi t)^2} \left(\alpha_C\,\gamma^2\,-\,\gamma_0^2\right) &=& - \  \frac{k'(D-p-3)}{\ell^2}\, (M_C t)^{-\,2\alpha_C}  \ .
\eea

There are now a few options:
\begin{itemize}
\item If $k=k'=0$ and $\gamma \neq 0$, the solution is
\beq
\alpha_A \ = \ \alpha_C \ = \ \frac{\gamma_0^2}{\gamma^2} \ , \qquad \alpha_\phi \ = \ - \ \frac{2}{\gamma} \ ,
\eeq
and the last of eqs.~\eqref{sys_alpha2} links $T$ and $M_\phi$ according to
\beq
\left( - 1 \ + \ (D-1)\,\frac{\gamma_0^2}{\gamma^2}\right) \ \equiv \ \left( - 1 \ + \ \frac{\gamma_c^2}{\gamma^2}\right)  \ = \ \frac{T\,\gamma^2 \,(D-2)}{16 M_\phi^2} \ .
\eeq
The Hamiltonian constraint is then identically satisfied,
and one recovers the LM solution. In this case, the three mass scales $M_A$, $M_C$ and $M_\phi$ can be identified after redefining the spatial and internal coordinates.
The solution is then
\bea
ds^2 &=& - \ dt^2 \ + \ \left(M_\phi \,t\right)^\frac{2 \gamma_0^2}{\gamma^2} \left(d \vec{x}^2 \ + \ d\vec{y}^2 \right) \ , \nonumber \\
e^\phi &=& \left(M_\phi \,t\right)^{-\,\frac{2}{\gamma}} \ .
\eea
This exact LM~\cite{lm} solution emerged in cases $(2a)_2$ and $(2d,\epsilon')$ with $\phi_1=0$.

\item Another option, if $k \neq 0$ and $k'=0$, is
\bea
&&\alpha_A \ = \  1 \ , \qquad \alpha_C \ = \ \frac{\gamma_0^2}{\gamma^2} \ , \qquad \alpha_\phi \ = \ - \ \frac{2}{\gamma} \ , \nonumber \\
&& \frac{T(D-2)}{16\,M_\phi^2} \left(\gamma_0^2\,-\,\gamma^2\right) \ = \   \frac{k\,p}{\left( \ell\,M_A\right)^2}  \ , \nonumber \\
&&   p \  +\ (D-p-2)\, \frac{\gamma_0^2}{\gamma^2} \ = \   \frac{T\,\gamma^2 \,(D-2)}{16 \, M_\phi^2}\ .
\eea
The last equation links $M_\phi$ and $T$, and implies that $T>0$, while the second links $M_\phi$, $M_A$ and $\ell$ and determines $k$ as the sign of $\gamma_0 - \gamma$, and the Hamiltonian constraint is then identically satisfied. 
The solution is then
\bea
ds^2 &=& - \ dt^2 \ + \ \ell^2 \left(M_A t\right)^2\, ds_{p+1,k}^2  \ + \ \left(M_\phi \,t\right)^\frac{2 \gamma_0^2}{\gamma^2} \ d\vec{y}^2\ , \nonumber \\
e^\phi &=& \left(M_\phi \,t\right)^{-\,\frac{2}{\gamma}} \ . \label{9.17}
\eea
In particular, if $k<0$ one obtains in this case a Milne-like spacetime.

\item Another option, if $k = 0$ and $k' \neq 0$, is
\bea
&&\alpha_A \ = \  \frac{\gamma_0^2}{\gamma^2}  \ , \qquad \alpha_C \ = \ 1 \ , \nonumber \\
&& \frac{T(D-2)}{16\,M_\phi^2} \left(\gamma_0^2\,-\,\gamma^2\right) \ = \   \frac{k'\,(D-p-3)}{\left( \ell\,M_A\right)^2}   \ , \nonumber \\
&&   D-p-3 \  +\ (p+1)\, \frac{\gamma_0^2}{\gamma^2} \ = \   \frac{T\,\gamma^2 \,(D-2)}{16 \, M_\phi^2}\ .
\eea
These conditions are obtained from the preceding ones interchanging $A$ and $C$, and thus $p$ with $D-p-3$. Now $k'$ is determined as the sign of $\gamma_0 - \gamma$. 
The solution is then
\bea
ds^2 &=& - \ dt^2  \ + \ \left(M_\phi \,t\right)^\frac{2 \gamma_0^2}{\gamma^2} \ d\vec{x}^2 \ + \ \ell^2 \left(M_C t\right)^2\, ds_{D-p-2,k'}^2  \ , \nonumber \\
e^\phi &=& \left(M_\phi \,t\right)^{-\,\frac{2}{\gamma}} \ .
\eea
This emerged as a special solution in the presence of tension and $k'$ in eqs.~\eqref{exactkpminus}. The preceding case of eq.~\eqref{9.17} can be obtained from this by interchanging $(k,k')$ and $(p,D-p-3)$, following the discussion in Section~\ref{sec:Tensionk}.

\item Moreover, if both $k$ and $k'$ are not zero, there are isotropic solutions with $\alpha_A=\alpha_C=1$ 
provided
\bea
&&   \frac{T\,\gamma^2}{16 \, M_\phi^2} \ = \ 1 \ , \nonumber \\
&& (D-2)\left[ \left(\frac{\gamma_0}{\gamma}\right)^2\,-\,1\right] \ = \ \frac{k\,p}{\left( \ell\,M_A\right)^2}  \ = \  \frac{k'(D-p-3)}{\left( \ell\,M_C\right)^2} \ , 
\eea
so that $T>0$, $k=k'$,  and their sign is determined by the sign of $\gamma_0-\gamma$. The Hamiltonian constraint is again identically satisfied, and the background reads
\bea
ds^2 &=& - \ dt^2  \ + \ \frac{\gamma^2\, t^2}{(D-2)\left|{\gamma_0}^2\,-\,\gamma^2\right| } \left[ p\, ds_{p+1,k}^2 \ + \ (D-p-3)\, ds_{D-p-2,k'}^2 \right] \ , \nonumber \\
e^\phi &=& \left(M_\phi \,t\right)^{-\,\frac{2}{\gamma}} \ .
\eea
These are the special solutions found in eqs.~\eqref{sol_gammanot0kkpT_CT}, whose properties are summarized in Table~\ref{tab:solsTkp13}.
\item If $T=0$ and at least one of the curvatures does not vanish, eqs.~\eqref{sys_alpha} demand that $\alpha_\phi=0$, and the system reduces to
\bea
\frac{\alpha_A}{t^2} \left[ -\,1 +(p+1)\alpha_A +(D-p-2)\alpha_C\right ] &=&   - \  \frac{k\,p}{\ell^2}\, (M_A t)^{-\,2\alpha_A} , \nonumber \\
\frac{\alpha_C}{t^2} \left[ -\,1 +(p+1)\alpha_A +(D-p-2)\alpha_C\right ] &=&  - \  \frac{k'(D-p-3)}{\ell^2}\, (M_C t)^{-\,2\alpha_C} \ . 
\eea
If both $k$ and $k'$ differ from zero, there is a solution with $\alpha_A=\alpha_C=1$, which demands that $k=k'=-1$, and then $M_A$ and $M_C$ are determined by
\beq
D-2 \ = \ \frac{p}{\left( M_A\,\ell \right)^2} \ = \ \frac{D-p-3}{\left( M_C\,\ell \right)^2} \ .
\eeq
The background then reads
\bea
ds^2 &=& - \ dt^2  \ + \ \frac{t^2}{(D-2)} \left[ p\, ds_{p+1,k=-1}^2 \ + \ (D-p-3)\, ds_{D-p-2,k'=-1}^2 \right] \ , \nonumber \\
e^\phi &=& e^{\phi_0} \ ,
\eea
and one recovers eqs.~\eqref{milnelike} that, as we pointed out at the end of Section~\eqref{sec:t0curvnot0}, describe a special exact solution for the two-curvature system in the $\rho \to \infty$ limit.
\end{itemize}
\newpage
\section{\sc Conclusions}
  \begin{table}[!ht]
 \begin{center}
  \scalebox{0.8}{
\begin{tabular}{ || c | c||c|c|| } 
 \hline\hline
 n & Case & Early $t$ & Late $t$ \\ [0.5ex] 
  \hline\hline
   & & & $T>0, \gamma \leq \gamma_0$\,: \, \emph{$(LM)$} \\
  &  $\mathbf{e^{-\,2\,C} \gg e^{\gamma\,\phi}}$ & K. & $T<0, \gamma < \gamma_c$\,: \, {$(\overline{II})$} \\ $1$   & \textbf{Early $k'$ Dom. (III)} & \eqref{acphik'pos}, \eqref{ineqsin} &   $T<0, \gamma = \gamma_c$\,: \  $(\overline{II})$  \\ & $\mathbf{\left(k'=  1\right)}$
 & \textbf{[f.p.]} &  $T>0, \gamma>\gamma_0$ \,: \, $(\overline{II})$ \\ 
 & & & $T<0, \gamma>\gamma_c$ \,: \, $(LM)$ \\ [0.5ex] \hline
 & & & $T>0, \gamma \leq \gamma_0$\,: \, \emph{$(LM)$} \\
   &  $\mathbf{e^{-\,2\,C} \gg e^{\gamma\,\phi}}$ & K. & $T<0, \gamma \leq \gamma_c$\,: \, $(\overline{II})$ \\ $2$  & \textbf{Early $k'$ Dom. (V)} & \eqref{acphik'pos}, \eqref{ineqsin} &  $T<0, \gamma = \gamma_c$\,: \ $(\overline{II})$ \\& $\mathbf{\left(k'= - 1\right)}$
 & \textbf{[f.p.]} & $T>0, \gamma>\gamma_0$ \,: (\emph{2B})  \\ 
 & & & $T<0, \gamma>\gamma_c$ \,: \, $(LM)$ \\ [0.5ex] \hline 
  $3$ &  $e^{-2C} \gg e^{\gamma\phi}$ & K. &  \\ & \textbf{Early $k'$ Dom. $(\overline{\mathbf{III}})$} & \eqref{acphik'pos}, \ \eqref{ineqsin2}  & $T>0$ or $T<0$ : \  $(\overline{II})$  \\ & $\mathbf{\left(k'=1\right)}$
 & \textbf{[f.p.]} &  \\ [0.5ex] 
  \hline 
  $5$ &  $\mathbf{e^{\gamma\phi} \gg e^{-2C}}$ & iso K. & iso $(\overline{II})$ ($k'=-1)$ \\ & \textbf{Early T$<0$ Dom. (0a)} & \eqref{asymptbeta0} &   \\ & $\mathbf{\left(\gamma=\gamma_c\right)}$
 & \textbf{[f.p.]} & $(\overline{II}) \  (k'=1)$ \\ [0.5ex] 
  \hline
   $6$ & $\mathbf{e^{\gamma\phi} \gg e^{-2C}}$ & K. constr.  \eqref{ineqgammaeq}  & iso $(\overline{II}) \ (k'=-1)$ \\ & \textbf{Early T$<0$ Dom. (0b)} & \eqref{gammac_init2} &  \\ & $\mathbf{\left(\gamma=\gamma_c\right)}$
 & $(\gamma_c\alpha_\phi + 2 \alpha_C<0)$ \textbf{[f.p.]} &  $(\overline{II}) \ (k'=1)$ \\ [0.5ex] 
  \hline
   $7$ & $\mathbf{e^{\gamma\phi} \gg e^{-2C}}$ & K. \ \eqref{ineqgammaeq}  & \emph{ $({2B})$} ($k'=-1)$ \\ & \textbf{Early T$>0$ Dom. (0b)} & \eqref{gammac_init2} &  \\ & $\mathbf{\left(\gamma=\gamma_c\right)}$
 & $(\gamma_c\alpha_\phi + 2 \alpha_C<0)$ \textbf{[f.p.]} & \emph{ $(\overline{II})$ ($k'=1)$} \\ [0.5ex] 
  \hline
   $8$ & $\mathbf{e^{\gamma\phi} \gg e^{-2C}}$ & iso K. \ \eqref{ineqgammaeq}  & \emph{ $({2B})$} ($k'=-1)$ \\ & \textbf{Early T$>0$ Dom.} $\mathbf{\overline{ (0b)}}$ & \eqref{criticalTpos} $(t \to - t)$ &  \\ & $\mathbf{\left(\gamma=\gamma_c\right)}$
 &  \textbf{[i.p.]} & \emph{ $(\overline{II})$  ($k'=1)$} \\ [0.5ex] 
  \hline
   $9$ & $\mathbf{e^{\gamma\phi} \gg e^{-2C}}$ & K. , fig.~\ref{fig:onea} & $(\overline{II})$  ($k'=1$) \\ & \textbf{Early T<0 Dom.} \textbf{(1a)} &  \eqref{solutiongammamore1}, \eqref{phi1vn} &  \\ & $\left(\gamma \leq \gamma_0\right)$
 & $(\gamma\,{\beta}_\phi^-+2 {\beta}_C^-<0)$ \textbf{[f.p.]} & \emph{ $(\overline{II})$} ($k'=-1$) \\ [0.5ex]  
   \hline
      $10$ & $\mathbf{e^{\gamma\phi} \gg e^{-2C}}$ & K. , fig.~\ref{fig:onea} &  $(\overline{II})$ ($k'=1$) \\ & \textbf{Early T<0 Dom.} \textbf{(1a)} &  \eqref{solutiongammamore1}, \eqref{phi1vn} &  \\ & $\left(\gamma_0<\gamma< \gamma_c\right)$
 & $(\gamma\,{\beta}_\phi^-+2 {\beta}_C^-<0)$ \textbf{[f.p.]} & $(\overline{II})$ ($k'=-1$) \\ [0.5ex]  
   \hline
   $11$ & $\mathbf{e^{\gamma\phi} \gg e^{-2C}}$ & K., fig.~\ref{fig:oneb} & $(\overline{II})$  ($k'=1$) \\ & \textbf{Early T>0 Dom.} \textbf{(1b)} &  \eqref{solutiongammamore1}, \eqref{phi1v1zetatnegn} &  \\ & $\left(\gamma>\gamma_c\right)$
 &  $(\gamma\,\widehat{\beta}_\phi^-+2 \widehat{\beta}_C^-<0)$ \ \textbf{[f.p.]} & \emph{ (2B)} ($k'=-1$) \\  [0.5ex] \hline
   $12$ & $\mathbf{e^{\gamma\phi} \gg e^{-2C}}$ &  &   $(\overline{II})$ $(k'=-1)$ \\ & \textbf{Early T>0 Dom.}  & $\mathbf{\overline{dS}}:$  \eqref{desitter} $(t\to -t)$ &    \\ & $\mathbf{\left(\gamma=0 \right)}$ $\mathbf{\overline{(2a)}_1}$ or $\mathbf{\overline{(2b)}_1}$ &  \textbf{[i.p.]}
 & $(\overline{II})$ $(k'=1)$ \\ [0.5ex] 
 \hline\hline
\end{tabular}}
 \end{center}
 \vskip 12pt 
 \caption{\small A first portion of the results obtained combining analytical and numerical indications on systems subject to both $T$ and $k'$. The numerical results concern the late-time behavior in the last column, and for example $(\overline{II})$ indicates the time reversal of $(II)$, which is the free Kasner behavior at early times with a finite future of Table~\ref{tab:solsTkp11}. Moreover, $LM$ $(\overline{LM})$ indicates the LM attractor with infinite (finite) future.}\vskip 12pt
 \label{tab:tab21}
 \end{table}

   \begin{table}[!ht]
 \begin{center}
  \scalebox{0.8}{
\begin{tabular}{ || c | c||c|c|| } 
 \hline\hline
 n & Case & Early $t$ & Late $t$ \\ [0.5ex] 
  \hline\hline
    $13$ & $\mathbf{e^{\gamma\phi} \gg e^{-2C}}$ &  &  \\ & \textbf{Early T>0 Dom.}  & K:  \eqref{Tnokpgammazeroc}\,, \ \eqref{sinetaless}&   $(dS)$ $(k'=1,-1)$  \\ & $\mathbf{\left(\gamma=0 \right)}$ \textbf{${(2b)}_1$} &  \textbf{[f.p.]}
 &  \\ [0.5ex]  \hline
    $14$ & \textbf{$e^{\gamma\phi} \gg e^{-2C}$} &  &   $(2B)$ $(k'=-1)$ \\ & \textbf{Early T>0 Dom.}  & LM:   \eqref{lm} &    \\ & $\mathbf{\left(\gamma_0<\gamma<\gamma_c \right)}$ $\mathbf{{(2a)}_2}$  &  \textbf{[f.p.]}
 & $(\overline{II})$ $(k'=1)$ \\ [0.5ex] 
 \hline
     $15$ & $\mathbf{e^{\gamma\phi} \gg e^{-2C}}$ &  &   \\ & \textbf{Early T>0 Dom.}  & $\overline{LM}$:   \eqref{lm} \ $(t \to - t)$ &   $(\overline{LM})$ \ $(k'=1,-1)$  \\ & $\mathbf{\left(0<\gamma<\gamma_0 \right)}$ $\mathbf{\overline{(2a)}}_2$  &  \textbf{[i.p.]}
 &  \\ [0.5ex]
  \hline
     $16$ & $\mathbf{e^{\gamma\phi} \gg e^{-2C}}$ &  &   $(LM)$ $(k'=-1)$ \\ & \textbf{Early T>0 Dom.}  & K:   \eqref{Tnokpgammaless} , \ \eqref{phi1vn}   & \\ & $\mathbf{\left(0<\gamma<\gamma_c \right)}$ $\mathbf{{(2b)}_2}$  &  \textbf{[f.p.]}
 & $(\overline{II})$ $(k'=1)$ \\ [0.5ex]
  \hline
       ${17}$ & $\mathbf{e^{\gamma\phi} \gg e^{-2C}}$ &  &   $(LM)$ $(k'=-1)$ \\ & \textbf{Early T>0 Dom.}  & $\overline{LM}$:  \eqref{Tnokpgammaless} , \ \eqref{phi1vn}  & \\ & $\mathbf{\left(0<\gamma<\gamma_0 \right)}$ $\mathbf{\overline{(2b)_2}}$  &  \textbf{[i.p.]}
 & $(\overline{II})$ $(k'=1)$ \\ [0.5ex]
  \hline
     $18$ & $\mathbf{e^{\gamma\phi} \gg e^{-2C}}$ &  &   \\ & \textbf{Early T<0 Dom.}  & LM:   \eqref{2cepsp}  & $(\overline{LM})$ $(k'=1,-1)$ \\ & $\mathbf{\left(\gamma>\gamma_c \right)}$ $\mathbf{{(2c)}}$  &  \textbf{[f.p.]}
 &   \\ [0.5ex]
   \hline
     $19$ & $\mathbf{e^{\gamma\phi} \gg e^{-2C}}$ &  &   $(\overline{LM})$ $(k'=-1)$ \\ & \textbf{Early T<0 Dom.}  & LM:   \eqref{neg_E_epsilonpos_lim}  &  \\ & $\mathbf{\left(\gamma>\gamma_c \right)}$ $\mathbf{{(2d,\epsilon', \forall \phi_1)}}$  &  \textbf{[f.p.]}
 & $(\overline{II})$  $(k'=1)$ \\ [0.5ex]
    \hline
     $20$ & $\mathbf{e^{\gamma\phi} \gg e^{-2C}}$ &  &   $(\overline{LM})$ $(k'=-1)$ \\ & \textbf{Early T<0 Dom.}  & K:   \eqref{neg_E_epsilonpos_lim} $(t \to - t)$ &  \\ & $\mathbf{\left(\gamma>\gamma_c \right)}$ $\mathbf{{(\overline{2d},\epsilon',\phi_1>0)}}$  &  \textbf{[f.p.]}
 & $(\overline{II})$  $(k'=1)$ \\ [0.5ex]
    \hline
     $21$ & $\mathbf{e^{\gamma\phi} \gg e^{-2C}}$ & K: \ fig.~\ref{fig:oneb} & \\ & \textbf{Early T<0 Dom.}  & \eqref{Tnokpgammaless}, \ \eqref{phi1v1zetatnegn}  &  $(\overline{LM})$ $(k'=1,-1)$\\ & $\mathbf{\left(\gamma>\gamma_c \right)}$ $\mathbf{{(2e,\epsilon'=-1)}}$  &  \textbf{[f.p.]}
 &  \\ [0.5ex]
     \hline
     $22$ & $\mathbf{e^{\gamma\phi} \gg e^{-2C}}$ & LM: &   $(\overline{II})$ \ $(k'=1)$ \\ & \textbf{Early T<0 Dom.}  & \eqref{Tnokpgammaless}, \ \eqref{phi1v1zetatnegn} $(t \to - t)$ & \\ & $\mathbf{\left(\gamma>\gamma_c \right)}$ $\mathbf{{(\overline{2e},\epsilon'=1)}}$  &  \textbf{[f.p.]}
 &   $(\overline{LM})$ \ $(k'=-1)$ \\ [0.5ex]
     \hline
     $23$ & $\mathbf{e^{\gamma\phi} \gg e^{-2C}}$ & LM: & $(\overline{II})$  \ $(k'=-1)$ \\ & \textbf{Early T<0 Dom.}  & \eqref{Tnokpgammaless}, \ \eqref{phi1v1zetatnegn} $(t \to - t)$ &  \\ & $\mathbf{\left(\gamma>\gamma_c \right)}$ $\mathbf{{(\overline{2e},\epsilon'=-1)}}$  &  \textbf{[f.p.]}
 &  $(\overline{II})$ \ $(k'=1)$ \\ [0.5ex]
 \hline\hline
\end{tabular}}
 \end{center}
 \vskip 12pt 
 \caption{\small A second portion of the results obtained combining analytical and numerical indications on systems subject to both $T$ and $k'$. The numerical results concern the late-time behavior in the last column, and for example $(\overline{II})$ indicates the time reversal of $(II)$, which is the free Kasner behavior at early times with a finite future of Table~\ref{tab:solsTkp11}. Moreover, $LM$ $(\overline{LM})$ indicates the LM attractor with infinite (finite) future.}\vskip 12pt
 \label{tab:tab22}
 \end{table}

This paper was devoted to extending the results of~\cite{ms21_susy,ms21_nonsusy} by allowing the presence of negative tension and curvatures $(k,k')$ in spatial and internal slices. In the resulting systems, reviewed in Section~\ref{sec:harmonic}, the cosmological evolution is driven by the three ingredients $(T,k,k')$, and thus by the associated exponential factors, with various types of dynamics reflecting the competition among their effects. A central outcome of this analysis is that a wide class of cosmological evolutions connects a small number of asymptotic regimes. Although exact solutions are available only in special cases, the combination of these results, scaling solutions, and numerical tests provides a coherent semi-quantitative picture of the generic dynamics. 

The resulting picture is summarized in Tables~\ref{tab:tab21} and \ref{tab:tab22} for the physically important class of systems with flat spatial slices, driven by tension and a non-trivial internal curvature, where exact solutions are generally unavailable. An important feature is that positive curvature and negative tension act as focusing agents that drive the system toward collapse. On the other hand, negative curvature and positive tension tend to prolong the evolution and can lead to Milne-like, Lucchin-Matarrese or de Sitter asymptotics. 

Among the eight families of exact solutions with negative tension summarized in Table~\ref{tab:solsTkp12}, only $(2d)$ with $\phi_1<0$ has one singularity, rather than two, while positive-tension solutions exhibit a vastly different pattern. Most examples driven by positive tension alone experience initial and final Kasner asymptotics, albeit with an infinite future, except for two cases with $\gamma<\gamma_c$ and ${\cal E} \geq 0$, which experience future LM asymptotics, and two cases with $\gamma=0$, which experience future $dS$ asymptotics. However, negative tension can also give rise to LM asymptotics, in addition to the more common final Kasner behavior. 

One of the recurring themes throughout the paper is the role of two special values for $\gamma$, the parameter characterizing the tadpole potential, already encountered in the analysis of the exact solutions. These values are
\be
\gamma_c \ = \ \frac{4 \sqrt{D-1}}{D-2} \ , \qquad \gamma_0 \ = \ \frac{4}{D-2} \ , \qquad 
\eeq
and separate qualitatively different regimes. $\gamma_c$ is related to the transition between climbing and descending behavior but, in the presence of curvature effects, $\gamma_0$ emerges as an equally important threshold, separating vastly different classes of cosmological evolutions.

The detailed arguments presented in Section~\ref{sec:t0curvnot0} show that, in the absence of tension, a positive curvature $k'$ (or $k$) links the early Kasner behaviors allowed to the well-defined final ones. On the other hand, a negative curvature links early Kasner behaviors to Milne-like future asymptotics. The adiabatic argument in Section~\ref{sec:adiabatic} lends further support to our general considerations. For systems driven by positive curvature $k'$, which evolve within a finite cosmic-time span, the inclusion of $k$ reduces the total duration $t_0$ if $k>0$ and increases it if $k<0$. 

A striking feature of all the cosmologies under scrutiny is indeed the ubiquity of asymptotic Kasner behavior, which emerges when all three ingredients $(T,k,k')$ play a subdominant role. Tension or curvatures typically connect two such Kasner behaviors to each other, or one limiting Kasner behavior to the scaling solutions of Section~\ref{sec:log_asympt} that are stable under perturbations, at least in one asymptotic regime. The different exact solutions thus provide explicit interpolations among these different regimes.

We have identified several exact solutions that emerge when only one of the three ingredients $(T,k,k')$ is present but continue to provide valuable information when one dominates the others. We also found additional types of special exact solutions that emerge when two or more ingredients play a comparable role. Most of them have instabilities, but when this is not the case (solutions $(2B)$, $(2E)$ and $(2F)$ in Table~\ref{tab:solsTkp13}), the common future asymptotics acts as an organizing structure for different regimes of the complete system. Indeed, although these special solutions arise for finely tuned relations among competing terms, our numerical tests indicate that stable ones act as organizing centers for broader classes of nearby cases, thereby extending their significance well beyond their exact domain of validity. 

The non-Kasner scaling solutions of Section~\ref{sec:log_asympt} provide a complementary class of asymptotic behaviors, which can act as genuine attractors of the cosmic-time dynamics and therefore play a distinguished role in organizing the late-time evolution. The emergence of a relatively small number of recurrent asymptotic regimes from systems driven by the three competing ingredients $(T,k,k')$ indicates that the underlying dynamics is considerably more structured than might have been anticipated and points towards a deeper organizing principle that remains to be uncovered. Kasner, Milne-like and Lucchin-Matarrese regimes provide the basic building blocks of the system, while tension and curvatures primarily determine how the cosmological evolution interpolates between them. Viewed broadly, the cosmological evolution is governed by the competition among three distinct driving mechanisms — tension, spatial curvature, and internal curvature — and most of the qualitative features that emerge can be traced to the changing hierarchy among them during the evolution. In this sense, the large variety of cosmologies discussed in this paper can be viewed as different interpolating trajectories among a comparatively small set of asymptotic building blocks.

Several issues deserve further investigation. To begin with, the stability analysis performed here is restricted to homogeneous perturbations and should be extended to the full spectrum of fluctuations, as in the cases discussed in~\cite{ivano} and\cite{dms_review}. An important open problem is indeed to extend the present dynamical-system analysis to the fully inhomogeneous case. This would make it possible to assess whether the asymptotic regimes identified here, and in particular the Milne-like and scaling attractors, survive beyond the homogeneous sector and thus play a genuine role in the low-energy dynamics of non-supersymmetric string vacua. Moreover, the numerical evidence we have gathered suggests the existence of a more compact dynamical classification than the one currently available. Finally, it would be interesting to clarify the links with the cosmological billiards inspired by supergravity and first highlighted in~\cite{damour}, or possibly with structures hinting at hidden integrability. More generally, the results obtained here suggest that non-supersymmetric string cosmologies driven by tension and curvature effects possess a surprisingly rich but well-structured dynamical organization, whose full mathematical significance remains to be understood.
Although we have concentrated throughout on cosmological models, the correspondence highlighted in Appendix~\ref{app:cosmo_comp} explains how to translate these results into the corresponding ones for interval compactifications.
 
\section*{\sc Acknowledgments}
\vskip 12pt
The work of A.S. was supported in part by INPS and INFN (IS GSS-Pi).

\newpage
\begin{appendices}
\section{\sc Summary of Exact Solutions}

In this appendix, we summarize in a few tables the main properties of the exact solutions discussed in the main body of the paper, with references to the equations where more details can be found, and complement the discussion in Section~\ref{sec:tnot0curv0}.

\subsection{\sc General Solutions with Only One Curvature}

 Table~\ref{tab:solsTkp11} summarizes the main features of the exact expanding cosmologies discussed in Sections~\ref{sec:t0curv0} and \ref{sec:t0curvnot0}.
\begin{table}[hbt!]
 \begin{center}
  \scalebox{0.8}{
\begin{tabular}{ ||c|c||c|c|| } 
 \hline\hline
  n. & Case & Early $t$ & Late $t$  \\ [0.5ex] 
  \hline\hline
   I &  $k'=0$ & F [i.p.] &  F  [i.f.] \\ 
 [0.5ex] 
  \hline
 II & $k'=0$ & K \eqref{param_theta}, [f.p.] &  K \eqref{param_theta},  [i.f.] \\ 
 [0.5ex] 
  \hline
  III & $k'>0$ & K \eqref{nudelta}, \eqref{acphik'pos}, [f.p.] &  K, \eqref{nudelta},\eqref{acphik'pos}, [f.f.]  \\
 [0.5ex] \hline
 IV & $k'<0,\ \rho=\infty$ & Milne \eqref{milne}, [f.p.] &  Milne \eqref{milne}, [i.f.]  \\ [0.5ex] 
  \hline
  V & $k'<0,\ \rho<\infty$ & K \eqref{acphik'pos}, [f.p.]  &  Milne \eqref{milne}, [i.f.]  \\ [0.5ex] 
 \hline\hline
\end{tabular}}
 \end{center}
 \vskip 12pt 
 \caption{\small Cosmologies driven only by the curvature $k'$.}\vskip 12pt
 \label{tab:solsTkp11}
 \end{table}

We restrict the attention to solutions with various choices of $k'$ and $k=0$: all other cases with various choices of $k$ and $k'=0$ can be obtained by interchanging $A$ and $C$ and $p$ and $D-p-3$. For brevity, we write K for Kasner, F for flat, f.p. for finite past, i.p for infinite past, f.f. for finite future and i.f for infinite future, and simply quote the equations where the results can be found.
 
 \newpage
\subsection{\sc General Solutions with Only Tension} \label{app:tension_only}

In this appendix, we elaborate on the main properties of the cosmologies driven by tension alone, which were the subject of Section~\ref{sec:tnot0curv0}, so that the interested reader can find more details on the classification. These results are referred to with the same labels as used in Section~\ref{sec:tnot0curv0}, complement those contained in~\cite{ms21_nonsusy}, and are summarized in Table~\ref{tab:solsTkp12}.
 \begin{table}[hbt!]
 \begin{center}
 \scalebox{0.8}{
\begin{tabular}{ ||c|c||c|c|c|c|| } 
 \hline\hline
  n. & Case & Sol. & Early $t$ & Late $t$ & $e^\phi$ \\ [0.5ex] 
  \hline\hline
  $(0a)$ & $\gamma=\gamma_c,T<0,\beta=0$ & \eqref{sol_gamma_c_beta0}  & iso K \ [f.p.] &  iso K \ [f.f.] & \eqref{asymptbeta0}  \\ [0.5ex]  \hline
  $(0b,T>0)$ & $\gamma=\gamma_c,\beta \neq 0$ & \eqref{gammac_init} & K \ [f.p.] &  iso K, [i.f.] &  \eqref{gammac_init2} \\ [0.5ex]  \hline $(0b,T<0)$ & $\gamma=\gamma_c,\beta \neq 0$ & \eqref{gammac_init} & K \ [f.p.]  & iso K [f.f.] & \eqref{gammac_init2}  \\ [0.5ex]  \hline
 $(1a)$ & $\gamma<\gamma_c,{\cal E}> 0, T<0$ & \eqref{solutiongammamore1}, \eqref{phi1vn} &  K\ [f.p.]  & K \ [f.f.]  &  \eqref{betaeta} \\ [0.5ex]  \hline
    $(1b,\epsilon'=-1)$ & $\gamma>\gamma_c,{\cal E}> 0, T>0$ & \eqref{solutiongammamore1}, \eqref{phi1v1zetatnegn} &  K\ [f.p.]  & K \ [i.f.]  &  \eqref{betaplusminus1b} \\ [0.5ex] \hline
        $(2a)_1$ &  $\gamma=0,{\cal E}= 0, T>0$ & \eqref{desitter} &  dS\ [i.p.]  & dS \ [i.f.]  &  \eqref{desitter} \\ [0.5ex] \hline
                $(2b)_1$ &  $\gamma=0,{\cal E}> 0, T>0$ & \eqref{Tnokpgammazeroc} &  K \ [f.p.]  & dS \ [i.f.]  &  \eqref{desitter} \\ [0.5ex] \hline
                $(2a)_2$ & $0<\gamma<\gamma_c,{\cal E}= 0, T>0$ & \eqref{lm} &  LM\ [f.p.]  & LM \ [i.f.]  &  \eqref{lm} \\ [0.5ex] \hline
                $(2b)_2$ &  $0<\gamma<\gamma_c,{\cal E}> 0, T>0$ & \eqref{Tnokpgammaless}, \eqref{phi1vn} &  K \ [f.p.]  & LM \ [i.f.]  &  \eqref{lm} \\ [0.5ex]  \hline
           $(2c,\epsilon'=1)$ & $\gamma>\gamma_c,{\cal E}< 0, T<0$ & \eqref{neg_E_epsilonpos},\eqref{phi1v1zetatnegn2} &  LM \ [f.p.]  & LM \ [f.f.]  &  \eqref{2cepsp} \\ [0.5ex]  \hline  
            $(2d,\epsilon'=\pm 1)$ & $\gamma>\gamma_c,{\cal E} = 0, T<0$ & \eqref{neg_E_epsilonpos_lim},\eqref{v1propphi1} &  LM\ [f.p.]  & K \ [f.f. $(\phi_1>0)$]  &  \eqref{betasmorethan} \\ [0.5ex]  \hline 
             $(2d,\epsilon'=\pm 1)$ &  $\gamma>\gamma_c,{\cal E} = 0, T<0$ & \eqref{neg_E_epsilonpos_lim},\eqref{v1propphi1} &  LM\ [f.p.]  & K \ [i.f. $(\phi_1\leq 0)$]  &  \eqref{betasmorethan} \\ [0.5ex]  \hline 
             $(2e,\epsilon'=1)$ &  $\gamma>\gamma_c,{\cal E} > 0, T<0$ & \eqref{Tnokpgammaless},\eqref{phi1v1zetatnegn} & K\ [i.p.]  & LM \ [f.f.]  &  \eqref{lm} \\ [0.5ex]  \hline  $(2e,\epsilon'=-1)$ & $\gamma>\gamma_c,{\cal E} > 0, T<0$ & \eqref{Tnokpgammaless},\eqref{phi1v1zetatnegn} &  K\ [f.p.]  & LM \ [f.f.]  &  \eqref{lm} \\ [0.5ex]  

 \hline\hline
\end{tabular}}
 \end{center} \caption{\small Cosmologies driven only by the tension $T$. For brevity, for each pair of solutions mapped into one another by time reversal, we included only one of them. We only display solutions that are initially expanding: their time reversals would be finally contracting. Moreover, the nomenclature [f.p], [i.p], [f.f], [i.f.] refers to the singularities of the dynamical system: we do not tell them apart from actual singularities, which explains the label [f.p.] in one  occurrence of dS.}\vskip 12pt
 \label{tab:solsTkp12}
 \end{table}

\begin{itemize}
\item[(1a)] In this case $\gamma<\gamma_c$, $T<0$ and ${\cal E}>0$, and the harmonic-gauge solution reads
\bea
ds^2 &=& \ - \ \frac{e^{- \, \gamma\left(\phi_1\,\tau+\phi_0\right)}\, d\tau^2}{\left[ \Delta\,\rho\, \cosh\left(\frac{\tau}{\rho} \right)\right]^{(D-1)\omega}} \ + \ e^{\frac{-\,\gamma\,\phi_1\,\tau}{D-1}}\,\frac{\left[e^{\frac{2(D-p-2)\tau}{D-1}}\, d \vec{x}^2 \ + \ e^{-\,\frac{2(p+1)\tau}{D-1}}\,d\vec{y}^2\right]}{\left[ \Delta\,\rho\, \cosh\left(\frac{\tau}{\rho} \right)\right]^\omega} \ , \nonumber \\
e^\phi &=& e^{\phi_1\,\tau\,+\,\phi_0} \, \left[\Delta\,\rho\, \cosh\left(\frac{\tau}{\rho} \right) \right]^\frac{(D-2)^2\gamma\,\omega}{16} \ , \label{solutiongammamore1}
\eea
with the parametrization of eqs.~\eqref{phi1vn}, where
\beq
\omega \ = \ \frac{2}{D-1} \ \frac{1}{1 \ - \ \left(\frac{\gamma}{\gamma_c}\right)^2}\label{omega_app} \ .
\eeq 
In the interesting case $\gamma=0$, where the potential reduces to a negative cosmological constant, these solutions become
\bea
ds^2 &=& \ - \ \frac{d\tau^2}{\left[ \Delta\,\rho\, \cosh\left(\frac{\tau}{\rho} \right)\right]^{2}} \ + \ \frac{e^{\frac{2(D-p-2)v_1\,\tau}{(D-1)}}\, d \vec{x}^2 \ + \ e^{-\,\frac{2(p+1)v_1\,\tau}{(D-1)}}\,d\vec{y}^2}{\left[ \Delta\,\rho\, \cosh\left(\frac{\tau}{\rho} \right)\right]^\frac{2}{(D-1)}} \ , \nonumber \\
e^\phi &=& e^{\phi_1\,\tau\,+\,\phi_0} \ , \label{Tnegokpgammazero}
\eea
with
\beq
\Delta \, = \, \sqrt{\frac{\left|T\right|(D-1)}{(D-2)}} \ , \quad \rho\,\phi_1 \,=\, \frac{2}{\gamma_c}\,\cos\eta\ , \quad \rho\,v_1 \, = \, \sqrt{\frac{(D-2)}{(p+1)(D-p-2)}}\, \sin \eta \, ,
 \label{solutiongammazetoTneg}
 \eeq
and can be exactly formulated in cosmic time. One thus finds
\bea
ds^2 \!\!&=&\!\! - \,dt^2  +  \left(\frac{\sin\Delta t}{2}\right)^\frac{2}{(D-1)} \left\{ \left[\tan\left(\frac{\Delta t}{2}\right) \right]^{\frac{2(D-p-2)\rho \,v_1}{D-1}}\!\!\!\! d\vec{x}^2 +  \left[\tan\left(\frac{\Delta t}{2}\right) \right]^{-\,\frac{2(p+1) \rho\, v_1}{D-1}} \!\!\!\! d\vec{y}^2\right\}  , \nonumber \\
e^\phi &=& e^{\phi_0} \left[\tan\left(\frac{\Delta t}{2}\right) \right]^{\frac{2\,\cos\eta}{\gamma_c}}  ,
\eea
where  $0<t<\frac{\pi}{\Delta}$. Note that this result is independent of $\rho$, a property that only holds for $\gamma=0$. The limiting behavior at early times is captured by
\bea
ds^2 &\sim& - \,dt^2 \, + \,  \left[ \left(\frac{\Delta t}{2}\right)^{\frac{2\left[1\,+\,(D-p-2) \rho\,v_1\right]}{D-1}}\!\!\! d\vec{x}^2\, + \, \left(\frac{\Delta t}{2}\right)^{\frac{2\left[1\,-\,(p+1) \rho\,v_1\right]}{D-1}} \!\!\! d\vec{y}^2\right] \, , \nonumber \\
e^\phi &\sim& \left(\frac{\Delta t}{2}\right)^{\frac{2\,\cos\eta}{\gamma_c}} \, , \label{limiting_gamma0_early}
\eea
while at late times
\bea
ds^2 &\sim& - \,dt^2 \, + \,  \left[ \left(\frac{\pi\,-\,\Delta t}{2}\right)^{\frac{2\left[1\,-\,(D-p-2) \rho\,v_1\right]}{D-1}}\!\!\! d\vec{x}^2\, + \, \left(\frac{\pi\,-\,\Delta t}{2}\right)^{\frac{2\left[1\,+\,(p+1) \rho\,v_1\right]}{D-1}} \!\!\! d\vec{y}^2\right] \, , \nonumber \\
e^\phi &\sim& \left(\frac{\pi\,-\,\Delta t}{2}\right)^{-\, \frac{2\,\cos\eta}{\gamma_c}} \, . \label{limiting_gamma0_late}
\eea
Both limits are once more of the free Kasner form discussed in Section~\ref{sec:t0curv0}.

For general values of $\gamma<\gamma_c$, one can deduce from eqs.~\eqref{solutiongammamore1} the asymptotic links between $\tau$ and the cosmic time $t$,
\beq
dt \ = \ e^{-\,\frac{\pm 1 \ + \ \frac{\gamma}{\gamma_c}\,\cos\eta}{1 \ - \ \left(\frac{\gamma}{\gamma_c}\right)^2 }\,\tau}\ d\tau
\eeq
which imply that all these cosmologies have a \emph{finite} past and a \emph{finite} future, with 
$\eta$-dependent singularities $t_0^\pm(\eta)$. The corresponding Kasner exponents, which are determined by
\bea
\beta_A^\pm &=& \ - \ \frac{2}{(D-1)}\,\frac{\left[ 1-\left(\frac{\gamma}{\gamma_c}\right)^2\right]\left(D-p-2\right)\rho\,v_1 \ \mp \ 1}{\left[ 1-\left(\frac{\gamma}{\gamma_c}\right)^2\right]\gamma\,\rho\,\phi_1 \ \pm \ 2} \ , \nonumber \\
{\beta}_C^\pm &=& \frac{1 \ - \ (p+1)\,{\beta}_A^\pm}{(D-p-2)}\ ,  \qquad 
\beta_\phi^\pm \ = \  - \ 2 \ \frac{\left[  1-\left(\frac{\gamma}{\gamma_c}\right)^2 \right] \rho\,\phi_1 \ \pm \ \frac{2\,\gamma}{\gamma_c^2}}{\left[  1-\left(\frac{\gamma}{\gamma_c}\right)^2 \right] \gamma\,\rho\,\phi_1 \ \pm \ {2}} 
\ ,  \label{alphasgammamore}
\eea
read
\bea
{\beta}_A^\pm(\eta) &=&\frac{1}{D-1}\left[1 \ - \ \frac{\sin\eta \,\sqrt{\frac{(D-2)(D-p-2)\left[1 \ - \ \left(\frac{\gamma}{\gamma_c}\right)^2\right]}{(p+1)}}}{\pm 1 \ + \ \frac{\gamma}{\gamma_c}\,\cos\eta}\right] \ , \nonumber \\
{\beta}_\phi^\pm(\eta) &=& - \ \frac{2}{\gamma_c}\, \frac{\cos\eta \ \pm \ \frac{\gamma}{\gamma_c}}{\frac{\gamma}{\gamma_c}\,\cos\eta \ \pm \ 1} \ , \label{betaeta}
\eea
with ${\beta}_C^\pm(\eta)$ related to ${\beta}_A^\pm(\eta)$ as in eqs.~\eqref{alphasgammamore}.
For $\gamma=0$, these results reduce to the exponents displayed in eqs.~\eqref{limiting_gamma0_early} and \eqref{limiting_gamma0_late}.

In all these cosmologies ${\beta}_A^\pm(\eta)$ and ${\beta}_C^\pm(\eta)$ have typically opposite signs. Moreover, ${\beta}_\phi^\pm(\eta)$ can have both signs, so strong-coupling and weak-coupling behaviors are possible at early and late times. The special value $\gamma_0$ for super-critical strings lies within this range.

For finite values of $\rho$, perturbations can induce small variations of $\eta$, and thus of $\phi_1$ and $v_1$ and also $t_0^\pm$. The relative effects on $t_0^\pm$ involve the ratio $\frac{\delta\,t_0^\pm}{t_0^\pm - t}$, signaling instabilities of these solutions near the ends, where however the effective theory is no longer valid. For this reason, in the following we ignore this type of problem. There are in principle other sources of instability, whenever small variations of the parameters have sizable effects. All these solutions are stable if the relative variations of the ${\beta}^\pm$ induced by small variations of $\eta$ are also small. The denominators in eqs.~\eqref{betaeta} never vanish, so instabilities can only be present when one of the ${\beta}_\phi^\pm$ vanishes.

\item[(1b, $\epsilon'$) ] In this case $\gamma>\gamma_c$, $T> 0$ and ${\cal E}>0$,
the background is still given in eqs.~\eqref{solutiongammamore1}, but the parametrization is now as in eqs.~\eqref{phi1v1zetatnegn}. The Kasner exponents are still determined by eqs.~\eqref{alphasgammamore}, whose denominator is now
\beq
{2\left[\mp \ 1 \ + \ \frac{\gamma}{\gamma_c}\,\epsilon'\, \cosh\zeta \right]} \ ,
\eeq
and has the sign of $\epsilon'$. Consequently, the link between $\tau$ and the cosmic time $t$,
\beq
dt \ = \ e^{-\,\frac{\mp 1 \ + \ \frac{\gamma}{\gamma_c}\,\epsilon'\,\cosh\zeta}{\left(\frac{\gamma}{\gamma_c}\right)^2 \ - \ 1}\,\frac{\tau}{\rho}}\ d\tau
\eeq
implies that these cosmologies have an infinite past and a finite future for $\epsilon'=1$, and a finite past and an infinite future for $\epsilon'=-1$, which is physically the more interesting option. Moreover, the Kasner exponents are in this case
\bea
\widehat{\beta}_A^\pm(\zeta) &=&\frac{1}{D-1}\left[1 \ - \ \frac{\sinh\zeta \,\sqrt{\frac{(D-2)(D-p-2)\left[\left(\frac{\gamma}{\gamma_c}\right)^2 \, - \, 1\right]}{(p+1)}}}{\mp 1 \ + \ \frac{\gamma}{\gamma_c}\,\epsilon'\,\cosh\zeta}\right] \ , \nonumber \\
\widehat{\beta}_C^\pm(\zeta) &=& \frac{1 \ - \ (p+1)\,\widehat{\beta}_A^\pm(\zeta)}{(D-p-2)}\ , \qquad
\widehat{\beta}_\phi^\pm(\zeta) \ = \ - \ \frac{2}{\gamma_c}\, \frac{\epsilon'\,\cosh\zeta \ \mp \ \frac{\gamma}{\gamma_c}}{\epsilon'\, \frac{\gamma}{\gamma_c}\,\cosh\zeta \ \mp \ 1} \ . \label{betaplusminus1b}
\eea
Note that $\widehat{\beta}_A^\pm(0)$ and $\widehat{\beta}_C^\pm(0)$ are both positive, independently of $\epsilon'$, while $\widehat{\beta}_\phi^\pm(0)$ has the sign of $\epsilon'$. For nonzero values of $\zeta$, $\widehat{\beta}_A^\pm(\zeta)$ and $\widehat{\beta}_C^\pm(\zeta)$ have typically opposite signs, with $\widehat{\beta}_A^\pm(\zeta)$ typically negative for $\zeta>0$. Consequently, when spacetime expands the internal space typically contracts, and vice versa, both in the past and in the future. Moreover, $\widehat{\beta}_\phi^\pm(\zeta)$ is positive for large enough values of $\zeta$ and can have both signs otherwise.

As we have seen, $\epsilon'$ determines the past and future durations of these backgrounds, and consequently:
\begin{itemize}
\item for $(\epsilon'=+1)$, with an infinite past and a finite future, $\widehat{\beta}_{A,C}^-(\zeta)>0$ indicate initial contractions and $\widehat{\beta}_{A,C}^-(\zeta)<0$ initial expansions. On the other hand, $\widehat{\beta}_{A,C}^+(\zeta)>0$ indicate final contractions and $\widehat{\beta}_{A,C}^+(\zeta)<0$ indicate final expansions. In a similar fashion, $\widehat{\beta}_\phi^-(\zeta)>0$ ($<0)$ indicates strong (weak) coupling at early times, while $\widehat{\beta}_\phi^+(\zeta)>0$ ($<0)$ indicates weak (strong) coupling at late times.

\item for $(\epsilon'=-1)$, with a finite past and an infinite future, $\widehat{\beta}_{A,C}^-(\zeta)>0$ indicate initial expansions and $\widehat{\beta}_{A,C}^-(\zeta)<0$ indicate contractions. On the other hand, $\widehat{\beta}_{A,C}^+(\zeta)>0$ indicate final expansions and $\widehat{\beta}_{A,C}^+(\zeta)<0$ indicate final contractions. In a similar fashion, $\widehat{\beta}_\phi^-(\zeta)>0$ ($<0)$ indicates weak (strong) coupling at early times, while $\widehat{\beta}_\phi^+(\zeta)>0$ ($<0)$ indicates strong (weak) coupling at late times.
\end{itemize}

\item[(2a) ] In this case $\gamma<\gamma_c$, $T> 0$ and ${\cal E}=0$, and the background reads
\bea
ds^2 &=& \ - \ \frac{e^{- \, \gamma\phi_0}\, d\tau^2}{\left( \Delta\,|\tau|\right)^{(D-1)\omega}} \ + \ \frac{d \vec{x}^2 \ + \ \,d\vec{y}^2}{\left( \Delta\,|\tau|\right)^\omega} \ , \nonumber \\
e^\phi &=& e^{\phi_0} \, \left( \Delta\,|\tau|\right)^\frac{(D-2)^2\gamma\,\omega}{16}  \ ,\label{gammalessTpE0}
\eea
where $-\infty<\tau<0$ and $\Delta$ is given in eq.~\eqref{solutiongammazetoTneg}. These results affords simple presentations in cosmic time, which depend on whether or not $\gamma$ vanishes.
If $\gamma=0$, the solution becomes a de Sitter background,
\bea
ds^2 &=& \!\! - \, dt^2 \, + \, e^{\frac{2\Delta \,t}{(D-1)}} \ \Big( d \vec{x}^2 \  +\   d\vec{y}^2\Big) \, . \nonumber \\
e^\phi &= &  e^{\phi_0} \ ,
\label{desitter}
\eea
while if $\gamma \neq 0$ the solution becomes
\bea
ds^2 &=& - \ dt^2 \ + \ \left(d\vec{x}^2 \ + \ d\vec{y}^2\right) \left(\Delta\,t\right)^\frac{32}{(D-2)^2\,\gamma^2} \ , \nonumber \\
e^\phi &=& e^{\phi_0} \, \left(\Delta\,t\right)^{-\,\frac{2}{\gamma}} \ , \label{lm}
\eea
which is known as the LM attractor~\cite{lm}, so that there is a discontinuous behavior at $\gamma=0$. Note that the conformal transformation 
\beq
ds^2 \ \rightarrow \ ds^2 \ e^{\frac{16\,\phi}{(D-2)^2\gamma}}
\eeq
gives rise to a flat metric. For the special value $\gamma=\gamma_0$ of eq.~\eqref{gamma_0} this is precisely the Weyl rescaling leading to the string frame, and one recovers the ``linear-dilaton'' solution of~\cite{lindil1,lindil2} for sub-critical strings, for which $T>0$.

\item[(2b) ] In this case $\gamma<\gamma_c$, $T> 0$ and ${\cal E}>0$, and the background reads 
\bea
ds^2 &=& \ - \ \frac{e^{- \, \gamma\left(\phi_1\,\tau+\phi_0\right)}\, d\tau^2}{\left[ \Delta\,\rho\, \sinh\left|\frac{\tau}{\rho} \right|\right]^{(D-1)\omega}} \ + \ e^{\frac{-\,\gamma\,\phi_1\,\tau}{D-1}}\,\frac{\left[e^{\frac{2(D-p-2)\tau}{D-1}}\, d \vec{x}^2 \ + \ e^{-\,\frac{2(p+1)\tau}{D-1}}\,d\vec{y}^2\right]}{\left[ \Delta\,\rho\, \sinh\left|\frac{\tau}{\rho} \right|\right]^\omega} \ , \nonumber \\
e^\phi &=& e^{\phi_1\,\tau\,+\,\phi_0} \, \left[\Delta\,\rho\, \sinh\left|\frac{\tau}{\rho} \right| \right]^\frac{(D-2)^2\gamma\,\omega}{16} \ , \label{Tnokpgammaless}
\eea
where $-\infty<\tau<0$ and $\phi_1$ and $v_1$ are parametrized as in eqs.~\eqref{phi1vn}.

The solution simplifies for $\gamma=0$ and becomes
\bea
ds^2 &=& \ - \ \frac{d\tau^2}{\left[ \Delta\,\rho\, \sinh\left|\frac{\tau}{\rho} \right|\right]^{2}} \ + \ \frac{e^{\frac{2(D-p-2)v_1\,\tau}{(D-1)}}\, d \vec{x}^2 \ + \ e^{-\,\frac{2(p+1)v_1\,\tau}{(D-1)}}\,d\vec{y}^2}{\left[ \Delta\,\rho\, \sinh\left|\frac{\tau}{\rho} \right|\right]^\frac{2}{(D-1)}} \ , \nonumber \\
e^\phi &=& e^{\phi_1\,\tau\,+\,\phi_0} \ , \label{Tnokpgammazero}
\eea
where now
\beq
\Delta \ = \ \sqrt{\frac{T(D-1)}{(D-2)}} \ , \qquad \rho\,v_1 \ = \ \sqrt{\frac{(D-2)}{(p+1)(D-p-2)}}\, \sin \eta \ . \label{deltarhov1}
\eeq
In this case, there is also a simple analytic expression in cosmic time $(0<t<\infty)$:
\bea
ds^2 \!\!\!&=& \!\!\! - \, dt^2  + \left[\frac{\sinh(\Delta \,t)}{2} \right]^\frac{2}{(D-1)} \! \left\{ \left[ \coth\left(\frac{\Delta \,t}{2}\right)\right]^{\frac{2(D-p-2)\rho v_1}{(D-1)}} \!\!\!\!\! d \vec{x}^2  + \left[ \coth\left(\frac{\Delta \,t}{2}\right)\right]^{-\,\frac{2(p+1)\rho v_1}{(D-1)}} \!\! \!\!\! d\vec{y}^2\right\} , \nonumber \\
e^\phi &=& e^{\phi_0}\, \left[\coth\left(\frac{\Delta \,t}{2}\right) \right]^\frac{2\,\cos\eta}{\gamma_c} \ . \label{Tnokpgammazeroc}
\eea

Note that, on account of eq.~\eqref{deltarhov1}, this one-parameter family of solutions is independent of $\rho$. Indeed, as can be seen from eqs.~\eqref{Tnokpgammaless}, only for $\gamma=0$ can the $\rho$ dependence be absorbed by coordinate changes. The early-time behavior is free Kasner-like, with
\bea
ds^2 \!\!\!&\sim& \!\!\! - \, dt^2 \, + \, \left(\Delta \,t\right)^\frac{2}{(D-1)} \! \left[ \left(\frac{\Delta \,t}{2}\right)^{-\,\frac{2(D-p-2)\rho v_1}{(D-1)}} \!\!\!\!\!\!\! d \vec{x}^2  + \left(\frac{\Delta \,t}{2}\right)^{\frac{2(p+1)\rho v_1}{(D-1)}} \!\!\!\!\!\!\! d\vec{y}^2\right] \, , \nonumber \\
e^\phi &\sim& \left(\frac{\Delta \,t}{2}\right)^{-\,\frac{2\,\cos\eta}{\gamma_c}} \ , \label{Tnokpgammazeroc_early}
\eea
while at late times these solutions approach the de Sitter behavior of eqs.~\eqref{desitter}.

For $\gamma \neq 0$, one can still characterize rather simply the limiting behaviors at early and late times. In all cases, the initial singularity lies finitely far in the past, since the combination that enters the link between $\tau$ and the cosmic time $t$ as $\tau \to - \infty$,
\beq
- \,\gamma\,\phi_1 \ + \ (D-1)\,\frac{\omega}{\rho} \ = \ \frac{2}{1 \ - \ \left(\frac{\gamma}{\gamma_c}\right)^2} \left(1 \ - \ \frac{\gamma}{\gamma_c} \, \cos\eta \right) \label{den_betas}
\eeq
is positive if $\gamma<\gamma_c$. Close to the initial singularity, one finds
\bea
ds^2 &\sim& - \ dt^2 \ + \ \left(\Delta t\right)^{2\,\beta_A^-}\, d\vec{x}^2 \ + \ \left(\Delta t\right)^{2\,\beta_C^-}\, d\vec{y}^2 \ , \nonumber \\
e^\phi &\sim& \left(\Delta t\right)^{\beta_\phi^-} \ , \label{2blatepast}
\eea
where $t$ approaches zero, $\Delta$ is now given in eq.~\eqref{Deltagamma} since $\gamma \neq 0$, and the three ``minus'' exponents are given in eqs.~\eqref{betaeta}.

On the other hand, as $\tau \to 0^-$, which lies in the infinite future, these solutions recover the late-time behavior of case (2a).

\item[(2c, $\epsilon'$) ] In this case $\gamma>\gamma_c$, $T< 0$ and ${\cal E}<0$, so that $W$ is given in eq.~\eqref{Wsin} and the parametrization of $\phi_1$ and $v_1$ is as in eqs.~\eqref{phi1v1zetatnegn2}. 
Consequently
\bea
ds^2 &=& \ - \ \frac{e^{- \, \gamma\left(\phi_1\,\tau+\phi_0\right)}\, d\tau^2}{\left[ \Delta\,\rho\, \sin\left(\frac{\tau}{\rho} \right)\right]^{-\,(D-1)|\omega|}} \ + \ e^{\frac{-\,\gamma\,\phi_1\,\tau}{D-1}}\,\frac{\left[e^{\frac{2(D-p-2)\tau}{D-1}}\, d \vec{x}^2 \ + \ e^{-\,\frac{2(p+1)\tau}{D-1}}\,d\vec{y}^2\right]}{\left[ \Delta\,\rho\, \sin\left(\frac{\tau}{\rho} \right)\right]^{-\, |\omega|}} \ , \nonumber \\
e^\phi &=& e^{\phi_1\,\tau\,+\,\phi_0} \, \left[\Delta\,\rho\, \sin\left(\frac{\tau}{\rho} \right)  \right]^{-\, \frac{(D-2)^2\gamma\,|\omega|}{16}} \ , \label{neg_E_epsilonpos}
\eea
with $0<\tau<\pi\,\rho$.

Making use of eq.~\eqref{omega}, one can see that the total cosmic time span,
\beq
t_0 \ = \ \int_{0}^{\pi\rho} d\tau \ e^{- \, \frac{\gamma}{2}\left(\phi_1\,\tau\,+\,\phi_0\right)}\ \left[ \Delta\,\rho\, \sin\left(\frac{\tau}{\rho} \right) \right]^{\frac{1}{2\,(D-1)|\omega|}} \ ,
\eeq
is finite. Close to the initial and final singularities, these solutions approach
\bea
ds^2 &=& \ - \ du^2  \ + \  u^\frac{32}{\left[\gamma(D-2)\right]^2}  \left(d\vec{x}^2 \ + \ d\vec{y}^2 \right) \ , \nonumber \\
e^\phi &=& e^{\phi^\pm} \left[\frac{\gamma^2\,u}{\gamma^2\,-\,\gamma_c^2} \right]^{-\,\frac{2}{\gamma}} \ , \label{2cepsp}
\eea
where $u$ is identified with the cosmic time $t$ at early times and with $t_0-t$ at late times and $\phi^\pm$ are constants. As $\rho \to \infty$, one recovers case (2a). Note that the solution with $\epsilon'=-1$ can be obtained via time reversal from the solution with $\epsilon'=1$.

\item[(2d, $\epsilon'$) ] In this case $\gamma>\gamma_c$, $T< 0$ and ${\cal E}=0$, and the parametrization is given in eqs.~\eqref{v1propphi1}
Consequently
\bea
ds^2 &=& \ - \ \frac{e^{- \, \gamma\left(\phi_1\,\tau+\phi_0\right)}\, d\tau^2}{\left( \Delta\,{\tau} \right)^{-\,(D-1)|\omega|}} \ + \ e^{\frac{-\,\gamma\,\phi_1\,\tau}{D-1}}\,\frac{\left[e^{\frac{2(D-p-2)\tau}{D-1}}\, d \vec{x}^2 \ + \ e^{-\,\frac{2(p+1)\tau}{D-1}}\,d\vec{y}^2\right]}{\left(\Delta\,{\tau} \right)^{-\, |\omega|}} \ , \nonumber \\
e^\phi &=& e^{\phi_1\,\tau\,+\,\phi_0} \, \left(\Delta\,{\tau} \right)^{-\, \frac{(D-2)^2\gamma\,|\omega|}{16}} \ , \label{neg_E_epsilonpos_lim}
\eea
with $0<\tau<\infty$ and $\Delta$ given in eq.~\eqref{Deltagamma}. These solutions have a finite past, and a finite future for $\phi_1>0$, but an infinite future for $\phi_1 \leq 0$. In particular, for $\phi_1=v_1=0$ they recover the LM background of eqs.~\eqref{lm}, in an unusual realization with $\gamma>\gamma_c$ and negative tension, which also captures, in all cases, their behavior near the initial singularity. On the other hand, as $\tau \to \infty$, which corresponds to a finite value of $t$ for $\phi_1>0$ and to an infinite value of $t$ for $\phi_1 \leq 0$, these cosmologies approach a Kasner-like behavior with
\bea
\beta_A^+ &=& \frac{1}{D-1}\left[1 \ - \ \epsilon'\,\sqrt{\frac{(D-2)(D-p-2)}{(p+1)}\left[1\ - \ \left(\frac{\gamma_c}{\gamma} \right)^2 \right]} \right]  \ , \nonumber \\
\beta_C^+ &=& \frac{1 \ - \ (p+1)\beta_A^+}{(D-p-2)}\ , \qquad 
\beta_\phi^+ \ = \  - \ \frac{2}{\gamma} \ . \label{betasmorethan}
\eea

\item[(2e, $\epsilon'$) ] In this case $\gamma>\gamma_c$, $T< 0$ and ${\cal E}>0$, and the background is as in eqs.~\eqref{Tnokpgammaless}, but with the parametrization of $\phi_1$ and $v_1$ as in eqs.~\eqref{phi1v1zetatnegn}.

The link between $\tau$ and $t$ at early cosmic times,
\beq
dt \ = \ e^{-\,\frac{ 1 \ + \ \frac{\gamma}{\gamma_c}\,\epsilon'\,\cosh\zeta}{\left(\frac{\gamma}{\gamma_c}\right)^2 \ - \ 1}\,\tau}\ d\tau
\eeq
implies that these cosmologies have an infinite past for $\epsilon'=1$, and a finite past for $\epsilon'=-1$, with a Kasner behavior in both cases. Moreover, there is always a future singularity at a finite cosmic time, where the system approaches the LM behavior of eq.~\eqref{lm}. 
\end{itemize}

  \newpage
\subsection{\sc Special Solutions}

The special exact results obtained for cosmologies driven by two curvatures or by tension and one curvature are summarized below. The entries in the table indicate whether the solution has finite past (f.p.) or infinite past (i.p.), and similarly for the future. In addition, it provides links to the equations and indicates whether the solutions are unstable ($--$), or unsatble in the past and stable in the future ($\mathrm{stab}_f$), and so on. For the rest, the notation is as in the preceding tables.
 \begin{table}[hbt!]
 \begin{center}
 \scalebox{0.78}{
\begin{tabular}{ |c|c||c|c|c|c|c|| } 
 \hline\hline
  n. & Case & Sol. & Early $t$ & Late $t$ & $e^\phi$ & Stab. \\ [0.5ex] 
  \hline\hline
   $(1A)$ & $C-A=c$ & & & & & \\ [0.5ex] & $(k,k',T,\rho)=(1,1,0,<\infty)$ & \eqref{specialkkppos}  & K [f.p.] &  K [f.f.] & \eqref{specialkkpposcosmic} &  -- \\ [0.5ex]  \hline
   $(1B)$ &  $C-A=c$ & & & & & \\ [0.5ex] 
   & $(k,k',T,\rho)=(-1,-1,0,<\infty)$ & \eqref{specialkkpneg}  & K [f.p.] &  Milne-like [i.f.] & \eqref{specialkkpnege},\eqref{milnelike} &  $\mathrm{stab}_f$ \\ [0.5ex] \hline
   $(1C)$ &  $C-A=c$ & & & & & \\ [0.5ex] 
   & $(k,k',T,\rho)=(-1,-1,0,\infty)$ & \eqref{milnelike}  & Milne-like [f.p.] &  Milne-like [i.f.] & \eqref{milnelike},\eqref{milnelike} & $\mathrm{stab}_{f}$ \\ [0.5ex] \hline
   $(2A)$ &  $\gamma\,\phi+ 2C=2c$ & & & & & \\ [0.5ex] 
   & $(k,k',\gamma,T,\rho)=(0,1,0,>0,\infty) $ & \eqref{gamma_0_t2}  & dS [i.p.] &  dS [i.f.] & \eqref{gamma_0_t2},\eqref{gamma_0_t2} &  -- \\ [0.5ex]
   \hline 
   $(2B)$ &  $\gamma\,\phi+ 2C=2c$ & & & & & \\ [0.5ex] 
   &  $(k,k',\gamma,T,\rho)=(0,1,<\gamma_0,>0,\infty)$  & \eqref{exactkpminus}  & LM+lin [f.p.] &  LM+lin [i.f.] & \eqref{exactkpminus},\eqref{exactkpminus} & -- \\ [0.5ex] 
   & $(k,k',\gamma,T,\rho)=(0,-1,>\gamma_0,>0,\infty)$ & & & & & $\mathrm{stab}_{f}$\\ [0.5ex]
   \hline 
   $(2C)$ &  $\gamma\,\phi+ 2C=2c$ & & & & & \\ [0.5ex] 
   & $(k,k',\gamma,T,\rho,\epsilon)=(0,1,0,>0,<\infty,1)$ & \eqref{gamma_0_t}  & K+const [f.p.] & dS+const [i.f.] & \eqref{gamma_0_t},\eqref{gamma_0_t} & -- \\  [0.5ex]
   \hline 
    $(2D)$ & $\gamma\,\phi+ 2C=2c$ & & & & & \\ [0.5ex] 
     &  $(k,k',\gamma,T,\rho,\epsilon)=(0,1,0,>0,<\infty,-1)$ & \eqref{gamma_0_t} & K+const [f.p.] &  dS+const [i.f.] & \eqref{gamma_0_t},\eqref{gamma_0_t} & -- \\  [0.5ex]
   \hline 
   $(2E)$ &  $\gamma\,\phi+ 2C=2c$ & & & & & \\ [0.5ex] 
   &  $(k,k',\gamma,T,\rho,\epsilon)=(0,1,<\gamma_0,>0,<\infty,1)$ & \eqref{Tkppos_gammaless}  & K [f.p.] &  LM+lin [i.f.] & \eqref{alphaskpone},\eqref{exactkpminus} & -- \\ [0.5ex] 
   &  $(k,k',\gamma,T,\rho,\epsilon)=(0,-1,>\gamma_0,<0,<\infty,1)$ & & & & & $\mathrm{stab}_{f}$ \\ [0.5ex]
   \hline 
    $(2F)$ &    $\gamma\,\phi+ 2C=2c$ & & & & & \\ [0.5ex] 
   & $(k,k',\gamma,T,\rho,\epsilon)=(0,1,<\gamma_0,>0,<\infty,-1)$ & \eqref{Tkppos_gammaless}  & K [f.p.] &  LM+lin [i.f.] & \eqref{alphaskpone},\eqref{exactkpminus} & -- \\ [0.5ex] 
   &  $(k,k',\gamma,T,\rho,\epsilon)=(0,-1,>\gamma_0,<0,<\infty,-1)$ & & & & & $\mathrm{stab}_f$ \\ [0.5ex]
    \hline 
    $(2G)$ & $\gamma\,\phi+ 2C=2c$ & & & & & \\ [0.5ex] 
   & $(k,k',\gamma,T,\rho,\epsilon)=(0,-1,0,<0,<\infty,1)$ & \eqref{metricgamma0-kp}  & K [f.p.] &  F [f.f.] & \eqref{specialkkpnege},\eqref{milnelike} & -- \\  [0.5ex]
   \hline 
   $(2H)$  & $\gamma\,\phi+ 2C=2c$ & & & & & \\ [0.5ex] 
     & $(k,k',\gamma,T,\rho,\epsilon)=(0,-1,0,<0,<\infty,-1)$ & \eqref{metricgamma0-kp}  & K+const [f.p.] &  K+const [f.f.] & \eqref{metricgamma0-kp},\eqref{metricgamma0-kp} & -- \\  [0.5ex]
   \hline 
   $(2I)$ & $\gamma\,\phi+ 2C=2c$ & & & & & \\ [0.5ex] 
   & $(k,k',\gamma,T,\rho,\epsilon)=(0,-1,<\gamma_0,<0,<\infty,1)$ & \eqref{neg_T}  & K [f.p.] &  K [f.f.] & \eqref{alphaskpone2},\eqref{alphaskpone2} & -- \\ [0.5ex] 
   & $(k,k',\gamma,T,\rho,\epsilon)=(0,1,>\gamma_0,<0,<\infty,1)$ & & & & & --\\ [0.5ex]
   \hline 
    $(2J)$ &  $\gamma\,\phi+ 2C=2c$ & & & & & \\ [0.5ex] 
   & $(k,k',\gamma,T,\rho,\epsilon)=(0,-1,<\gamma_0,<0,<\infty,-1)$ & \eqref{neg_T} & K [f.p.] &  K [f.f.] & \eqref{alphaskpone2},\eqref{alphaskpone2} & -- \\ [0.5ex] 
   & $(k,k',\gamma,T,\rho,\epsilon)=(0,1,>\gamma_0,<0,<\infty,-1)$ & & & & & -- \\ [0.5ex]
 \hline\hline
\end{tabular}}
 \end{center}
 \vskip 12pt

 \caption{\small Special cosmologies driven by $k$ and $k'$ or by $k'$ and $T$. When subscripts are present, the stability holds at the corresponding end but not at the other: for example, $\mathrm{stab}_f$ means unstable in the past and stable in the future. The entry ``--'' indicates that the solution is unstable at both ends.}\vskip 12pt
 \label{tab:solsTkp13}
 \end{table}
 \newpage

 The special exact results obtained for cosmologies driven both by curvatures and also by tension are summarized below. The notation is as in the preceding table.
 \begin{table}[hbt!]
 \begin{center}
 \scalebox{0.85}{
\begin{tabular}{ ||c|c||c|c|c|c|| } 
 \hline\hline
  n. & Case & Sol. & Early $t$ & Late $t$ & Stab. \\ [0.5ex] 
  \hline\hline
   $(3A)$ & $A \ \& \ 
   C$ const. & & & & \\ [0.5ex] &  $(k,k',\gamma,T)=(1,1,0,>0) $ & \eqref{kkpTgamma0} & lin $\phi$ [i.p.] &  lin $\phi$ [i.f.]  &  -- \\ [0.5ex]  \hline
     $(3B)$ & $ 
   C$ const. & & & & \\ [0.5ex] &  $(k,k',\gamma,T)=(-1,-1,0,<0) $ & \eqref{adsm1m1} & [f.p.] &  [f.f.] &  -- \\ [0.5ex]  \hline
        $(3C)$ & $ 
   C$ const. & & & & \\ [0.5ex] &  $(k,k',\gamma,T)=(1,1,0,>0) $ & \eqref{ds11} & [i.p.] &  [i.f.] &  stab \\ [0.5ex]  \hline
           $(3D)$ & $ 
   C$ const. & & & & \\ [0.5ex] &  $(k,k',\gamma,T)=(-1,1,0,>0) $ & \eqref{dsm11} & [f.p.] &  [i.f.] &  $\textrm{stab}_f$ \\ [0.5ex]  \hline
   $(3E)$ & $\gamma\,\phi+2 C=2 w, \ C-A = y$ & & & & \\ [0.5ex] 
  &  $(k,k',\gamma,T)=(1,1,<\gamma_0,>0) $ & \eqref{sol_gammanot0kkpT_CT}  & lin. [f.p.] &  lin. [i.f.] & -- \\ [0.5ex] 
   & $(k,k',\gamma,T)=(-1,-1,>\gamma_0,>0)$ & & & & $\mathrm{stab}_f$ \\ [0.5ex]
 \hline\hline
\end{tabular}}
 \end{center}
 \vskip 12pt

 \caption{\small Special Cosmologies driven by $k$, $k'$ and $T$. When subscripts are added, the stability holds at the corresponding end but not at the other: for example, $\mathrm{stab}_f$ means unstable in the past and stable in the future. The entry ``--'' indicates that the solution is unstable at both ends. The details refer to the dynamical systems: for instance case $(3B)$ is $AdS$, and the singularities at finite past and future are due to the coordinate systems.}\vskip 12pt
 \label{tab:solsTkp14}
 \end{table}
\newpage

\section{\sc On Cosmologies and Compactifications} \label{app:cosmo_comp}

Although we focused throughout on cosmological solutions, all the results in this paper can be directly translated into corresponding ones for compactifications, with
\bea
ds^2 &= & e^{2A(r)} \, \ell^2\, ds_{p+1,k}^2 \ + \ e^{2B(r)}\, dr^2 \ + \ e^{2C(r)}\,  \ell^2\, ds_{D-p-2,k'}^2 \ , \nonumber \\
e^\phi &=& e^{\phi(r)} \ ,
\eea
where here $ds_{p+1,k}^2$ indicates a manifold of Lorentzian, rather than Euclidean, signature.

The natural link between the two setups would rest on the analytic continuation $t \to i r$. However, starting from of the preceding cosmological solutions, with
\bea
ds^2 &= & - \ e^{2B(\tau)}\, d\tau^2 \ + \ e^{2A(\tau)} \, \ell^2\, ds_{p+1,k}^2 \ + \ e^{2C(\tau)} \, \ell^2\, ds_{D-p-2,k'}^2 \ , \nonumber \\
e^\phi &=& e^{\phi(\tau)} \ ,
\eea
the structure of the second-order equations and of the Hamiltonian constraint imply that one simply that one can simply replace $\tau$ with $r$, retaining the same functions $A$, $B$, $C$ and $\phi$ and the same parametrizations, while simply mapping the triple according to
\beq
(T,k,k') \ \rightarrow \ (-T,-k,-k') \ .
\eeq
In this fashion, a cosmological solution driven by a positive tension $T$ maps into a compactification driven by a negative $T$, and similarly for the curvatures. The same link holds between cosmic-time solutions and compactifications in Gaussian coordinates. Therefore, finite (infinite) past map into intervals whose left end lies at a finite (infinite) distance, and similarly for far future and right end of the compactification interval. In this sense, the results in this paper also add to the cases previously discussed in~\cite{ms21_susy,ms21_nonsusy,mrs1,mrs2}.

\end{appendices}

\end{document}